\documentclass{aa}  

\usepackage{graphicx,color,url}
\usepackage{natbib}
\usepackage{txfonts}
\newcommand{\kms}{km\,s$^{-1}$}
\begin{document}

   \title{Survival rates of planets in open clusters: The Pleiades, Hyades, and Praesepe clusters}


   \author{M. S. Fujii
          \inst{1}
          \and
          Y. Hori\inst{2}\fnmsep\inst{3}
          }

   \institute{Department of Astronomy, Graduate School of Science, The University of Tokyo, 7-3-1 Hongo, Bunkyo-ku, Tokyo 1130033, Japan\\
              \email{fujii@astron.s.u-tokyo.ac.jp}
         \and
            Astrobiology Center, 2-21-1 Osawa, Mitaka, Tokyo 1818588, Japan \\
            \and
         National Astronomical Observatory of Japan, 2-21-1 Osawa, Mitaka, Tokyo 1818588, Japan\\
             \email{yasunori.hori@nao.ac.jp}
             }

   \date{Received November xx, 2018; accepted xxxx xx, 2018}

 
  \abstract
   {In clustered environments, stellar encounters can liberate planets from their host stars via close encounters. Although the detection probability of planets suggests that the planet population in open clusters resembles that in the field, only a few dozen planet-hosting stars have been discovered in open clusters.}
   {We explore the survival rates of planets against stellar encounters in open clusters similar to the Pleiades, Hyades, and Praesepe and embedded clusters.}
   {We performed a series of $N$-body simulations of high-density and low-density open clusters, open clusters that grow via mergers of subclusters, and embedded clusters. We semi-analytically calculated the survival rate of planets in star clusters up to $\sim$1\,Gyr using relative velocities, masses, and impact parameters of intruding stars.}
   {Less than 1.5\,\% of close-in planets within 1\,AU and at most 7\,\% of planets with 1--10\,AU are ejected by stellar encounters in clustered environments after the dynamical evolution of star clusters.
   If a planet population from 0.01--100\,AU in an open cluster initially follows the probability distribution function of exoplanets with semi-major axis ($a_{\rm p}$) between 0.03--3\,AU in the field discovered by RV surveys ($\propto a_{\rm p}^{-0.6}$), the PDF of surviving planets beyond $\sim10$\,AU in open clusters 
   can be slightly modified to $\propto a_{\rm p}^{-0.76}$.
   The production rate of free-floating planets (FFPs) per star is 0.0096--0.18, where we have assumed that all the stars initially have one giant planet with a mass of 1--13\,$M_{\rm Jup}$ in a circular orbit. The expected frequency of FFPs is compatible with the upper limit on that of FFPs indicated by recent microlensing surveys. Our survival rates of planets in open clusters suggest that planets within 10\,AU around FGKM-type stars are rich in relatively-young ($\lesssim$10--100\,Myr for open clusters and $\sim$1--10\,Myr for embedded clusters), less massive open clusters, which are promising targets for planet searches.}
   {}
   \maketitle
%

\section{Introduction}

Advances in optical and infrared detection capabilities over the past 20 years have allowed us to find small planets with masses or radii comparable to the Earth and directly image giant planets in wide orbits beyond 10\,AU.
Since the first discovery of an exoplanet in 1995, the existence of nearly 4,000 exoplanets has been reported (\url{http://exoplanet.eu}). 

Most stars are believed to be born in clustering environment \citep[e.g.,][]{1993prpl.conf..245L,2000AJ....120.3139C,2003ARA&A..41...57L}.
Except for some open clusters, such clustering stars would become field stars. Are there any differences in the fraction of planet-harboring stars between the field and open clusters?
A number of surveys have monitored stars in young, metal-rich open clusters in order to explore the planet population in star clusters. 
However, the discovery of only approximately ten planets in four open clusters, 
the Hyades \citep{2007ApJ...661..527S,2013arXiv1310.7328Q,2016ApJ...818...46M}, Praesepe (Beehive) \citep{2012ApJ...756L..33Q, 2016A&A...588A.118M} \citep{2012ApJ...756L..33Q}, M67 \citep{2014A&A...561L...9B,2016A&A...592L...1B,2017A&A...603A..85B}, and IC\,4651 \citep{2018A&A...619A...2D}, has been disclosed by RV surveys. A planetary candidate previously reported in NGC~2423 \citep{2007A&A...472..657L} may be a false positive \citep{2018A&A...619A...2D}.
RV surveys in other open clusters, NGC\,2516, NGC\,2422 \citep{2016AJ....152....9B}, and NGC\,6253 \citep{2011A&A...535A..39M}, ended in the non-detection of planet signals.

Except for possible transiting planets in the Praesepe \citep{2008AJ....135..907P}, transit photometry surveys in other thirteen open clusters below failed to find convincing dips produced by planets in light curves:
NGC~6819 \citep{2003MNRAS.340.1287S}, NGC~7789 \citep{2005MNRAS.359.1096B},
NGC~6940 \citep{2005MNRAS.360..791H}, NGC~6633 \citep{2005MNRAS.360..703H}, 
NGC~6791 \citep{2003A&A...410..323B, 2005AJ....129.2856M, 2007A&A...470.1137M},
NGC~7086 \citep{2006AJ....132.2309R}, NGC~2158 \citep{2006AJ....131.1090M}, 
NGC~1245 \citep{2006AJ....132..210B},  NGC~2660, NGC~6208 \citep[e.g.,][]{2005PASP..117..141V},
NGC~2362 \citep{2008MNRAS.387..349M}, M37 \citep{2009ApJ...695..336H}, and
Trumpler 37 \citep{2014arXiv1403.6031E,2014arXiv1403.6020E}.

With {\it Kepler} and {\it K}2 missions, a handful of planets were newly detected in four star clusters: Kepler-66b and 67b in NGC~6811 \citep{2013Natur.499...55M}, five planets (K2-25\,b, K2-136\,b,c,d, and HD\,283869\,b) in the Hyades \citep{2016ApJ...818...46M,2018AJ....155....4M,2018AJ....155..115L,2018AJ....155...10C}, eight planets (K2-95\,b, K2-100\,b,101\,b,102\,b,103\,b,104\,b, and K2-264\,b,c) and one planetary candidate (EPIC\,211901114\,b) in the Preasepe \citep{2016AJ....152..223O, 2017AJ....153...64M,2017AJ....153..177P,2018AJ....156..195R,2018arXiv180901968L}, and K2-231\,b in Ruprecht 147 \citep{2018AJ....155..173C}.

The current status of planet searches in open clusters is suggestive of the rarity of planets orbiting stars that reside in open clusters.
In contrast, \citet{2011ApJ...729...63V} indicated that low detection probabilities of planets in distant open clusters cause the apparent lack of hot Jupiters, compared to the field. \citet{2013Natur.499...55M} suggested that both the orbital properties and the frequency of planets in open clusters are consistent with those in the field of the Milky Way.
\citet{2017A&A...603A..85B} also found that the occurrence rate of hot Jupiters in open clusters is slightly higher than that in the field. Previous planet searches indicated that planets discovered in open clusters appear to have properties in common with those found around the field stars.

\begin{figure}
\centering
\includegraphics[width=\hsize]{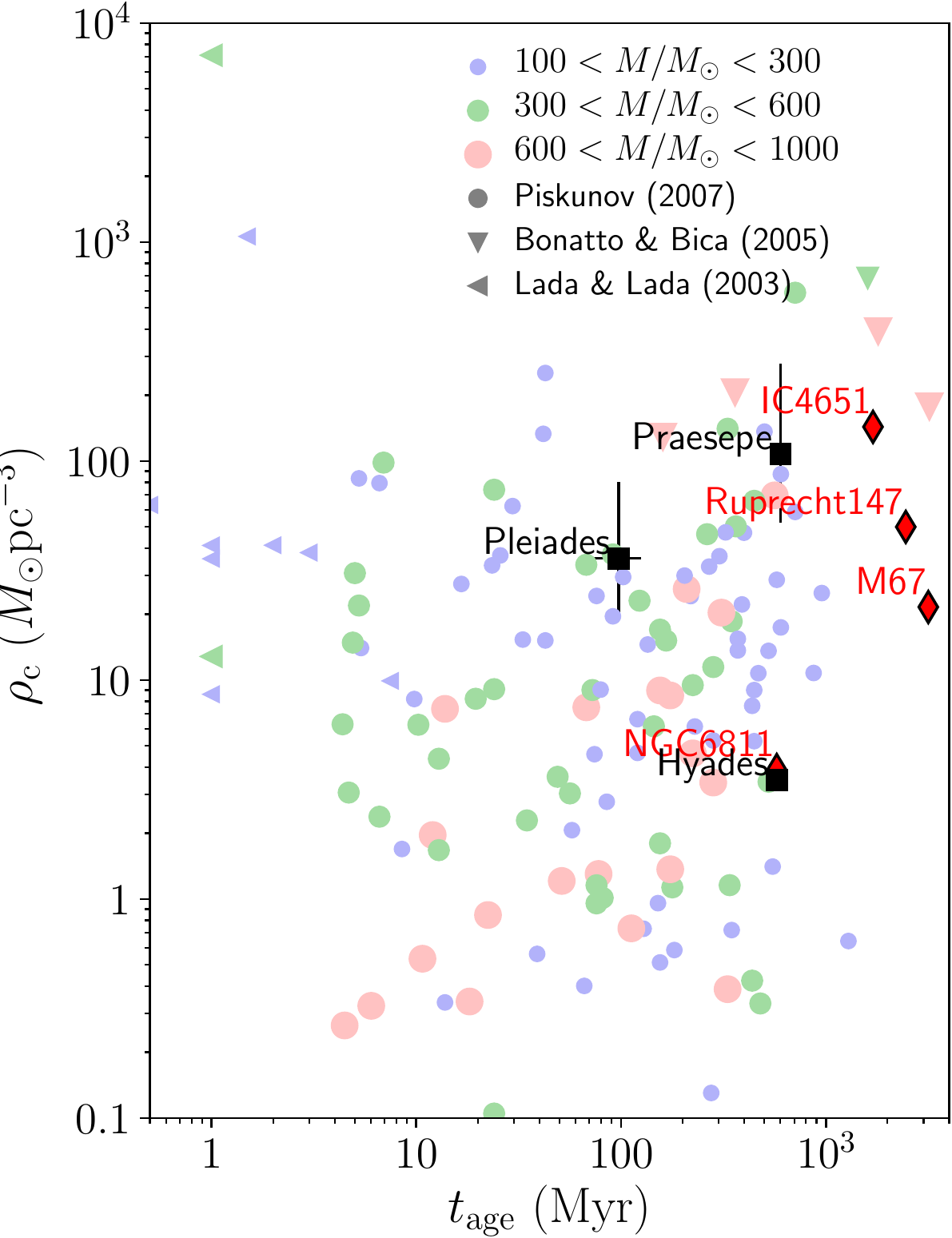}
\caption{Relationship between central densities of open clusters ($\rho_{\rm c}$) and their ages ($t_{\rm age}$). The data are from \citet{2003ARA&A..41...57L} for left-pointing triangles, \citet{2005A&A...437..483B} for down-pointing triangles, and \citet{2007yCat..34680151P} and \citet{2002A&A...386..187M} for circles and red diamonds (planet-harboring clusters). Each color represents the total mass of each open cluster: blue (100--300\,$M_\odot$), green (300--600\,$M_\odot$), and pink (600--1000\,$M_\odot$). Our target clusters and planet-harboring clusters are shown by filled black squares and red diamonds, respectively. 
\label{fig:core_dens_evolution_obs}}
\end{figure}

It is interesting to investigate whether 
stellar encounters can destroy planetary systems in open clusters.
The stellar encounter rate with a distance of $d$ in a star cluster is approximately written as the inverse the time interval between encounters ($t_{\rm enc}$): 
	\begin{eqnarray}
		\frac{1}{t_{\rm enc}} &=& 4\sqrt{\pi} n v d^2 \left(1 + \frac{GM_\star}{v^2 d} \right)\\
		\label{eq_enc}
		&\sim& 1.7 \times 10^{-6} \left(\frac{n}{10^2\,{\rm star\,pc^{-3}}}\right)
			\left(\frac{v}{1\,{\rm km\,s^{-1}}}\right) \left(\frac{d}{{10\,\rm AU}}\right)^2 \nonumber \\
		& \times & \bigg[ 1 + 89 \left(\frac{M_\star}{1\,M_\odot}\right) \left(\frac{v}{1\,{\rm km\,s^{-1}}}\right)^{-2} 
			\left(\frac{d}{{10\,\rm AU}}\right)^{-1} \bigg]\,\,\,{\rm Myr}^{-1},
	\end{eqnarray}
where $n$ is the number density of stars, $v$ is the velocity dispersion of stars, and $M_\star$ is the mass of individual stars in the star cluster \citep{2008gady.book.....B}. Here, we assume a single-component cluster and therefore all star has the same mass. Open clusters have typically central densities of $1$--$10^2\,M_\odot~\mathrm{pc}^{-3}$, ages of $1$--$10^3$\,Myr (see Figure\,\ref{fig:core_dens_evolution_obs}), and a low velocity dispersion of $\leq 1\,\mathrm{km~s}^{-1}$. For open clusters, close encounters within 10\,AU from a host star are expected to occur on timescales of $>100$\,Myr. 
With a stellar mass function and structure of star clusters, however, it is difficult to estimate the close encounter rate precisely using only such an analytic calculation.

Dynamical studies on planetary systems in open clusters started from the origin of Jovian planets with high eccentricities; for example, 16~Cygni Bb in a hierarchical triple star system. Stellar encounters in young clusters can rearrange the orbital configuration of gas giants. \citet{1997A&A...326L..21D} applied $N$-body simulations to study the dynamical evolution of planetary systems including Jupiter-like planets in open clusters. Few Jovian planets on an initially circular obit have eccentric orbits after a four-body interaction between two singles stars and a planetary system. \citet{1998ApJ...508L.171L} also found that even though encounters of binary stars is taken into account, ejection of planets from the systems rarely happens in open clusters.

In fact, ejection rates of planets depend on properties of star clusters. Production rates of free floaters in star clusters change in response to the velocity dispersion and initial stellar density for dense clusters \citep{2001MNRAS.322L...1S}. More massive clusters with larger velocity dispersion enhance orbital disruption of gas giant planets.  In globular clusters, a high ejection rate of planets was obtained from simulations \citep{2002ApJ...565.1251H}. 
Using $N$-body and Monte-Carlo simulations,
\citet{2009ApJ...697..458S} showed that disturbing encounters are rare for planets in open clusters such as the Hyades, whereas \citet{2013ApJ...778L..42P} claimed that $\sim 26$\,\% of stars lose their planets in the Pleiades due to tidal interactions.
 In more recent studies, \citet{2013MNRAS.433..867H} and \citet{2017MNRAS.470.4337C} investigated the ejection of planets from multiplanet systems in open clusters. These results also suggested inner planets ($<10$\,AU), which RV surveys can detect, tend to survive. 
 Especially, short-period gas giants with semi-major axes of $< 0.1$\,AU such as hot Jupiters are unlikely to be stirred up in both young open and globular clusters \citep{2001MNRAS.324..612D,2001MNRAS.322..859B}. In addition,  \citet{2016ApJ...816...59S} claimed that a mutli-giant planet system in star clusters may have a higher frequency of hot Jupiters as a result of dynamical encounter of stars.
 \footnote{With the exception of the pulsar planet, PSR B1620-26b in a globular cluster M4 \citep{1993Natur.365..817B}, photometric surveys in three mature globular clusters, NGC~6397, $\omega$~Centauri, and 47 Tucanae, have failed to find close-in gas giants \citep{2000ApJ...545L..47G,2008ApJ...674.1117W,2012A&A...541A.144N}. 
The frequency of hot Jupiters is less than several percentages around various stars: M-type \citep{2013A&A...549A.109B}, F, G, K-type \citep[e.g.,][]{2011arXiv1109.2497M}, and (evolved) A-type stars \citep{2010ApJ...721L.153J}. Non-detection of hot Jupiters that can survive in the three globular clusters might imply that planet formation can be inhibited in a metal-poor environment and/or in the presence of massive stars, for example, due to intense stellar XUV irradiations.}.

Another possibility for the lack of planets is the failure of planet formation. Disk stripping induced by stellar encounters removes building blocks for gas giant formation. 
\citet{2001MNRAS.322..859B} suggested that stellar encounters within 10\,AU that truncate tidally a planet-forming disk should occur in globular clusters with $>10^5$~stars~pc$^{-3}$. \citet{2014A&A...565A.130B} derived a relation between the encounter distance and final disk size, following the results of their test particle simulations, and examined the disk truncation due to stellar encounters \citep{2015A&A...577A.115V,2016ApJ...828...48V,2016MNRAS.457..313P,2018ApJ...868....1V}. These results suggested that in dense clusters, a significant number of protoplanetary disks can be disrupted by stellar encounters.

Not only the dynamical effect but far-ultraviolet (FUV) and extreme ultraviolet (EUV) radiation from massive stars can disrupt protoplanetary disks in a dense environment of star clusters \citep{2004ApJ...611..360A,2010MNRAS.402.2735E,2010ApJ...723L.113Y}. \citet{2006ApJ...641..504A} suggested that the effect of photoevaporation
is relatively small in small clusters with members of 100--1000. \citet{2018MNRAS.478.2700W} also estimated the disruption of protoplanetary disks in dense clusters with a number density of $10^4$\,pc$^{-3}$ and found that in such a dense environment, photoevaporation due to FUV irradiations works stronger than dynamical encounters. In addition, there is a well-known positive correlation between the frequency of giant planets and stellar metallicity in the field \citep{2005ApJ...622.1102F,2010PASP..122..905J}. Planet formation might have been quenched there in the globular clusters under extremely metal-poor environments rather than open clusters.

We investigate whether stellar encounters on planetary systems in a clustered environment can influence dynamical stability and detectability of exoplanets in open clusters, using {\it N}-body simulations.
Most of previous studies assumed a spherical star cluster under virial equilibrium with a central density similar to those of observed open clusters and modeled the cluster as an equal-mass system \citep{2009ApJ...697..458S,2013ApJ...778L..42P}.
The initial state of open clusters, however, may be neither spherical nor virial equilibrium \citep{2010MNRAS.407.1098A}.
For example, \citet{2012MNRAS.419.2448P} considered a fractal initial distribution of stars that mimics star-forming regions. They found that $\sim10$\,\% of planets at 30\,AU were liberated from their host stars in such substructured clusters. 
\citet{2015MNRAS.453.2759Z} also performed $N$-body simulations starting from fractal distributions of stars and found that at most $\sim 30$\,\% of planets at 100\,AU are ejected from their host planets.
Similarly, \citet{2012ApJ...753...85F} suggested that the merger scenario of a subclustered initial condition is preferable for the observed young massive clusters. 
We therefore consider three types of cluster models (single open clusters, embedded clusters, and merged clusters). For single open cluster models, we adopt classical low-density models similar to \citet{2009ApJ...697..458S}, high-density, and subclustered models that reproduce present-day properties of our target clusters. 

In this paper, we perform a series of $N$-body simulations of star clusters, including stellar evolution, and examine survival rates of planetary systems against stellar encounters in the Pleiades, Hyades, and Praesepe-type clusters. 
This work extends previous studies on the fate of planetary systems in a star cluster by long-term {\it N}-body simulations of open clusters and estimate the survival fraction of planets for each spectral type of stars in open clusters.
We focus on three known clusters because a large amount of data such as stellar positions for individual stars are available for them. We consider planetary systems with semi-major axes of 0.1--1000\,AU under three types of cluster models, which have a stellar population based on the initial mass function: single open clusters, embedded clusters, and merged subclusters. In the next section, we describe each cluster model and the treatment of stellar close encounters with planetary systems. In Section 3, we present the dynamical evolution of the star clusters during 600\,Myr and evaluate stellar encounter rates that cause orbital disruption of planetary systems. 
We also show ejection rates of planets from the system as functions of the orbital separation of planets and stellar types.
Based on a realistic distribution of planets derived from RV and direct-imaging surveys, 
we discuss the occurrence rates of free-floating planets and the survival rates of planetary systems in star clusters. In Section 4. We also discuss prospects for Doppler and photometric surveys of planet-hosting stars in nearby open clusters, including the Pleiades. We summarize our results in the last section.

\section{Methods}
We perform a series of $N$-body simulations of single and multiple
star clusters in order to investigate effects of close encounters on 
the orbital configuration of planetary systems in star clusters. Specifically,
we focus on typical open clusters,
which have characteristics similar to those of the present-day Pleiades,
Hyades, and Praesepe (see Table~\ref{tb:obs} for their physical properties),
and embedded clusters \citep{2003ARA&A..41...57L}. 

We summarize our models and methods in this study.
\begin{enumerate}
	\item We consider open clusters with two initial central densities ($16$ and $1.6 \times 10^4\,M_\odot\,{\rm pc}^{-3}$)
		and a growing cluster via the merger of subclusters. All these models dynamically evolve to a surface density distribution similar to observed open clusters.
		We also consider an embedded cluster which has the mass enclosed within $\sim$1\,${\rm pc}$ of $\sim$100\,$M_\odot$ \citep{2003ARA&A..41...57L}.
	\item Following the initial mass function (IMF) of stars \citep{2001MNRAS.322..231K},
		we randomly assign stellar masses to individual objects in a star cluster.
	        The IMF extends from $0.08\,M_\odot$ to 15, 10, and $8\,M_{\odot}$ for open-cluster, merger, and embedded-cluster models, respectively.
	\item We numerically calculate the position and velocity of an individual star in a clustered environment.
		The retention or ejection of a planet orbiting a host star is evaluated semi-analytically based on relative velocities, masses, and impact parameters of intruding stars.          
	\item We estimate the semi-major axis distribution of planets that survived stellar encounters and the number of produced free-floating planets, where we assume that each star initially has a single planet following the mass-semi-major axis distribution of exoplanets derived from both RV and direct-imaging surveys.

\end{enumerate}
In order to test our methods, we also performed simulations for a Plummer
model \citep{1911MNRAS..71..460P} with a single-mass 
component of $1\,M_{\odot}$, which is a model adopted in previous work
by \citet{2009ApJ...697..458S}.
In this study, the orbital evolution of both planets and stars were
calculated in $N$-body simulations. We compared our result with theirs
and confirm that the results are consistent (see Appendix \ref{Plummer}
for the details).

\subsection{Models}

We adopt initial conditions which evolve to star clusters similar 
to Pleiades at 100\,Myr and Hyades and Praesepe at 600\,Myr. 
We also model embedded clusters in accordance with
a recent hypothesis that open clusters
are merger remnants of smaller clusters \citep{2012ApJ...753...85F,
2012ApJ...754L..37S,2015MNRAS.449..726F,2015PASJ...67...59F} in which most of stars were born \citep{2003ARA&A..41...57L}.

\subsubsection{Open cluster models}
For open cluster models, we adopt King's model \citep{1966AJ.....71...64K} with $W_0=3$.
We consider two models with different initial core densities:
$\rho_{\rm c} =$ $1.6\times 10^4$ and $16\,M_{\odot}\,{\rm pc}^{-3}$
which we hereafter call the two models "w3-hd" (high density) and "w3-ld" (low density), respectively.
The core density of open clusters, including three clusters of our interest, 
is typically $1 - 10^2\,M_{\odot}\,{\rm pc}^{-3}$ (see Figure~\ref{fig:core_dens_evolution_obs}).
However, some recent studies suggested that open clusters once had a higher core density and then, their core densities decreased to the level as low as those in well-known open clusters via the dynamical evolution \citep{2012A&A...543A...8M,2016ApJ...817....4F}.
We therefore adopt an initially dense model (model w3-hd) as well as the low density model (model w3-ld). 
The total mass of these models is $1015\,M_{\odot}$.
We assume that the clusters are initially in virial equilibrium.

\begin{figure}
\centering
\includegraphics[width=\hsize]{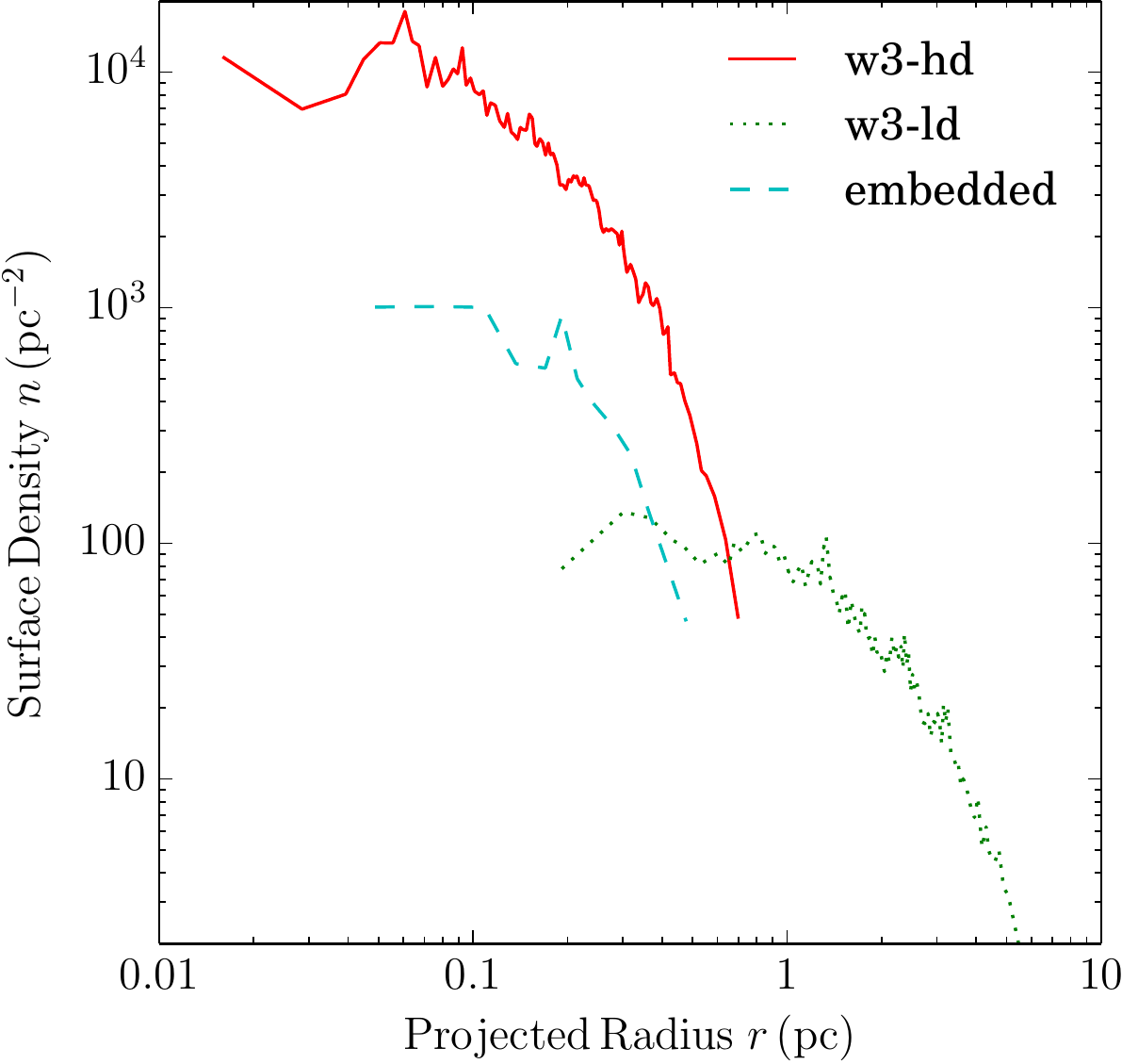}
\caption{Radial profiles of the initial surface number density in our cluster models. 
\label{fig:density_prof_init}}
\end{figure}

The initial density of model w3-hd is comparable to that of young massive clusters such as NGC 3606 with $> 6\times 10^4\,M_{\odot}\,{\rm pc}^{-3}$ \citep{2008ApJ...675.1319H}.
The known densest open cluster in the Milky Way, the Arches cluster,  
has $\sim 10^5\,M_{\odot}\,{\rm pc}^{-3}$ \citep{2002A&A...394..459S,2010ARA&A..48..431P}.  Theoretically, \citet{2012A&A...543A...8M} suggested that open clusters initially had a high density, while \citet{2014MNRAS.445.4037P} claimed that such a high density is not necessary for all open clusters and that subclustered initial conditions fit to the observed open clusters better. 
In Figure \ref{fig:density_prof_init}, we present the initial profiles of the surface number density in our open cluster models. For comparison with observations, we show the surface number density rather than the volume density. The core surface density of models w3-hd and w3-ld are $10^4$ and $100M_{\odot}$\,pc$^{-2}$, respectively.

\begin{table*}
\caption{Data from Observations \label{tb:obs}}
\centering
\begin{tabular}{lcccccccccc}
\hline \hline
Name           & $M\, (M_{\odot})$      &$r_{\rm c}\, ({\rm pc})$  & $r_{\rm h}\, ({\rm pc})$  & 
$r_{\rm t}\, ({\rm pc})$ & Age (Myr) &$t_{\rm rlx}\,({\rm Myr})$\tablefootmark{a} &$D$ (pc)\tablefootmark{b}  &
Ref. \\
\hline
Pleiades & 720--950 & 1.3--2.1 & 1.9--2 & 16--19  & 70--125& 15 & 125 & 1, 2, 3, 4 \\ 
Hyades & $435$ & 3.1 & 4.1 & 10.5  & 500-650 & 35 & 46.5 & 5, 6 \\
Praesepe & $\sim600$ & 0.8--3.5 & 3.9--4 & 11.8--16 &  600--900 & 39 & 170 & 1, 3, 7\\
\hline
\end{tabular}
\tablefoot{
$r_{\rm c}$, $r_{\rm h}$, and $r_{\rm t}$ are the core, half-mass, and tidal radii of each cluster.
\tablefoottext{a}{The relaxation times, $t_{\rm rlx}$, are derived from the current mass ($M$) and $r_{\rm h}$.}
\tablefoottext{b}{Distance from the Sun.}}
\tablebib{
(1) \citet{2012A&A...543A...8M} and references therein; (2) \citet{2008ApJ...678..431C}; (3) \citet{2007yCat..34680151P}
(4) \citet{1998A&A...329..101R};
(5) \citet{2011A&A...531A..92R}; (6) \citet{2009A&A...497..209V};
(7) \citet{2002AJ....124.1570A}}

\end{table*}

\begin{table*}
\caption{Simulation Models \label{tb:models}}
\centering
\begin{tabular}{lccccccccccc}
\hline \hline

Model           & $M$  &$r_{\rm c}$ &
$r_{\rm h}$  & $r_{\rm t}$ &
$W_{0}$  &
$\rho_{\rm c}$  &
$N$          & $m_{\rm max}$    &
$t_{\rm rlx}$ & $t_{\rm rlx,c}$ &
$N_{\rm run}$
\\
           & $ (M_{\odot})$    &
$({\rm pc})$  & $({\rm pc})$  & $({\rm pc})$ &
 &
$(M_{\odot}\,{\rm pc}^{-3})$  &    &
  & $(M_{\odot})$    &
$({\rm Myr})$ & $({\rm Myr})$ 
\\
\hline
\multicolumn{12}{c}{Embedded cluster model} \\
embedded & 122 & 0.21 & 0.27 & 1.0 & 3 & $2.3\times 10^3$  & 256 & 8 & 2.1 & 1.1 & 56 \\
\multicolumn{12}{c}{Open cluster model} \\
w3-ld & 1015 & 2.3 & 2.9 & 11    & 3 & 16             & 2048 & 15 & 123 & 46 & 20 \\
w3-hd & 1015 & 0.22 & 0.28 & 1.1 & 3 & $1.6\times 10^4$  & 2048 & 15 & 3.7 & 1.3 & 10 \\ 
merger-r1 & 976 & - & - & 1       & - & - & 2048 & 10 & - & - & 5 \\ 
merger-r2 & 976 & - & -  & 2        & -  & - & 2048 & 10 & - & - & 6 \\ 
\hline
\end{tabular}
\end{table*}

\subsubsection{Merger models}

Recent studies on star cluster formation suggested that open clusters may have grown via mergers of subclusters \citep{2012ApJ...753...85F,2012ApJ...754L..37S,2015MNRAS.449..726F} or fractal distribution of stars \citep{2014MNRAS.438..620P,2014MNRAS.445.4037P}.
The existence of multiple clumps has been observed in star forming regions  \citep{2004MNRAS.348..589C,2017MNRAS.465.1889G}.
For example, $\rho$ Oph main cloud has six identified clumps, in which pre-stellar cores with dozens of solar masses are included
\citep{1998A&A...336..150M,2007MNRAS.379.1390S}. The typical size of the clumps is $\sim0.3$\,pc and the distance among them is $<1$\,pc.
If stars successfully form in each clump, the cluster system likely evolves to the ensemble of embedded clusters. We therefore set up merger models (merger-r1 and -r2) from multiple subclusters (embedded clusters).

We model an embedded cluster with the total mass of $122\,M_{\odot}$ using a King's model \citep{1966AJ.....71...64K} with $W_{0}=3$.
The half-mass and tidal radius are 0.27\,pc and 1\,pc, respectively. 
The initial central density of our embedded cluster model is $2.3\times 10^3\,M_{\odot}\,{\rm pc}^{-3}$. This value is equivalent to the highest density
of observed embedded clusters (see Appendix \ref{apdx_star}). 
The embedded cluster model is also initially in virial equilibrium.
Our embedded cluster model is named "embedded" hereafter.
All parameters are summarized in Table \ref{tb:models}, and the initial surface density profile of model embedded is presented in Fig. \ref{fig:density_prof_init}.

We deploy randomly eight subclusters (model embedded) with zero relative velocity within
a sphere with radius of 1\,pc or 2\,pc for models merger-r1 and merger-r2, respectively.
Since merger events occur after the core collapse is triggered inside individual subclusters,
both models resemble the so-called "late-merger'' ones \citep{2013MNRAS.430.1018F}.
These merger models have a total mass comparable to those of open-cluster models.

\subsubsection{Initial mass function}
We randomly assign masses to individual stars irrespective of their initial positions. The stellar population in our cluster models follows Kroupa IMF with a lower mass cut-off of $0.08\,M_{\odot}$ \citep{2001MNRAS.322..231K}.
The IMF has an upper mass cut-off of 8, 15, and $10\,M_{\odot}$ for embedded cluster, open cluster, and merger models, respectively. This mass spectrum yields a mean stellar mass of $\sim 0.5\,M_{\odot}$, 
which is slightly higher than that estimated in observed clusters, $0.36\,M_{\odot}$ \citep{2006MNRAS.365.1333W}.

\citet{2006MNRAS.365.1333W} suggested that $m_{\rm max} \sim 30\,M_\odot$
in a star cluster with mass of $10^3\,M_{\odot}$.
Since the number of star with mass of $\sim 10M_{\odot}$ is expected to be at most unity
in open cluster models with $N = 2048$,
we adopt half or one third of their value (10 or 15\,$M_{\odot}$) as the upper cut-off of stellar mass.
For embedded cluster model with mass of $10^2\,M_{\odot}$, we use a smaller value ($8\,M_{\odot}$) as the upper cut-off of stellar mass.
The resulting number of stars for each spectral type (M, K, G, F, A, and B stars) is
$N_{\rm M}=1493$, $N_{\rm K}=293$, $N_{\rm G}=28$, $N_{\rm F}=60$, $N_{\rm A}=53$, and
$N_{\rm B}=71$ for open cluster models, respectively. 
The embedded cluster model has $N_{\rm M}=187$, $N_{\rm K}=37$, $N_{\rm G}=9$,
$N_{\rm F}=8$, $N_{\rm A}=7$, and $N_{\rm B}=8$.
Here, we classify stellar spectral types based on stellar masses:
M: $0.08 < m < 0.45$, K: $0.45 < m < 0.8$, G: $0.8 < m < 1.04$, F: $1.04 < m < 1.4$, 
A: $1.4 < m < 2.1$, and B: $2.1 < m < 16$, where $m$ is the mass of stars in units of $M_{\odot}$. We note that there is no O-type star in our simulations.

Although we adopted the upper-mass limit for the IMF based on \citet{2006MNRAS.365.1333W} in order to avoid the uncertainty of stellar evolution models for massive stars, this assumption may be vulnerable, if more massive stars statistically can form in small clusters. Actually, observations have found low-mass clusters including O-type stars \citep{2010ApJ...725.1886L} and relatively isolated O-type stars in star forming regions \citep{2012A&A...542A..49B}. Massive O-type stars can form stellar-mass black holes \citep{2000MNRAS.315..543H}, and they can eject planets from their surrounding stars. While such massive stars and stellar-mass black holes usually stay in the cluster core due to the mass segregation, most of low-mass stars, which mainly host planets, are located in cluster halo. Therefore, the lack of O-type stars in our simulation would not affect much the ejection of planets from low-mass stars.

We assume that the primordial binary population is null, because the planet formation around stars with a companion strongly depends on the orbital parameters and mass ratios of the binary system rather than stellar encounter in star clusters.
Observationally, companion frequencies in open clusters are estimated to be $\sim65$\,\% for FGK-type stars and $\sim 35$\,\% for 0.1--0.5$M_{\odot}$ stars \citep{2013ARA&A..51..269D}. Since the typical orbital separation of observed FGK-type binaries is 45\,AU \citep{2013ARA&A..51..269D}, it is smaller than typical distances of stellar encounters which are expected to occur in open clusters. Around such binaries, formation processes of circum-binary planets would be suppressed rather than dynamically disrupted. In addition, gravitational binding of these binaries is so strong (i.e., hard binary) that they dynamically behave as a single star with the binary mass to the other stars \citep{2009PASJ...61..721T}. In such a case, the dynamical evolution of star clusters seems not to be much different from that without primordial binaries. Thus, we consider cluster models with no primordial binaries for simplicity. 

\subsubsection{Relaxation time}
The dynamical evolution of star clusters such as core collapse proceeds on a timescale of the two-body relaxation.
The half-mass relaxation time is given as
\begin{eqnarray}
t_{\rm rh} = \frac{0.138Nr_{\rm h}^{3/2}}{G^{1/2}M^{1/2}\ln \Lambda},
\label{eq:t_rlx}
\end{eqnarray}
where $N$ is the number of stars, $r_{\rm h}$ is the half-mass radius,
and $M$ is the total mass of the cluster. The Coulomb logarithm is 
written as $\ln \Lambda=\ln \gamma N$, and $\gamma \sim 0.1$ 
for the cluster with single-mass components \citep{1987degc.book.....S}.

The core-collapse time ($t_{\rm cc}$) is known to be characterized by the relaxation timescale, specifically, $t_{\rm cc}\sim$15--20$t_{\rm rh}$ for single-component clusters\citep{1987degc.book.....S}. However, mass segregation in multiple mass-component systems accelerates the core collapse. \citet{2004ApJ...604..632G} suggested that the relaxation time of the cluster core ($t_{\rm rlx, c}$) rather than the half-mass one can be a good indicator that characterizes the dynamical evolution of star clusters. The core relaxation time is given by
\begin{eqnarray}
  t_{\rm rlx, c} = \frac{0.065 v_{\rm c}^3}{G^2m^2n_{\rm c}\ln \Lambda _{\rm c}},
\end{eqnarray}
where $v$ is the velocity dispersion, $m$ is the mean stellar mass, 
$n$ is the number density of stars.
The subscript $c$ indicates the value in the core and 
$\Lambda _{\rm c}=\gamma_{\rm c}N$. 
Following the description in \citet{2004ApJ...604..632G},
we find that $\gamma_{\rm c}=0.015$, $mn_{\rm c}=\rho_{0}$, and
$v_{\rm c}=\sigma_{0}$, where $\sigma_0$ follows 
$r^2_{\rm c}\simeq 9\sigma_{0}^2/(4\pi G\rho_{0})$ \citep{1996MNRAS.279.1037G}.
\citet{2004ApJ...604..632G} suggested that $t_{\rm cc}$ is also scaled by
$(m_{\rm max}/\mu)^{-1.3}$, where $\mu$ is the mean stellar mass and $m_{\rm max}$ is the maximum stellar mass. For our models, $\mu \simeq 2$--30. \citet{2014MNRAS.439.1003F} found a similar relation that
$t_{\rm cc} \propto (m_{\rm max}/\mu)^{-1}$.
These results give the expected core collapse time 80, 2.5, and 1\,Myr for models w3-ld, w3-hd, and embedded, respectively. Models w3-hd and embedded collapse are expected to collapse before the supernova explosion which causes a significant mass loss of the cluster (see Appendix \ref{apdx_star}).
The core and the half-mass relaxation times of our models are summarized in Table \ref{tb:models}.

\subsection{$N$-body simulations}

We perform $N$-body simulations of stellar dynamics in open clusters, using a sixth-order Hermite scheme
with individual time-steps \citep{2008NewA...13..498N}. 
We adopt an accuracy parameter $\eta$ \citep{2008NewA...13..498N} 
of 0.13--0.2, maintaining less than 
$\sim0.1$\,\% of the energy error of the system throughout the entire simulations, that is 600\,Myr for the open cluster
models and 170\,Myr for the embedded cluster model. 
Since small clusters are expected to be tidally disrupted within 100\,Myr and most of the encounter occurred in the first a few Myr, the dynamical evolution of the embedded cluster is no longer simulated after 170\,Myr. 

We take into account collisions between stars and effects of stellar evolution in the same way as \citet{2000MNRAS.315..543H}, assuming the solar metallicity. We adopt a sticky-sphere approach to describe collision events between stars using the zero-age main-sequence (ZAMS) radii of stars \citep{2000MNRAS.315..543H}. 
We found that collision events between stars rarely happen in our simulations, specifically, a few collisions only in the high-density model.
We assume that stars lose their masses at the end of their main-sequence (MS) stages and evolve into compact objects (neutron stars or white dwarfs) with their core masses (see also Appendix \ref{apdx_star} for the details).
Most of stars in our cluster models are low-mass objects (M,F,G-type stars) whose lifetimes during the MS stage are over 1\,Gyr, and B-type and A-type stars occupy only $\sim 6\,\%$ of stellar components in our cluster models.
Consequently, the choice of the stellar metallicity has a minor impact on the dynamical evolution of open clusters discussed in this study.

We ignore the orbital decay and expansion and ejection of planets associated with stellar evolution.
As a star evolves from a MS stage to a giant phase,
it loses its angular momentum via mass loss driven by a stellar wind.
A ram pressure of the stellar wind, however, has little effect on the orbital evolution of planets \citep[e.g.,][]{1998Icar..134..303D}. 
Stars with masses of $\lesssim 2.3M_\odot$ undergo the helium flash at the red-giant branch and their outer envelope expand rapidly. 
The strong tidal force acted on planets orbiting such evolved stars causes 
the engulfment of short-period planets by their host stars.
Meanwhile, the reduced gravity of an inflated star pushes outer planet outward 
\citep[e.g.,][]{2009ApJ...705L..81V,2011ApJ...737...66K}.
In fact, the lack of short-period planets, the so-called the period valley,
is seen around evolved A-type stars in the field \citep{2007ApJ...665..785J, 2008PASJ...60.1317S}. However, the lifetimes of stars with masses of $\lesssim 2.3\,M_\odot$ are comparable to or longer than 1\,Gyr, namely, our simulation time. Furthermore, the stellar mass loss has non-negligible effects on the stability of planetary 
systems around more massive stars
\citep[e.g.,][]{2011MNRAS.417.2104V,2013MNRAS.430.3383V} and multiple star 
systems \citep{2012MNRAS.422.1648V} in this study. 
If supernova explosions occur in the aftermath of dying stars, 
planets can be readily ejected under the severe circumstances 
\citep{2011MNRAS.417.2104V}. However, such desperate events can rarely happen in 
our open cluster models because of the scarcity of O, B-type stars. Any 
velocity kick is not taken in account during the stellar evolution either.

Since our modeled clusters stay within the Jacobi radius in the Milky Way galaxy,
the tidal field of the background galaxy in our simulations is not included.
Here, we consider a star cluster with the orbital velocity of $V_{\rm c}=220\,$\kms\,near the location of the Sun ($\sim 8$\,kpc) away from the Galactic center.
The density within the Jacobi radius 
in an isothermal halo with a constant circular velocity $V_{\rm c}$ is 
written as a function of distance from the Galactic center $R_{\rm G}$ (see Appendix B in \citet[][]{2011MNRAS.413.2509G} for the details),
	\begin{equation}
		\rho_{\rm J} = \left( \frac{3}{2\pi G} \frac{V_{\rm c}^2}{R_{\rm G}^2} \right) 
			\simeq 5.376 \left( \frac{R_{\rm G}}{{\rm kpc}} \right)^{-2}\,{\rm g\,cm^{-3}},
		\label{eq:rho_J}
	\end{equation}
where $r_{\rm J}$ is the Jacobi radius.
Using Eq.\,(\ref{eq:rho_J}) and $\rho_{\rm J}=M/(4\pi r_{\rm J}^3/3)$,
we obtain $r_{\rm J} = 14\,{\rm pc}$ for $R_{\rm G}=8$\,kpc and $M=10^3\,M_{\odot}$
($r_{\rm J} = 6.6\,{\rm pc}$ for $M=100\,M_{\odot}$),
where $M$ is the mass enclosed within $r_{\rm J}$.
Our clusters are initially confined within this Jacobi radius.
As the halo of a star cluster expands with time, the outskirts of a cluster halo overflow the Jacobi radii. 
The stars beyond the Jacobi radius are tidally stripped from the cluster. 
In our simulations, the fraction of such escaping stars is expected to be less than 10\% (see Appendix \ref{tidal} for the details).

\subsection{Ejection of planets}

We identify the ejection of planets in a semi-analytically way, based on the relative velocity, mass, and impact parameter of nearby passing stars.
Since it is too computationally expensive to simultaneously simulate the orbital evolution of planets in a dynamically-evolving cluster, we did not integrate 
the orbits of planets in this study.

We examine the maximum distance at which 
a planet can dwell in its original habitat after the encounter.
Gravitational interactions between a star-planet system and passing objects are approximated by the impulse force, if
the separation between a planet and its host star is small enough as compared to the impact parameters of passing stars. This distant-tide approximation is valid for close encounters seen in open clusters. 

Analytic approaches given in \citet{2008gady.book.....B} allow us to estimate the change in the energy of an individual planetary system per close encounter. 
The change in the velocity of a star-planet system due to one close encounter of a passing star is given by
\begin{equation}
\Delta {\boldmath v} = \frac{2GM_\star}{{\boldmath b}^2 {\boldmath V}} {\boldmath b},
\end{equation}
where $G$ is the gravitational constant, $\Delta \boldmath v$ is the change in the velocity of the planetary system, $M_\star$ is the mass of the passing star, ${\boldmath V}$ is the relative velocity between the passing star and the planetary system, and ${\boldmath b}$ is the impact parameter.
The change in the energy per unit mass of the star-planet system, $\Delta E$, due to a close encounter is given by 
\begin{equation} 
	\Delta E = \frac{2G^2 M^2_{\star}}{V^2 b^4}  a_{\rm p}^2,
\end{equation}
where $a_{\rm p}$ is the semi-major axis of the planet.
Also, the relative energy change of the planetary system is written as 
\begin{eqnarray}
\frac{\Delta E}{|E|} = \frac{4GM_\star^2 a_{\rm p}^3}{V^2b^4M_{\rm tot}},
\label{eq:de}
\end{eqnarray}
where $|E|$ is the energy of an individual planetary system per unit mass, $E = -GM_{\rm tot}/2a_{\rm p}$ and 
$M_{\rm tot}$ is the total mass of a planetary system, that is the sum of the planet's and its host star's mass. 
Mass ratios of observed planets to their host stars as small as $\lesssim 0.01$, except for the case of planetary systems around brown dwarfs.
The effect of stellar encounters in this study is insensitive to the mass of a planet, that is to say, 
$M_{\rm tot} \approx M_{\rm s}$, where $M_{\rm s}$ is the mass of a planet-hosting star.

We assume that $\Delta E/|E| = 1$ leads to the orbital disruption of planetary 
systems. From Eq.\,(\ref{eq:de}), we obtain
\begin{eqnarray}
a^*_{\rm p} = \left( \frac{V^2b^4M_{\rm tot}}{4GM_\star^2} \right)^{\frac{1}{3}}.
\label{eq:a}
\end{eqnarray}
We measure $b$ and $V$ for each close encounter of the nearest neighbor star. 
We record the minimum value of $a^*_{\rm p}$, which is defined as $a_{\rm{min}}$. 
Planetary systems continuously undergo the passage of stars beyond the
closest distance.
However, the cumulative energy change of an individual planetary system
due to weaker encounters is smaller than the energy change 
caused by the closest encounter.
Thus, we consider that planets within $a_{\rm{min}}$ can survive. 

\section{Results}

We simulated the dynamical evolution of our open cluster models during $\sim 1$\,Gyr.
The embedded cluster model was not calculated after 170\,Myr, because it has evolved to a low-density cluster at 170\,Myr and then it is expected to be tidally disrupted.  
We carried out several runs for each cluster model with different random seeds for the positions and velocities of individual particles in order to achieve better statistics:
$N_{\rm run} = 56$ for the embedded cluster, $N_{\rm run} = 20$ for open cluster models, and $N_{\rm run} = 5$--6 for the merger models because of their higher calculation costs, in terms of CPU time. 
Hereafter, errors denote standard deviations of a full set of runs ($N_{\rm run}$).

\subsection{Structure of star clusters after dynamical evolution}
While model w3-hd experienced core-collapse at a few Myr and the cluster density decreases with time, the core density of model w3-ld did not change much during 1\,Gyr (see Appendix \ref{dynamical_ev} for the time evolution of the core radius and density).  The outer region of model w3-ld, however, extended as well as model w3-hd. In merger models, subclusters merged within a few Myr and formed a cluster (see Appendix \ref{merger} for the snapshots). This merger timescale is consistent with previous $N$-body simulations modeling the formation of open clusters starting from a fractal distribution of stars \citep{2014MNRAS.438..620P}. 

\begin{figure*}
\centering
\includegraphics[width=0.45\hsize]{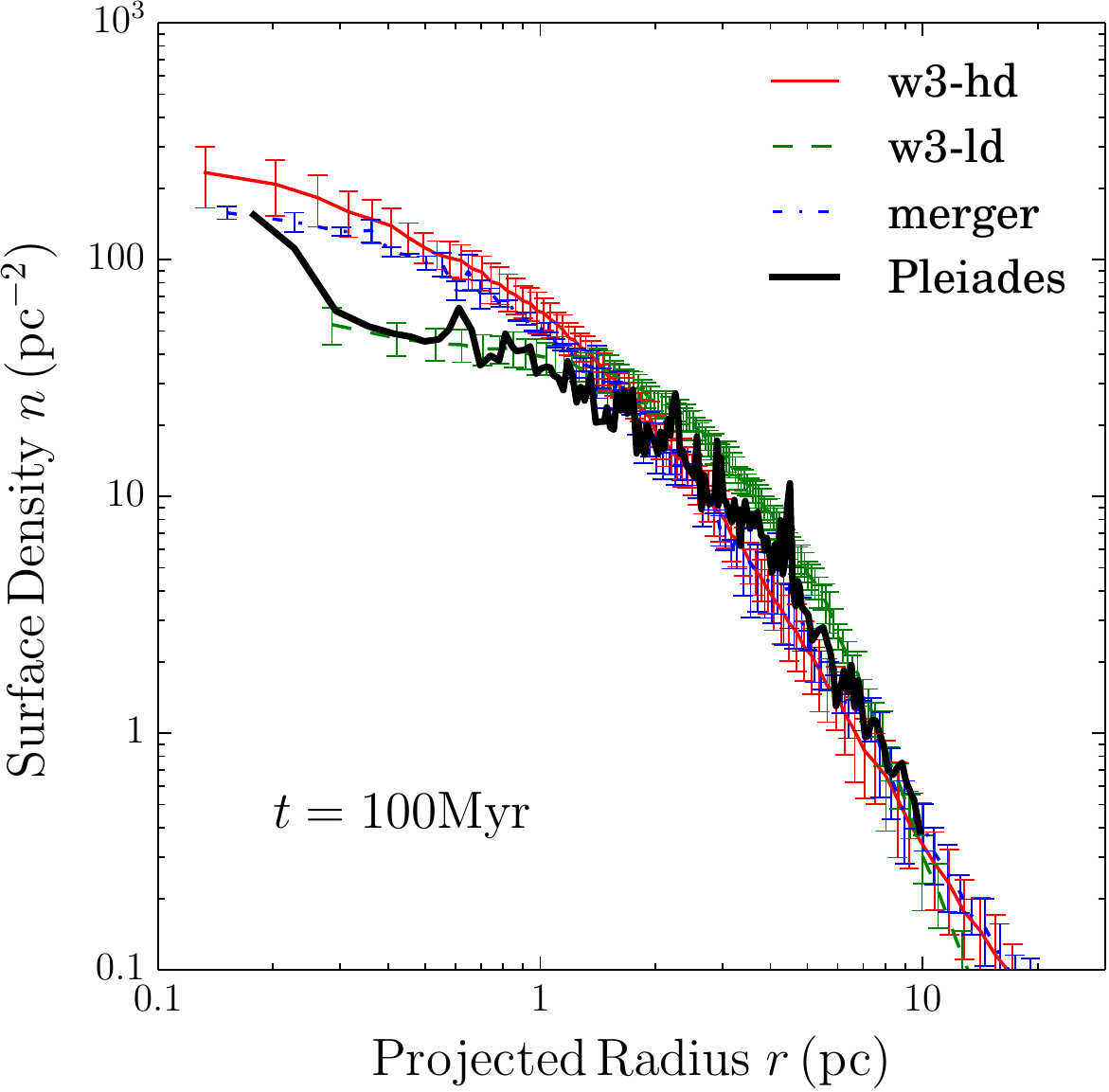}
\includegraphics[width=0.45\hsize]{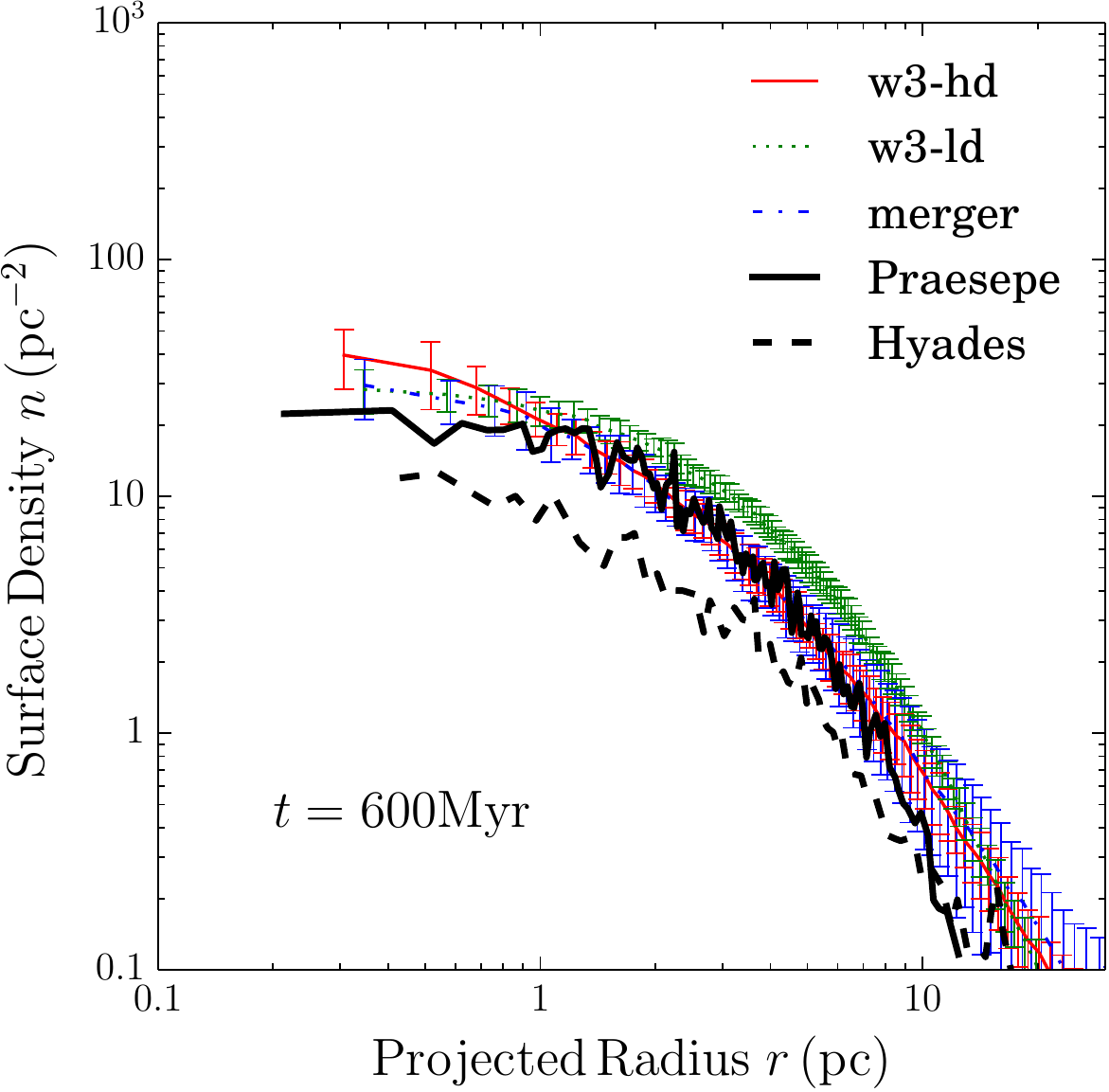}\\
\caption{Surface number density profiles as a function of projected radius
  of open-cluster models at $t = 100$\,Myr (left) and $t = 600$\,Myr (right). 
  Black curves show radial profiles of the surface number of density derived
  from two-dimensional observational data of stellar positions in the
  Pleiades (left; solid), Praesepe (right; solid), and Hyades (right; dashed).
Error bars indicate the run-to-run variations.
\label{fig:density_prof_2k}}
\end{figure*}

We calculated the surface density profile of the models using the method of \citet{1985ApJ...298...80C},
in which the local densities are calculated from the six nearest neighbors. 
We estimated the core densities and the core radii of
the simulated clusters and observed open clusters; the Pleiades, Hyades, and Praesepe.
Figure \ref{fig:density_prof_2k} shows these results at $t=100$ and 600\,Myr as old as the ages of the Pleiades, Hyades, and Praesepe (see Table \ref{tb:obs}).
We also show the surface density profiles of the planet-hosting open clusters in Figure \ref{fig:density_prof_2k}.
We measured the core radii of the Pleiades, Hyades, and Praesepe using the stellar positions in the three open clusters from open catalogs provided by VizieR; \citet{2007ApJS..172..663S,2009yCat..21720663S} for 
the Pleiades, \citet{2007AJ....134.2340K} for the Hyades, and 
\citet{2011A&A...531A..92R} and \citet{2011yCat..35319092R} for the Praesepe.
We obtained the core radius of 1.7\,pc, 2.6\,pc, and 2.4\,pc for the Pleiades, Hyades, and Praesepe, respectively. We note that these values are comparable to those derived from King's model fitting in previous studies, 1.3--2.1\,pc, 3.1\,pc, and 0.8--3.5\,pc (see Table \ref{tb:obs}).
Thus, we confirmed that the surface density profiles of our cluster models are similar to those of the Pleiades, Hyades, and Praesepe regardless of the initial stellar distributions of the cluster models.

\subsection{Ejection rates of planets in star clusters}

During each simulation, we recorded the minimum value of $a^*_{\rm p}$ (see Eq.\,(\ref{eq:a})) for each star.
We define this value as $a_{\rm min}$. 
We assume that planets inside $a_{\rm min}$ can survive, otherwise
they are immediately ejected by stellar encounters.
Thus, we obtain the rates of stellar encounters as a function of time and the semi-major axis of planets ($a_{\rm p}$).

\begin{figure*}
\centering
\includegraphics[width=0.4\hsize]{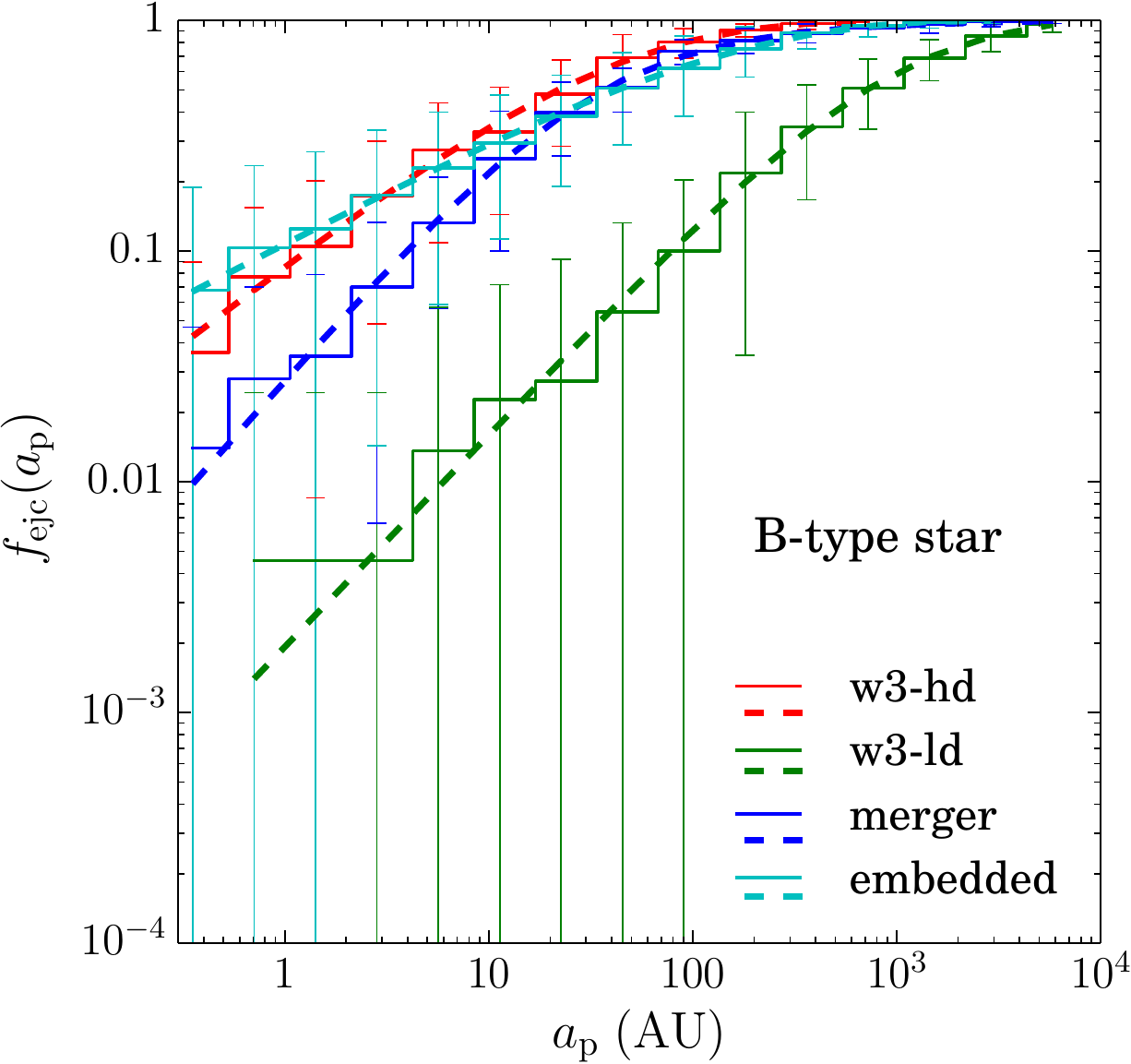}\includegraphics[width=0.4\hsize]{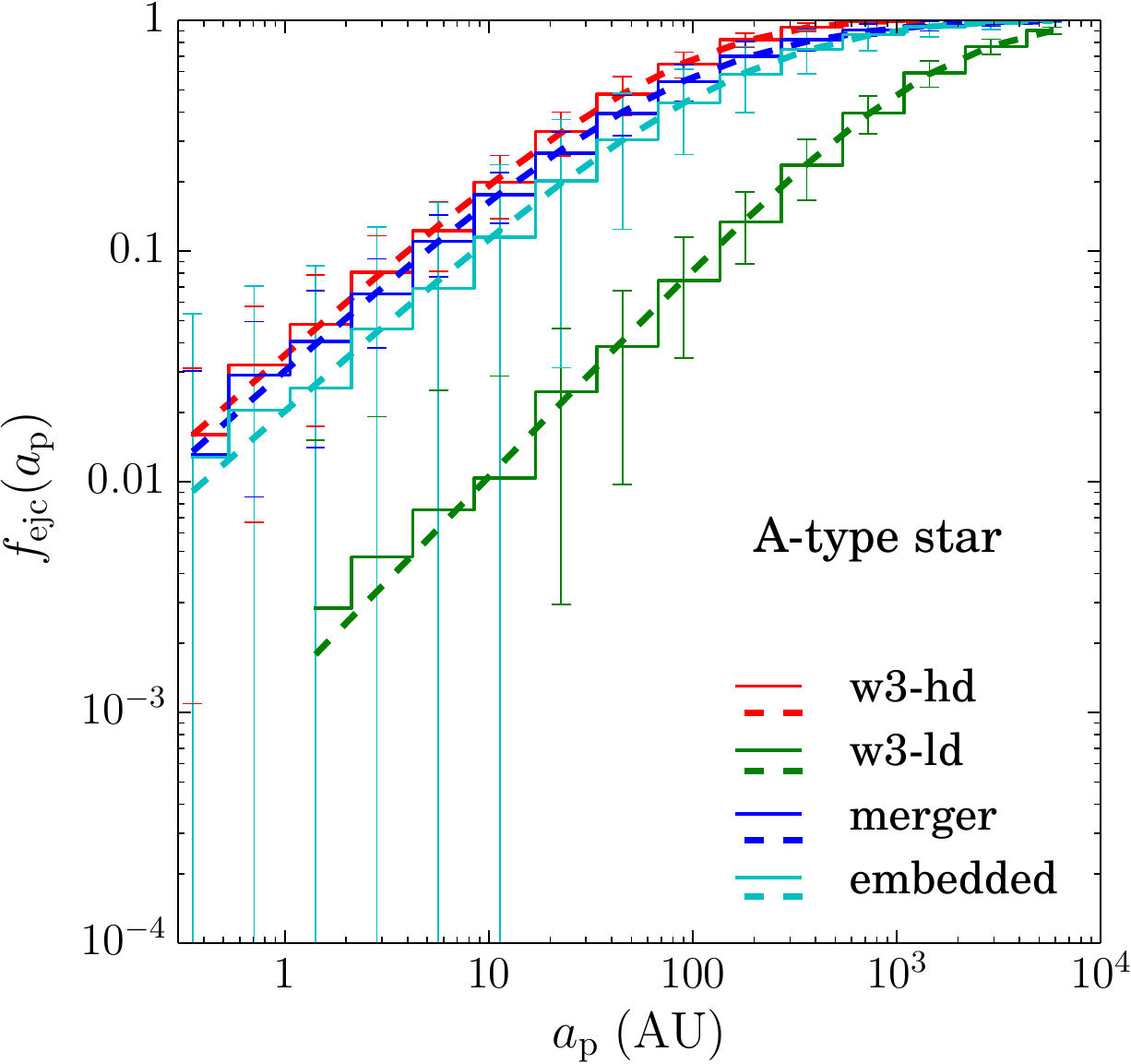}\\
\includegraphics[width=0.4\hsize]{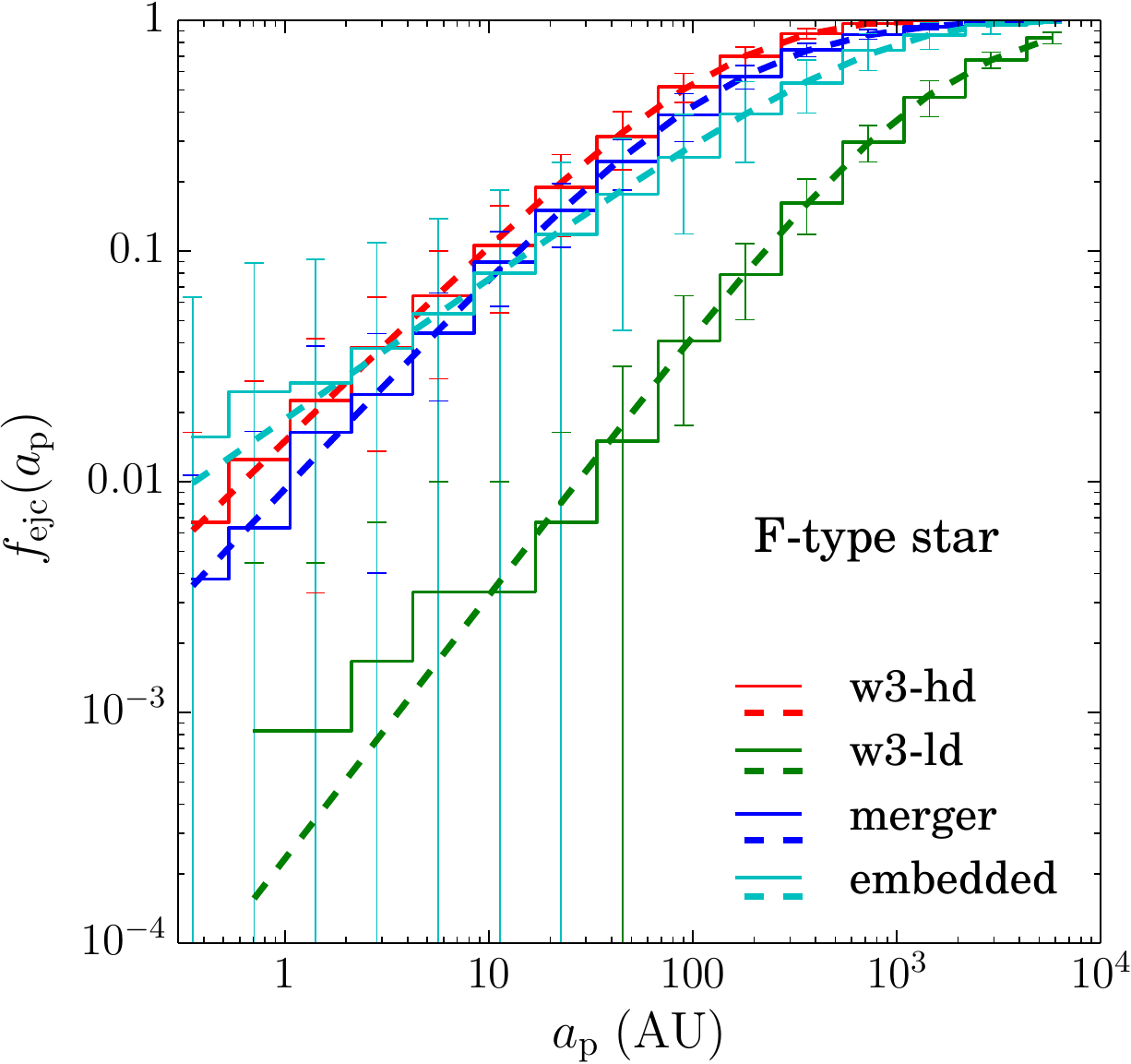}\includegraphics[width=0.4\hsize]{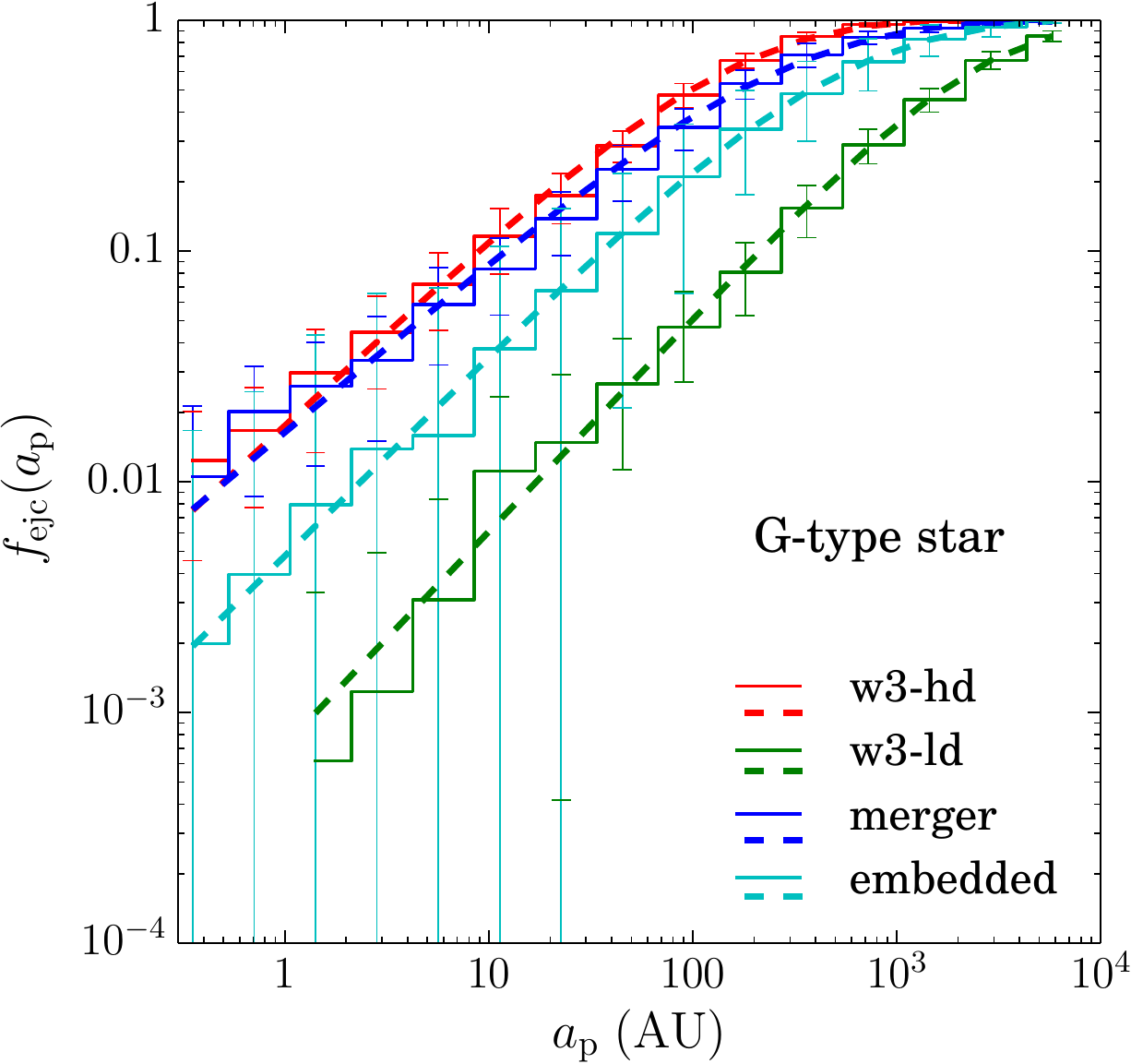}\\
\includegraphics[width=0.4\hsize]{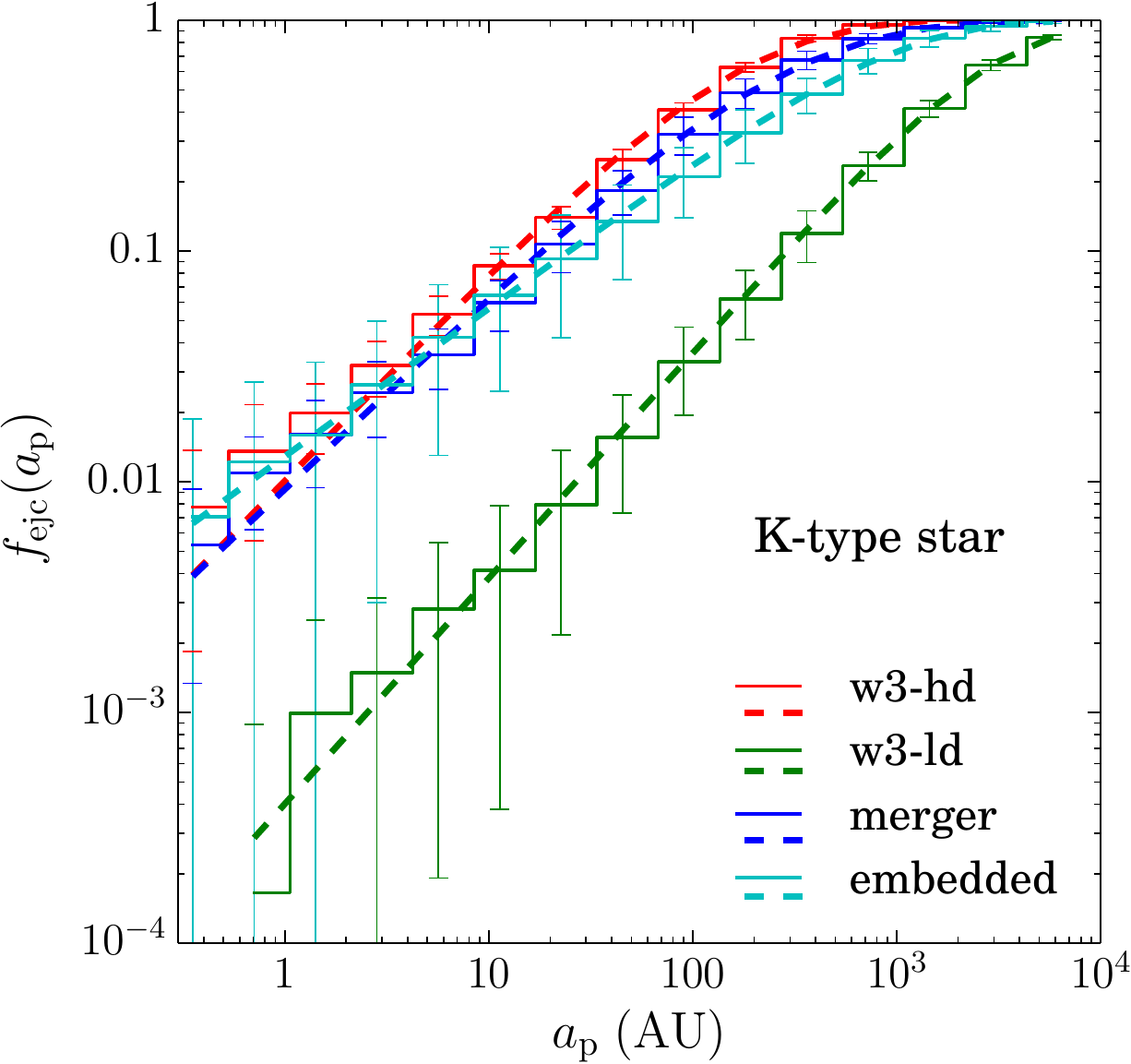}\includegraphics[width=0.4\hsize]{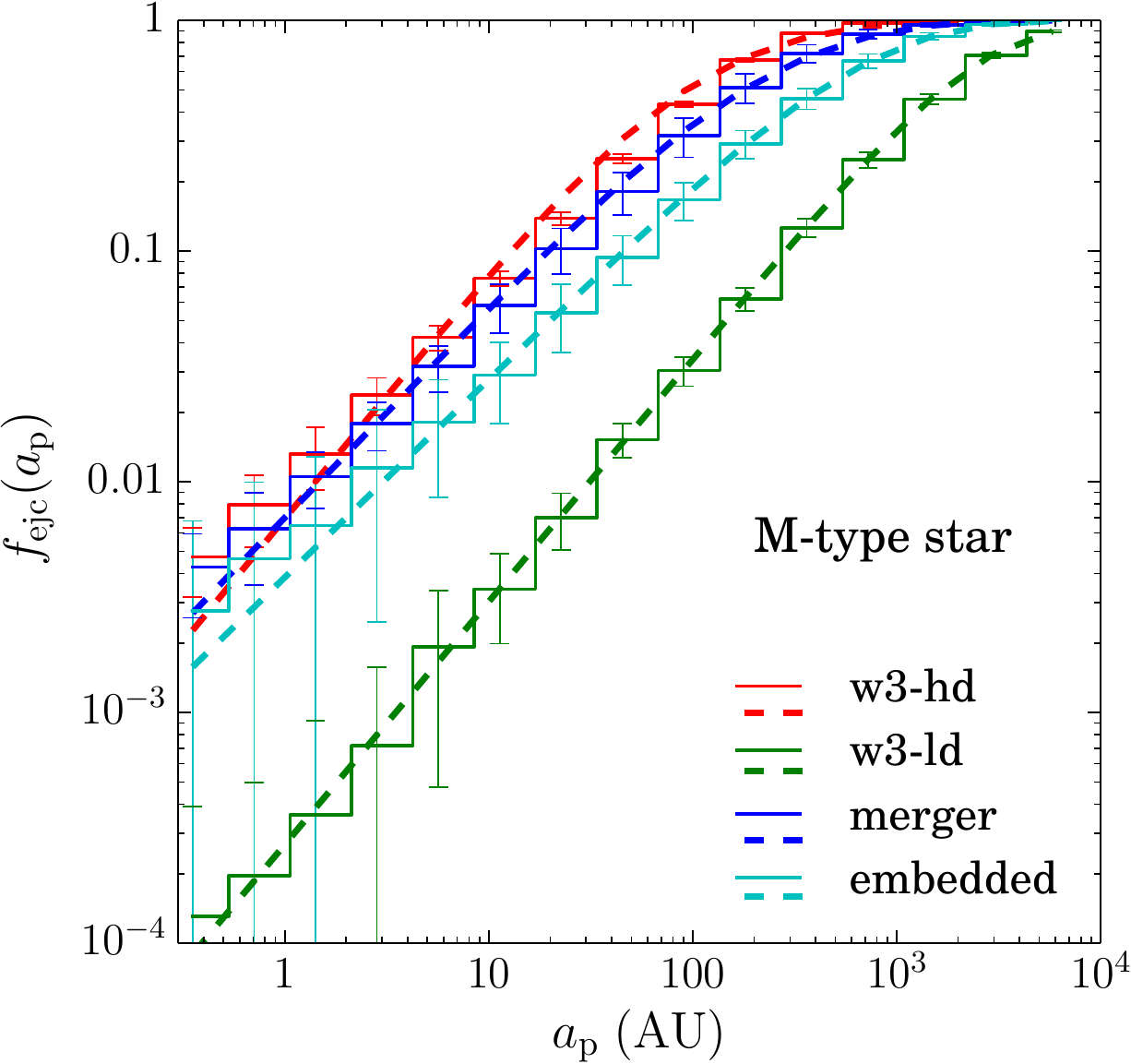}\\
\caption{Ejection rates of planet as a function of the semi-major axis
  of the planet ($a_{\rm p}$) orbiting each spectral type of star at $t=170$\,Myr
  for the embedded-cluster model and at $t=600$\,Myr for the other models.
  Error bars indicate run-to-run variations (standard deviations).
  Dashed curves are the fitting function given by Eq. (\ref{eq:esc_frac}).
  We classify stellar spectral types according to stellar mass as follows:
  M: $0.08 < m < 0.45$, K: $0.45 < m < 0.8$, G: $0.8 < m < 1.04$,
  F: $1.04 < m < 1.4$ A: $1.4 < m < 2.1$, and B: $2.1 < m < 16$,
  where $m$ is the mass of the (planet-hosting) star in units of $M_{\odot}$.
  \label{fig:f_a_min_type}}
\end{figure*}

The ejection rates of planets depend on 1) the semi-major axis of planets,
2) mass of the host star, and 3) elapsed time. Figure \ref{fig:f_a_min_type} shows the ejection rates of planets orbiting each spectral type of star ($f_{\rm ejc}(a_{\rm p})$)
as a function of $a_{\rm p}$ at $t=170$\,Myr for the embedded cluster and at $600$\,Myr for the open cluster models. 
The ejection rate of planets with $a_{\rm p}$ is the cumulative fraction of planets further than $a_{\rm min}$ that are expected to be ejected due to the stellar encounters.
As seen in the Figure \ref{fig:f_a_min_type}, the ejection rates of planets within 10\,AU
is only a few percent for all the models. This fact means that most of exoplanets observed by both RV and transit surveys can survive even in high-density clusters.

\begin{figure}
\centering
\includegraphics[width=\hsize]{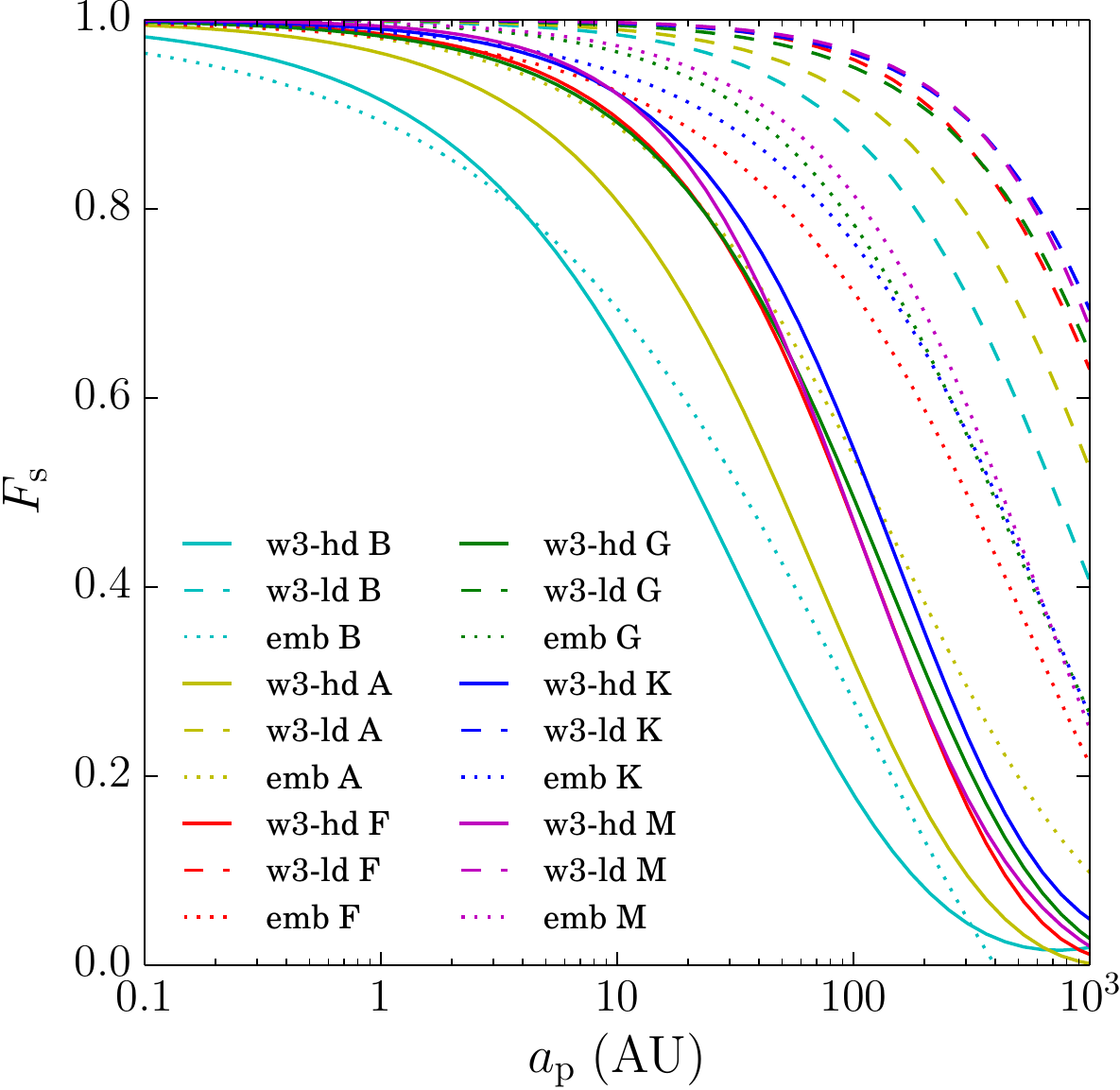}
\caption{Survival fraction of planets around each spectral type of star as a function of $a_{\rm p}$ ($F_{\rm s}=1-f(a_{\rm p})$; see Eq. \ref{eq:esc_frac}) for w3-hd and w3-ld models at $t=600$\,Myr and the embedded cluster model at $t=170$\,Myr.}
\label{fig:F_survive}
\end{figure}

We found a power-law function that reproduces the ejection rates of planets around each spectral type of star in each cluster model:
\begin{eqnarray}
f(a_{\rm p})=Aa_{\rm p}^{\beta}/(C+Da_{\rm p}^{\delta}).
\label{eq:esc_frac}
\end{eqnarray}
The best-fit parameters for models w3-hd, w3-ld, and embedded are summarized in Table \ref{tb:parameters}. We note that the survival fraction of planets as a function of $a_{\rm p}$ is defined as $F_{\rm s}(a_{\rm p})=1-f(a_{\rm p})$ (see Figure \ref{fig:F_survive} for $F_{\rm s}$).
Stars in a low-density environment lose only a few percent of their planets, whereas planets at $\sim 10$\,AU ($\sim 100$\,AU) are lost around $\sim 10\,\%$ (the half) of stars in the densest environment.
The fraction of surviving planets becomes significantly
smaller as the stellar mass increases.

\begin{table*}
\caption{The best-fit parameters for ejection rates of planets\label{tb:parameters}}
\centering
\begin{tabular}{lccccc}
\hline \hline

Models  & $A$   &$\beta$   & $C$ & $D$ & $\delta$ \\
\hline
B                 &               &       &           &       &         \\
w3-hd             & 0.139 & 0.685 & 1.57 & 0.0663 & 0.780 \\
w3-ld             & 0.00177 & 0.922 & 0.919 & 0.00101 & 0.970 \\
embedded          & 0.0108 & 0.466 & 0.0982 & 0.00256 & 0.616 \\
\hline
A                 &               &       &           &       &         \\
w3-hd             &  0.136 & 0.776 & 3.78 & 0.0602 & 0.874 \\
w3-ld             &  0.00271 & 0.908 & 2.08 & 0.000669 & 1.05 \\
embedded          & 0.00415 & 0.776 & 0.202 & 0.00277 & 0.816 \\
\hline
F                &               &       &           &       &         \\
w3-hd             &  0.0761 & 0.859 & 5.05 & 0.0252 & 0.993 \\
w3-ld             &  $2.34\times 10^{-5}$ & 1.14 & 0.101 & $3.55\times 10^{-5}$ & 1.09 \\
embedded          &  0.00875 & 0.612 & 0.468 & 0.000858 & 0.845 \\
\hline
G                 &               &       &       &       &        \\
w3-hd             & 0.111 & 0.815 & 6.32 & 0.0440 & 0.921 \\
w3-ld             & 0.000921 & 0.926 & 1.28 & 0.000127 & 1.12 \\
embedded          &  0.0696 & 0.868 & 14.5 & 0.0415 & 0.915 \\
\hline
K                 &               &       &       &       &                  \\
w3-hd             & 0.252 & 0.905 & 24.9 & 0.105 & 1.01 \\
w3-ld             & $6.15\times 10^{-5}$ & 0.979 & 0.153 & $2.73\times10^{-6}$ & 1.29 \\
embedded          & 0.0129 & 0.643 & 1.00 & 0.000872 & 0.916 \\
M         &               &       &       &       &                  \\
\hline
w3-hd             &  0.00551 & 1.08 & 0.790 & 0.00450 & 1.10 \\
w3-ld         &    $7.65\times 10^{-5}$  &   1.05 & 0.286 & $3.59\times 10^{-6}$ & 1.37  \\
embedded          &  0.106 & 0.860 & 27.6 & 0.0229 & 1.02 \\
\hline
FGKM         &               &       &       &       &                  \\
w3-hd       &  0.0132 & 1.01 & 1.71 & 0.00730 & 1.07 \\ 
w3-ld       &  $8.92\times 10^{-5}$ & 1.03 & 0.285 & $4.11\times 10^{-6}$ & 1.34 \\
embedded    &  0.00503 & 0.762 & 0.769 & 0.000598 & 0.981 \\
\hline
\end{tabular}
\end{table*}

As seen in Figure \ref{fig:f_a_min_type}, the ejection rates of planets depend on the stellar mass. One possible reason for that is the mass segregation. Since individual stars in a star cluster attempt to equalize their kinetic energy with surrounding objects, massive stars lose their energy and sink to the dense cluster center due to the mass segregation. As a result, massive stars such as B- and A-type stars experience close encounters more frequently than low-mass stars do. However, \citet{2016MNRAS.459L.119P} showed that in open clusters formed from subclusterd initial conditions, massive stars have a velocity dispersion similar to that of low-mass stars. This suggests that mass segregation may not be the only reason for the high ejection rates of planets from massive stars. Another possibility is gravitational focusing. 
If we assume that the velocity dispersion does not depend on the stellar mass, the encounter timescale for a certain distance is roughly anticorrelated with the planet-hosing star's mass in the gravitational focusing regime.
As is shown in Eq. (\ref{eq_enc}), the close encounter timescale depends on the mass of a planet-hosting star.
Thus, stellar encounters closer than $\sim 10$\,AU in typical open clusters should be gravitationally focused (see eq. (\ref{eq_enc})).

\begin{figure*}
\centering
\includegraphics[width=0.45\hsize]{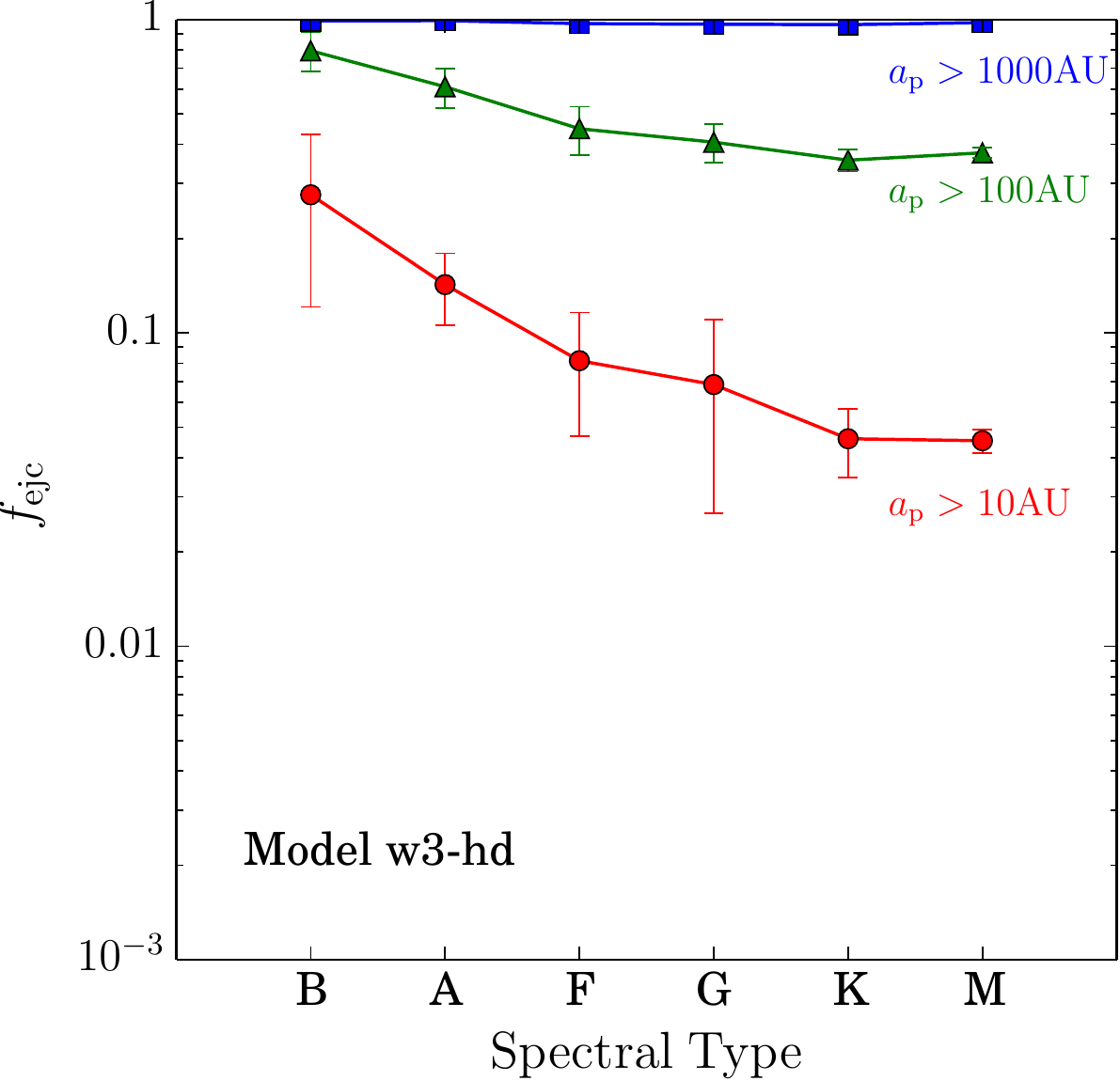}\includegraphics[width=0.45\hsize]{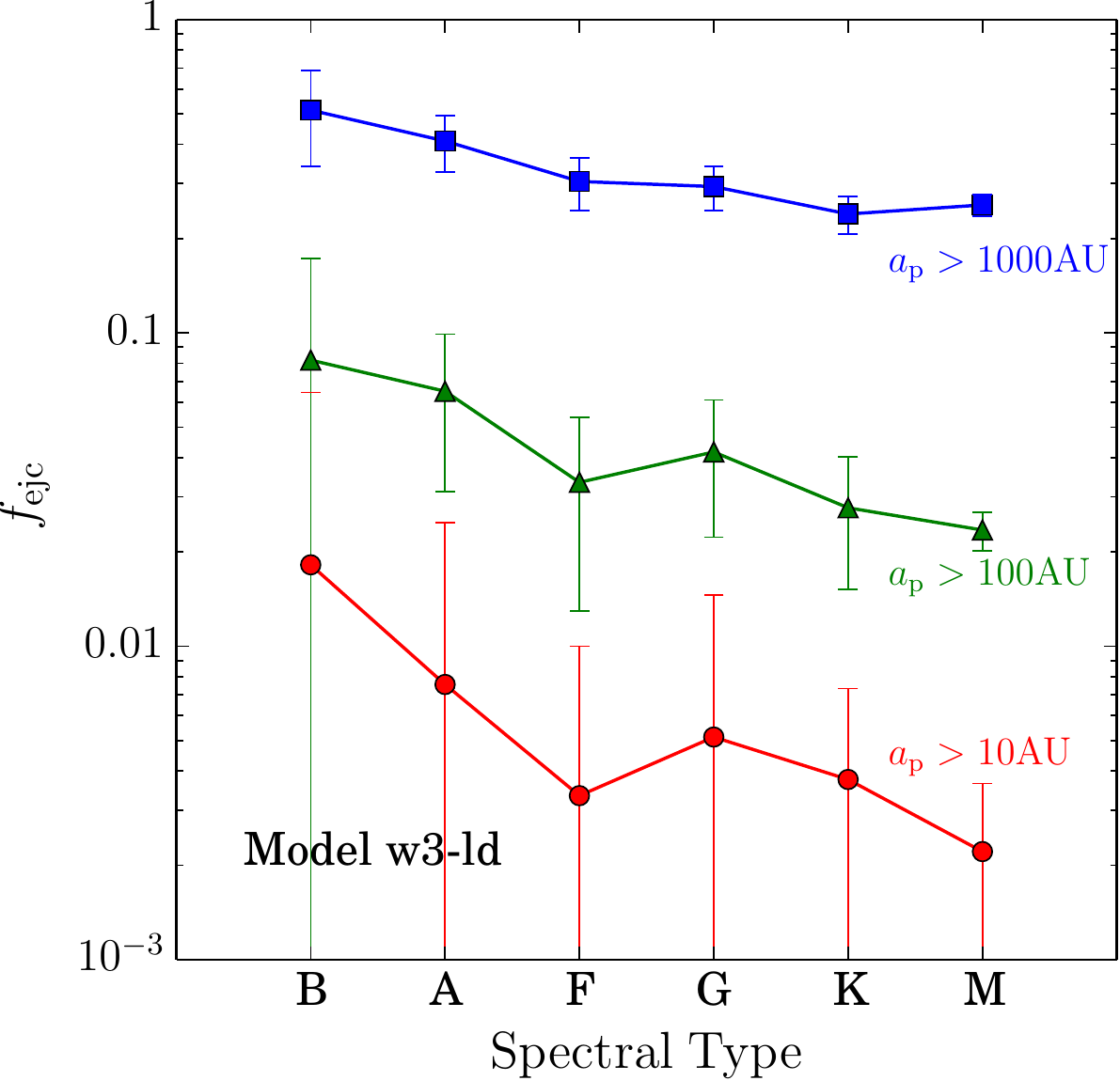}\\
\includegraphics[width=0.45\hsize]{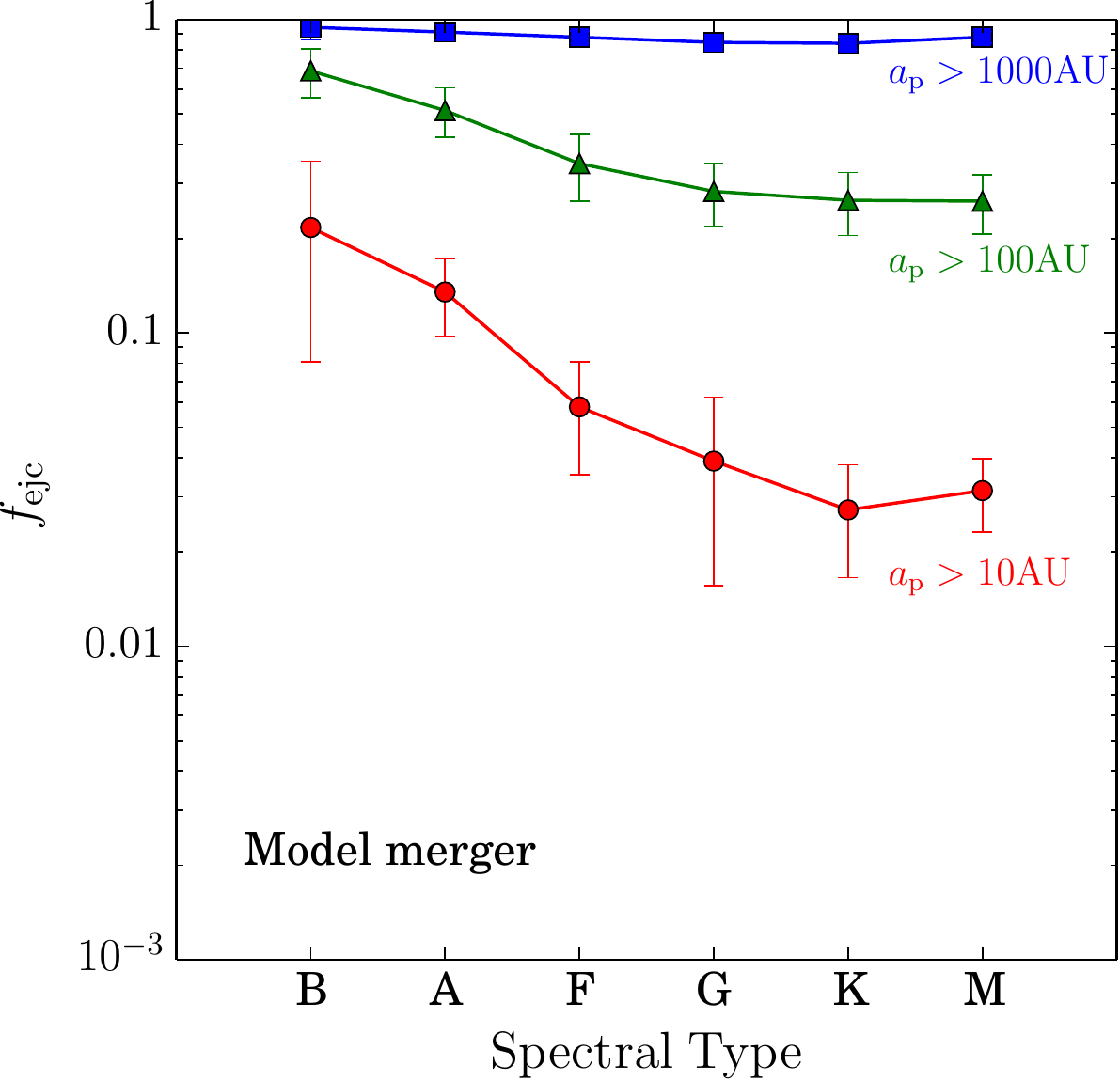}\includegraphics[width=0.45\hsize]{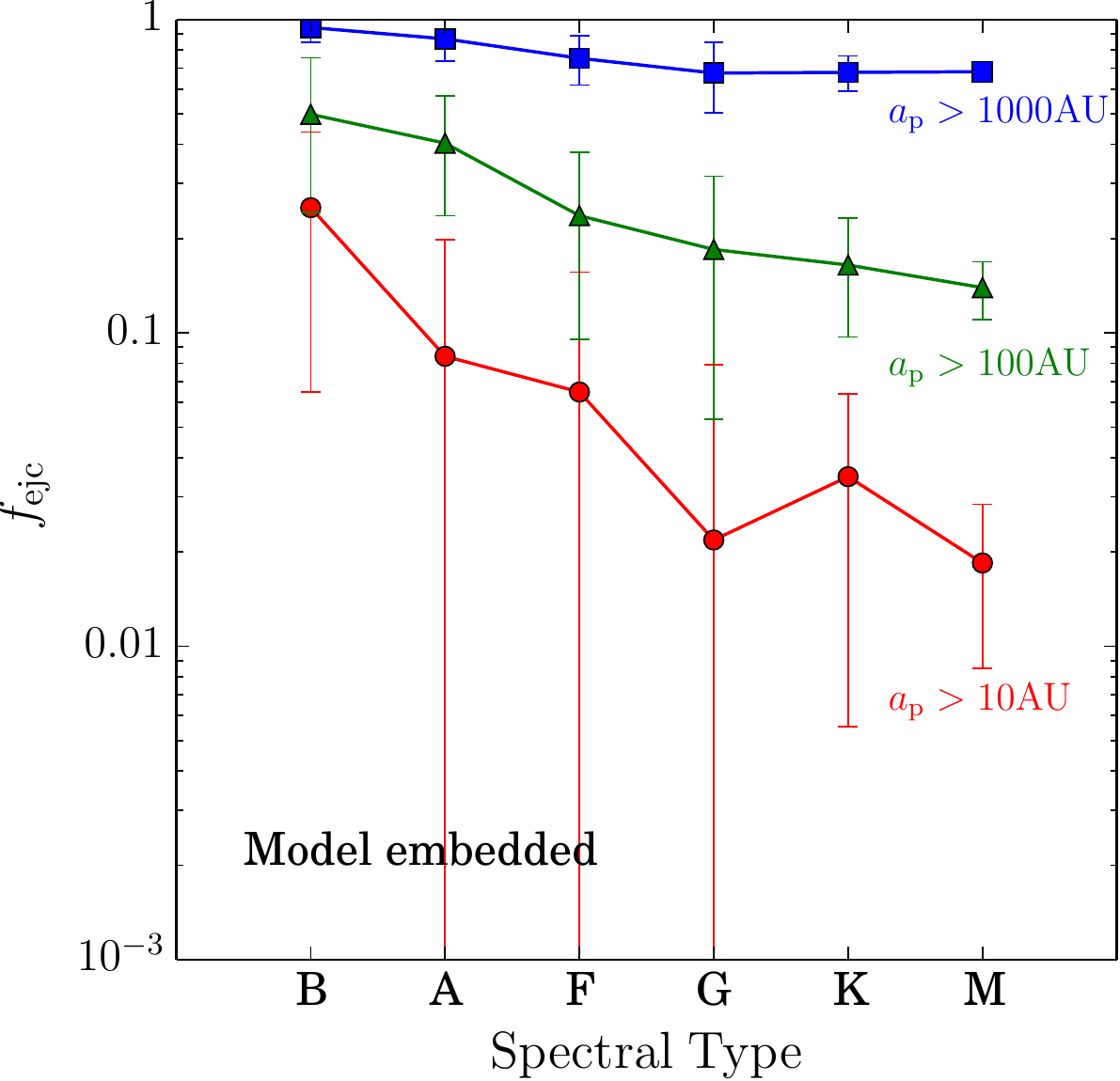}\\
\caption{Ejection rates of planets with $a_{\rm p} >10$, 100, and 1000\,AU around
each spectral type of star at $t=600$\,Myr ($170$\,Myr for embedded clusters). 
Error bars indicate run-to-run variations (standard deviations). Large error bars in ejection rates of planets around massive stars  
are attributed to the rarity of massive stars, as seen in the IMF.
\label{fig:survival_type}}
\end{figure*}

Figure \ref{fig:survival_type} presents the ejection rates of planets at 10, 100, and 1000\,AU around each spectral type of stars at $t=600$\,Myr for the open cluster models and $t=170$\,Myr for the embedded cluster model. 
The ejection rate of planets within 10\,AU for B-stars ($>2M_{\odot}$) is roughly an order of magnitude higher than that for M-stars (0.08--0.45$M_{\odot}$). This is consistent with the expectation from the gravitational focusing effect.
However, the frequency of orbital disruption of planets with $a_{\rm p} < 10$\,AU even around B-type stars ends in at most 10\,\%.
Planetary systems around low-mass stars such as M dwarfs are stirred up less frequently even in dense open clusters. We confirm that
close-in planets within 1\,AU can rarely be ejected from any types of stars in open clusters \citep{2001MNRAS.324..612D,2001MNRAS.322..859B}.

\begin{figure*}
\centering
\includegraphics[width=0.45\hsize]{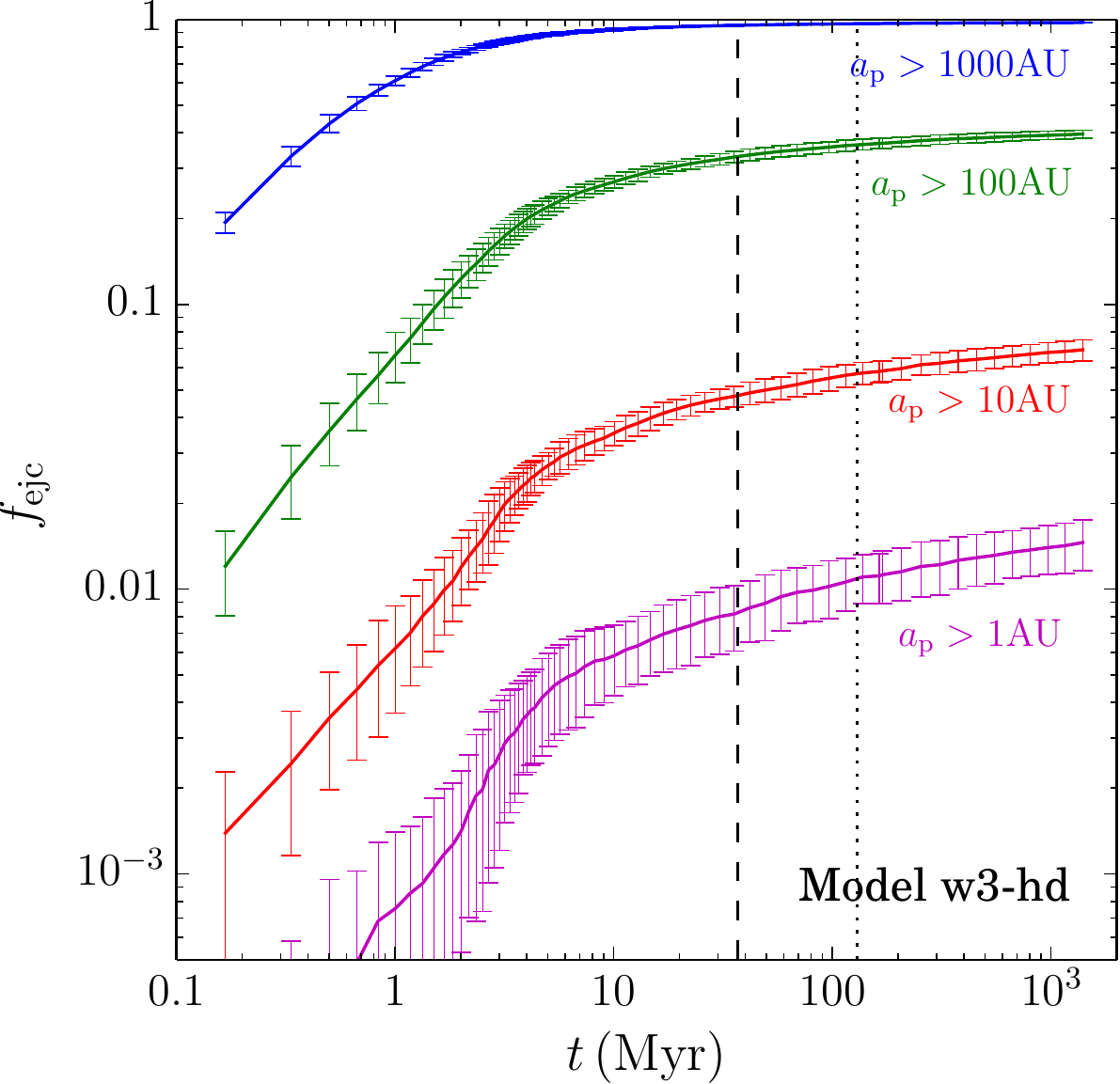}\includegraphics[width=0.45\hsize]{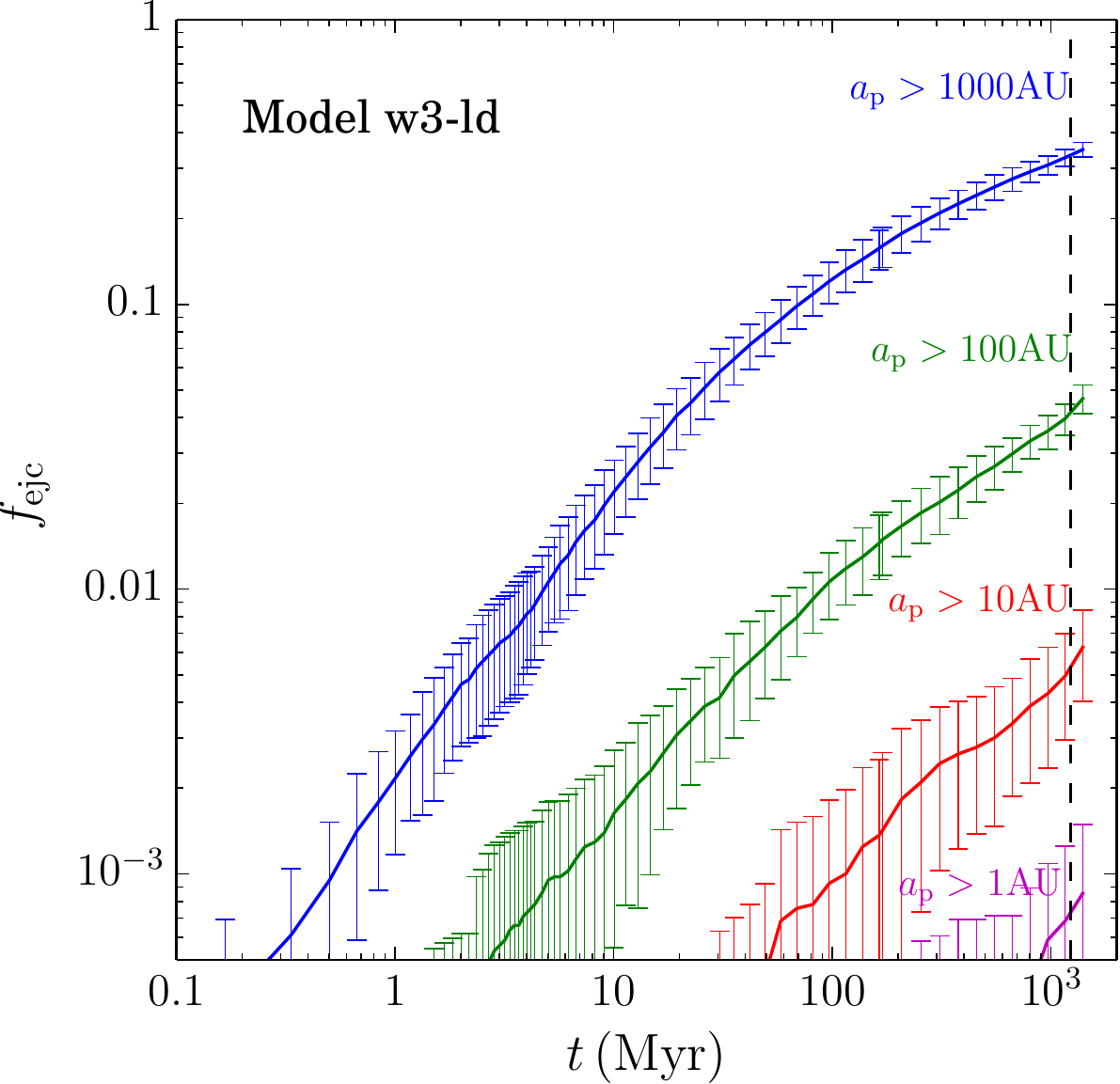}\\
\includegraphics[width=0.45\hsize]{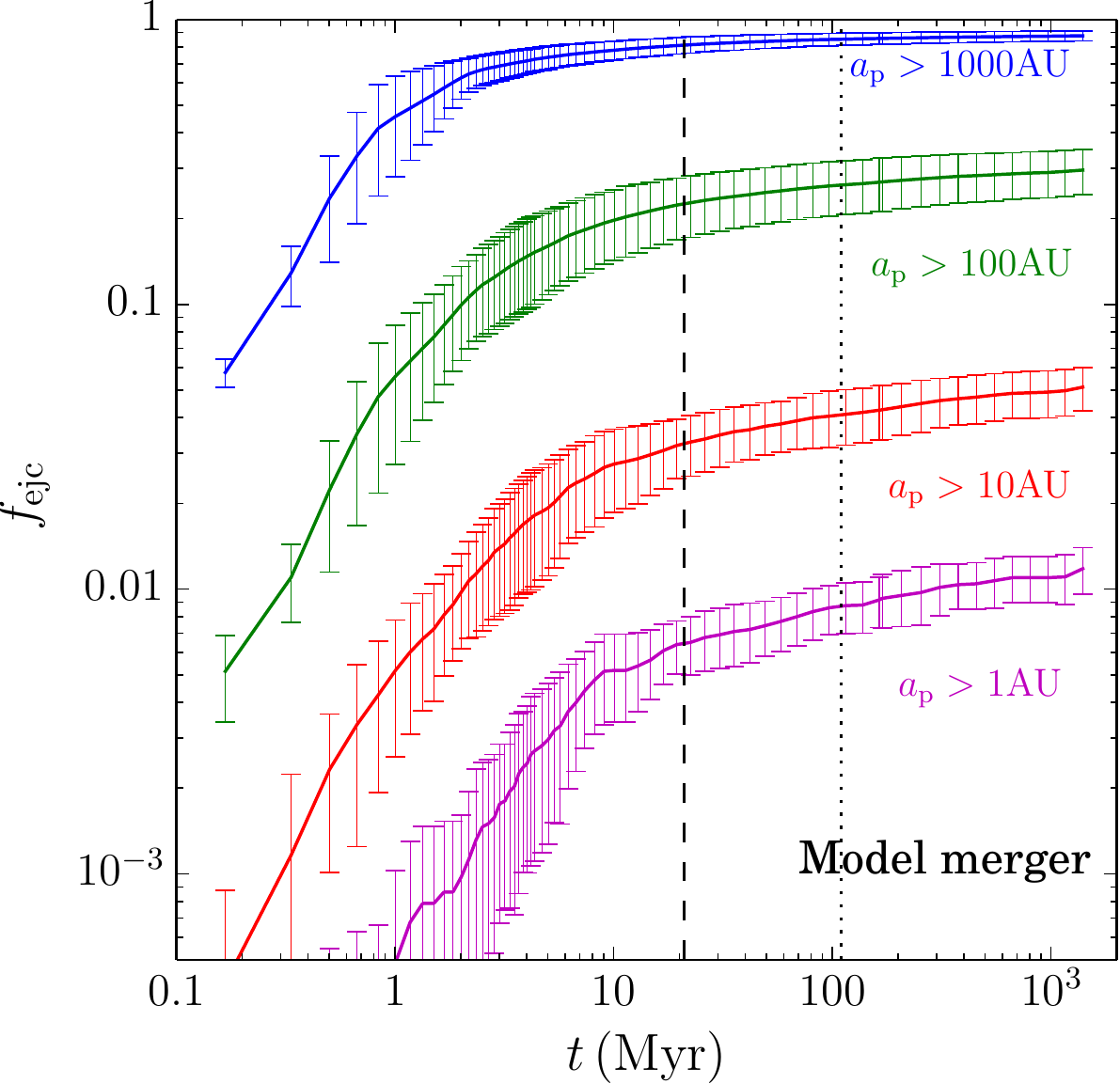}\includegraphics[width=0.45\hsize]{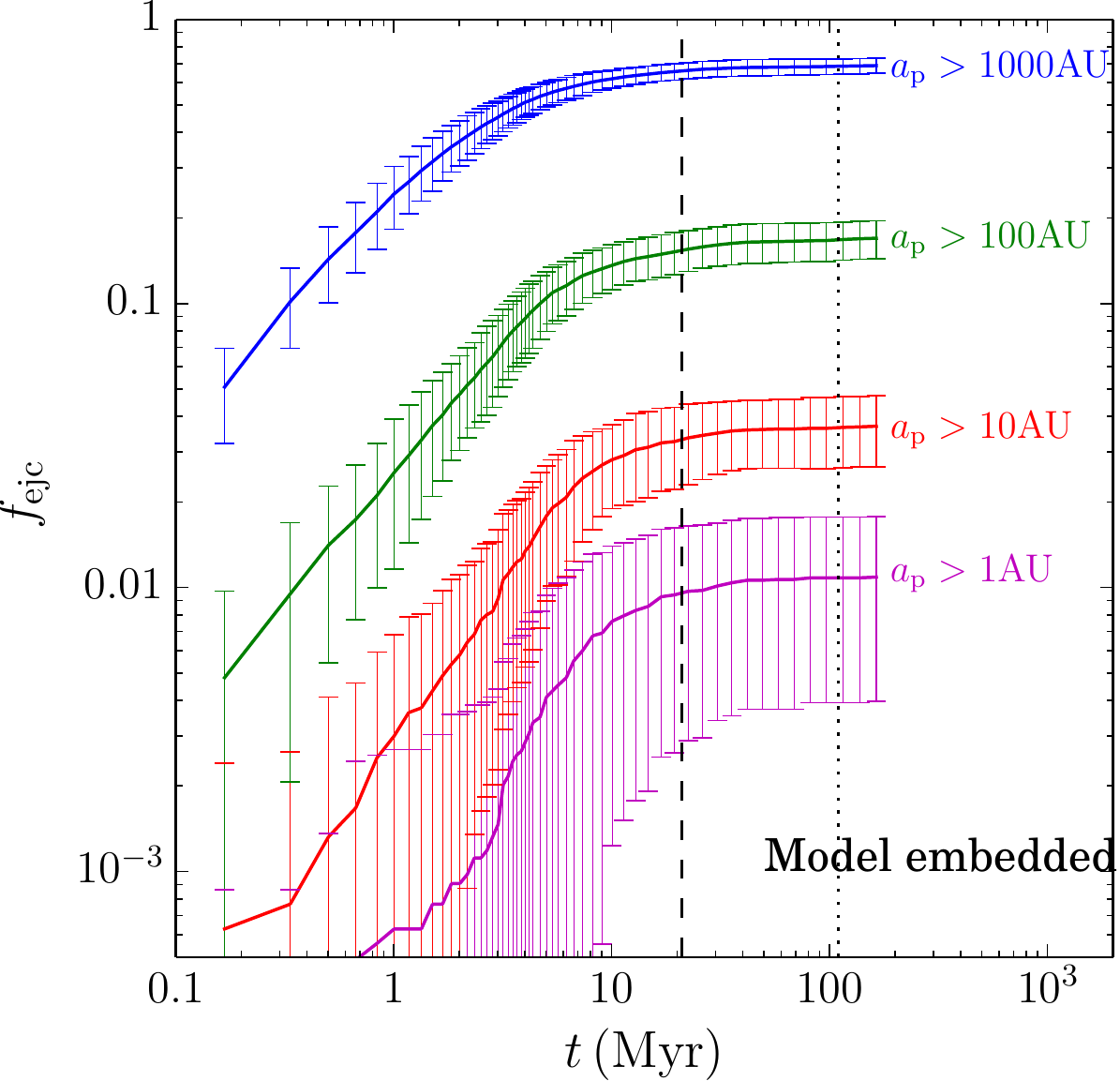}\\
\caption{Time evolution of ejection rates ($f_{\rm ejc}$)
of planets with $a_{\rm p} >$1, 10, 100, and 1000\,AU.
Error bars indicate run-to-run variations (standard deviations). 
Vertical dotted and dashed lines indicate 
$100\,t_{\rm rlx,c}$ and $10t_{\rm rlx}$, respectively. 
For the merger model, we adopt $t_{\rm rlx,c}$ and $t_{\rm rlx}$ of the subcluster.}
\label{fig:survival_rate}
\end{figure*}

The ejection rate of planets increases as the core collapse proceeds in a star cluster. Once the cluster density decreases after the core collapse, the ejection rate of planets saturates.
Figure \ref{fig:survival_rate} presents the time evolution of the
ejection rates of planets with $a_{\rm p} >$ 10, 100, and 1000\,AU.
Stellar encounter rates initially increase rapidly on the time-scale of 
their core relaxation time ($t_{\rm rlx,c}$); in the merger case, $t_{\rm rlx,c}$ of subclusters.
In all the models, most of encounters
occur by $10\,t_{\rm rlx,c}$ (the vertical dotted lines in Figure \ref{fig:survival_rate}). 
Although our criteria for disruptive encounters of passing stars are simplified, we confirmed 
that the results are consistent with 
previous simulations in which both stars and planets are treated as 
$N$-body systems \citep{2012MNRAS.419.2448P}; 
after 10\,Myr, 10\,\% of the planets originally orbiting at 30\,AU 
from their host stars became unbound.

\subsection{Planet frequency in open clusters}

As shown in the previous section, the survival rate of planets in open clusters is sufficiently high. On the other hand, a significant fraction of outer planets can be lost by stellar encounters in open clusters, especially in initially dense star clusters.
Here, we investigate how the expected planet-semi-major axis distribution in open clusters can be different from the observed one in the field.

\begin{figure}
\centering
\includegraphics[width=\hsize]{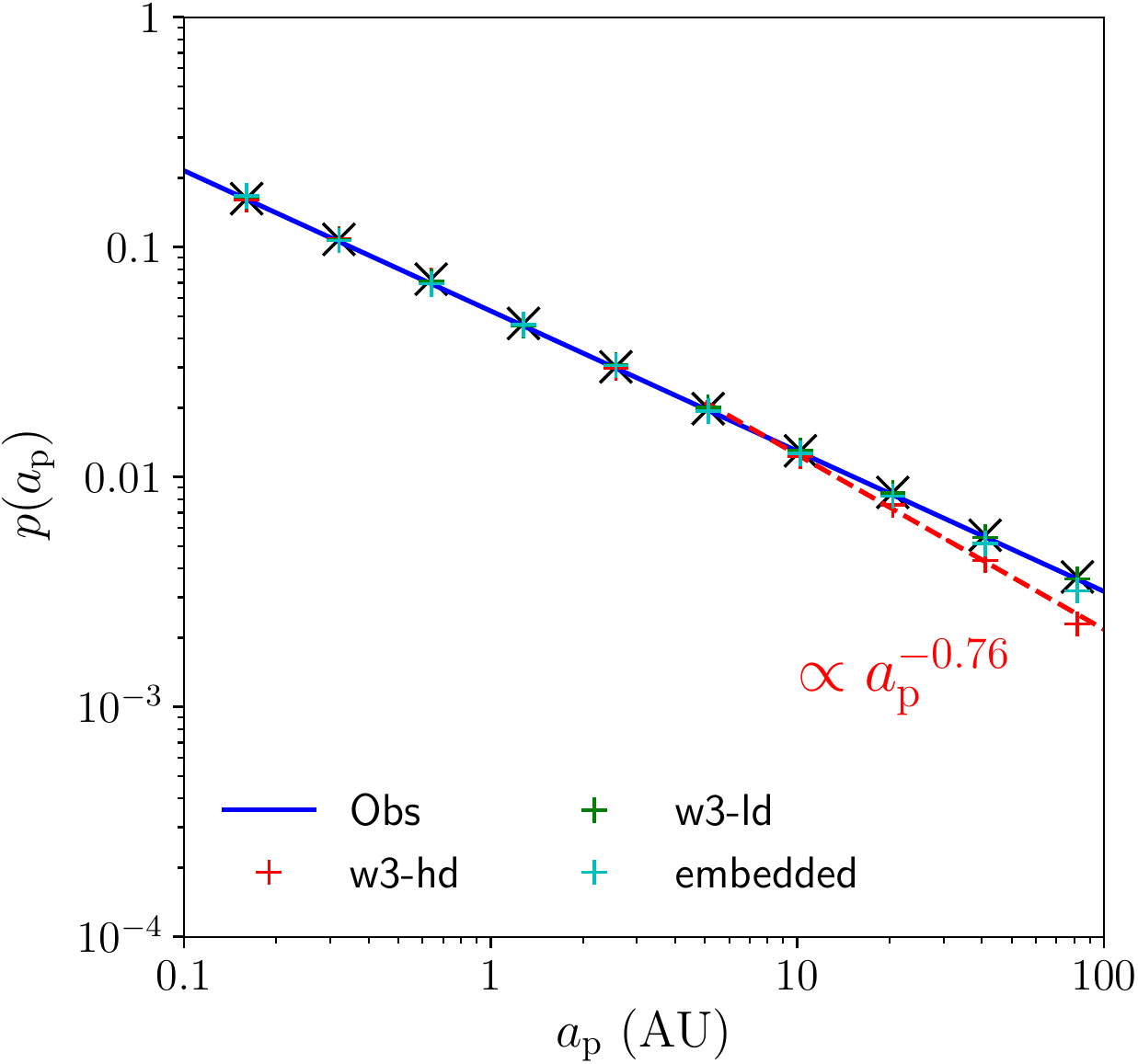}
\caption{Probability distribution function (PDF) of planets in star clusters (red, green, and cyan crosses are results of models w3-hd, w3-ld, and embedded, respectively.
The blue line shows the PDF obtained from observed planets between 0.03 and 3\,AU (see Eq. \ref{eq_pdf}). The red dashed line indicates a power-law distribution following $\propto a_{\rm p}^{-0.76}$, which is obtained from a fitting to model w3-hd for 10--100\,AU.}
\label{fig:p_dist}
\end{figure}

Based on the observed planet distribution in the field,
we estimate fractions of potentially-surviving planets at 1 -- 10\,AU around each 
spectral type of star in open clusters. Populations of low-mass planets beyond 1\,AU are not currently available because of observational difficulties in detecting them.
We focus on giant planets with $0.1 - 13\,M_{\rm Jup}$ 
between 1 -- 10\,AU in a single-planet system. We note that this assumption is 
prone to underestimate occurrence rates of planets. We consider planets orbiting FGKM-type
stars because our cluster models contain only one or two A- and B-type stars. We adopt a fitting 
formula of period distributions of planets with mass in the range of $0.3 - 10\,M_{\rm Jup}$ 
at 2--2,000\,days around FGK-type stars published in \citet{2008PASP..120..531C}:
\begin{eqnarray}
	& p(M_{\rm p}, a_{\rm p}) \left(= \frac{dN}{dM_{\rm p} da_{\rm{p}}}\right) \nonumber \\
		& =  1.03 \times 10^{-2} 
		\left(\frac{M_{\rm p}}{1\,M_{\rm Jup}}\right)^{-1.31}
		\left(\frac{a_{\rm p}}{{1\,\rm AU}}\right)^{-0.61},
	\label{eq_pf}
\end{eqnarray}
where $p$ is the probability distribution function (PDF), $N$ is the number of planets, and $M_{\rm p}$ 
and $a_{\rm p}$ are the planetary mass and semi-major axis. 
We assume that the mass-period distribution of planets orbiting M-dwarfs are the same power-law function for FGK-type stars as well as \citet{2008PASP..120..531C}.
Following the fact that 10.5\,\% of solar-type stars have a planet with mass of
0.3--10$\,M_{\rm Jup}$ at 0.03--3\,AU \citep{2008PASP..120..531C},
we determine a normalization constant of the PDF of inner planets with 0.1 to 13$\,M_{\rm Jup}$. Extrapolating Eq.(\ref{eq_pf}) 
from 0.1 to 13$\,M_{\rm Jup}$, we obtain 
\begin{eqnarray}
p_{\rm inner}(a_{\rm p}) = 5.28\times 10^{-2} \left(\frac{a_{\rm p}}{{1\,\rm AU}}\right)^{-0.61},
\label{eq_pdf}
\end{eqnarray}
for inner planets at 0.03--3\,AU.
The PDF of the inner planets is shown by the blue line in Figure \ref{fig:p_dist}. 

Planets frequency beyond 10\,AU is observationally less completed.
Recent results of direct-imaging surveys suggest a low frequency of wide-orbit gas planets \citep{2014ApJ...786....1B,2016A&A...594A..63G,2016PASP..128j2001B}.
Compiling the data of direct imaging surveys by Subaru/HiCIAO (SEEDS), Gemini North/NIRI, and Gemini South/NICI, \citet{2014ApJ...786....1B} 
derived a probability distribution of substellar companions ($>$ 5\,$M_{\rm Jup}$) 
within $\sim$\,10 -- 100\,AU as a power-law function of semi-major axis and planetary 
mass whose indices are in agreement with RV results within $1\sigma$ \citep{2008PASP..120..531C}:
\begin{eqnarray}
	 & p(M_{\rm p},a_{\rm p}) \left(= \frac{dN}{dM_{\rm p} da_{\rm p}}\right) \nonumber \\
		& =  (1.0\pm0.4) \times 10^{-3} 
		\left(\frac{M_{\rm p}}{1\,M_{\rm Jup}}\right)^{-0.7}
		\left(\frac{a_{\rm p}}{10\,{\rm AU}}\right)^{-0.8},
	\label{eq_pdf2}
\end{eqnarray}
where $p$ is the PDF,
and $M_{\rm p}$ and $a_{\rm p}$ are the planetary mass and semi-major axis, respectively. 
This PDF indicates that the frequency of distant planets decreases with increasing 
planetary mass and semi-major-axis.
Integrating Eq. (\ref{eq_pdf2}) from 1 to 13\,$M_{\rm Jup}$, we obtain
\begin{eqnarray}
 p_{\rm outer}(a_{\rm p}) = (3.86\pm1.54) \times 10^{-3}\left(\frac{a_{\rm p}}{10\,{\rm AU}}\right)^{-0.8},
\label{eq:pdf_distant}
\end{eqnarray}
for planets outer than 10\,AU. 
Note that the observed radial frequencies of inner- and outer-planets are not connected.

In order to obtain the planet frequency after stellar encounters in star clusters,
we assume that (i) all FGKM-type stars initially has planets between 0.01--100\,AU following the planet frequency of Eq. (\ref{eq_pdf}) and that (ii) planets can escape at the ejection rates of planets given by
Eq. (\ref{eq:esc_frac}). We use the best-fit parameters for ejection rates of planets given in Table \ref{tb:parameters}.

The PDF of planets after stellar encounters in models w3-ld, w3-hd, merger, and embedded, and the results are shown in Figure \ref{fig:p_dist}.
We found that stellar encounters change the planet distribution beyond $\sim10$\,AU;
the planet frequency between 10--100\,AU after stellar encounters follows $a_{\rm p}^{-0.76}$ for model w3-hd, which is close to the observed distribution of wide-orbit planets (see Eq. (\ref{eq:pdf_distant})), and $a_{\rm p}^{-0.62}$ and $a_{\rm p}^{-0.65}$ for models w3-ld and embedded, respectively.
Thus, the planets frequency beyond 10\,AU in star clusters can be shaped by stellar encounters.

\subsection{Occurrence rate of free-floating planets\label{ffp}}

Recent planet surveys revealed the existence of unbound or free-floating planetary-mass objects 
\citep[e.g.,][]{2000Sci...290..103Z,2011Natur.473..349S,2012A&A...548A..26D,2013ApJ...777L..20L,2017Natur.548..183M}. 
The results obtained by microlensing surveys toward the Galactic Bulge 
suggest that the frequency of free-floating Jovian planets are $\lesssim 0.25$ planets per main-sequence star 
\citep{2017Natur.548..183M}. Several dynamical processes have been proposed in order to explain the origin of free-floating planets (FFPs) in our Galaxy: planet-planet scattering \citep{2012MNRAS.421L.117V}, direct formation in grobulettes \citep{2007AJ....133.1795G, 2015MNRAS.446.1098H}, and the aftermath of evolved stars such as supernova explosions 
\citep{2012MNRAS.422.1648V}.

In this study, we have shown that stellar encounters in a star cluster can make wide-orbit planets beyond 10\,AU gravitationally unbound (see Figs.\, \ref{fig:survival_type} and \ref{fig:survival_rate}). 
We estimate the production rates of free-floating gas giants in star clusters from the ejection rate of planets given by Eq. (\ref{eq:esc_frac}). 
We suppose gas giants with masses of 1 -- 13\,$M_{\rm Jup}$ at semi-major axes of 0.01 -- 100\,AU. 
Directly-imaged planets on wide orbits tend to have no sibling, except for HR\,8799.
Therefore, we do not consider any multiple giant planet systems. We also assume that 
all FGKM-type stars initially harbor one gas giant planet between 0.01 and 100\,AU
following $f(a_{\rm p})\propto a_{\rm p}^{-0.61}$ (blue line in Figure \ref{fig:p_dist}).
We generate a population of FGKM-type stars following a Kroupa IMF with an upper- and lower-mass cut-off of 0.08 and 2.1\,$M_{\odot}$.
We calculated the ejection fraction of planets using the initial distribution of planets given by Eq. \,(\ref{eq_pdf}). We found the production rates of FFPs per
FGKM-type star to be 0.184, 0.00960, and 0.0677 for models w3-hd, w3-ld, and embedded.
 
If most of stars are born in star clusters and the initial core density of the clusters are 
relatively high, $\sim 0.1$ FFPs per main-sequence star are produced by stellar encounters. Since the expected frequency of FFPs is compatible with the upper limit on the frequency of FFPs indicated by microlensing surveys \citep{2017Natur.548..183M}, 
stellar encounters in star clusters can be one of promising sources for FFPs in the field.
We also estimated the case when planets are distributed between 0.01 and 30\,AU. In any cases, the production rate of FFPs per FGKM-star becomes less (see Table \ref{tb:ffp} for more details).

\begin{table*}
\caption{Production rate of free-floating planets per FGKM-type star\label{tb:ffp}}
\begin{tabular}{lcccc}
\hline \hline
 & \multicolumn{2}{c}{$n_{\rm p}=1$}      & \multicolumn{2}{c}{$p(a_{\rm p}) = 5.28\times 10^{-2} ({a_{\rm p}}/{{1\,\rm AU}})^{-0.61}$}  \\
Models   & $0.01<a_{\rm p}<100$\,AU   &  $0.01<a_{\rm p}<30$\,AU  &  $0.01<a_{\rm p}<100$\,AU   &  $0.01<a_{\rm p}<30$\,AU  \\
\hline
w3-hd & 0.184 & 0.0675 & 0.123 & 0.0276 \\
w3-ld & 0.00960 & 0.00270 & 0.00640 & 0.00110 \\
embedded & 0.0677 & 0.0260 & 0.0451 & 0.00945 \\
\hline
\end{tabular}
\end{table*}

\section{Discussion}

\subsection{Is the Pleiades a barren land for planets?}

A nearby young open cluster, the Pleiades \citep[70 -- 125\,Myr, 136\,pc:][]{2014Sci...345.1029M}, has the mass, size, and [Fe/H] similar to those of M67 with five known gas giants \citep{2017A&A...603A..85B}.
However, no planet orbiting stars is seen so far in the Pleiades \citep{2013PASJ...65...90Y}
with exception for that free-floating planetary-mass candidates \citep{2014A&A...568A..77Z} and a substellar companion to the dusty Pleiades star, HD\,23514, were reported \citep{2012ApJ...748...30R}.
Light-curve analyses of 1,014 Pleiades candidate stars in the Campaign 4 field observed by
the Kepler spacecraft in the two-wheel mode, the so-called $K2$ mission, have recently reported
the absence of the planet candidates \citep{2016MNRAS.457.2877G}.
These facts invoke three possibilities to explain the deficiency of stars harboring planets in the Pleiades; (i) the Pleiades is a real desert island of planets, (ii) the incompleteness of planet surveys, for example, a short-monitoring time of the $K2$ mission, 
and (iii) planets were ejected from host stars in a hostile environment at the early stage or planet-hosting stars ran away from the Pleiades. 

We revisit non-detection of planets in the Pleiades open cluster.
The Pleiades has a core radius and central density similar to our open cluster models (see also Figure \ref{fig:core_evolution}). 
Our cluster models suggest that the Pleiades is dynamically mature enough that destructive events of planetary system due to close encounters ceases (see Figure \ref{fig:survival_rate}). Distant planets beyond 100\,AU are expected to be ejected from their host stars in the Pleiades, while inner planets can survive against frequent stellar encounters. 
Compared to two planet-hosting clusters, the Hyades and Praesepe, the Pleiades may used to have an initially high stellar density, and therefore the wide-orbit planets in the Pleiades may be depleted more severely than those in the other two clusters.
However, there should be no significant differences in the survival rates of close-in planets between the Pleiades and two planet-hosting clusters, even though we assume a high-density cluster model as the initial condition of the Pleiades. Different initial states at the formation stage of open clusters have little effect on the detectability of close-in planets. In addition, since the Pleiades is not a metal-poor environment but a slightly super-solar metallicity. The Pleiades is a favorable site for planet formation, albeit its [Fe/H] is lower than those of the planet-hosting stars in the Hyades 
and Praesepe \citep{2009AJ....138.1292S}.
Thus, our results suggest that the Pleiades is a fertile land for planets and that non-detection of planets orbiting stars in the Pleiades could
be due to the incompleteness of planet surveys. Nevertheless, if intensive planet surveys in the future
failed to find any planets in the Pleiades, planet formation might have been inhibited via some mechanism, which would not have happened in the Hyades and Praesepe.

\subsection{Toward planet surveys: detectability of planets in star clusters}

Planets within 10\,AU around their host stars likely survive over the typical lifetimes of star clusters. In general, massive planets around massive stars are hardly ejected from the system because of their strong gravitational binding. However, massive stars such as B- and A-type stars are not dominant members in both an embedded and an open clusters according to the IMF. Planetary systems around massive stars which fall into dense regions due to the mass segregation also undergo violent encounters with passing stars (see Fig.\,\ref{fig:survival_type}). Provided that production rates of planets in clustered environments are comparable to those in the field, short-period massive planets around Sun-like stars and low-mass stars rather than massive ones are promising targets for planet searches by RV measurements and transit photometry.

Detection efficiency of planets depends on physical properties of a star cluster (mass, age, and [Fe/H]) and the distance from the Earth. A positive correlation between the frequency of giant planets and stellar metallicity indicates that metal-rich clusters are suitable to planet surveys. Star clusters older than the relaxation time ($t_{\rm rlx}$) is dynamically mature and therefore no more close encounters are expected: $t_{\rm rlx}$ is typically $\sim$ 10--100\,Myr for low-density open clusters, $\sim$ 1 -- 10\,Myr for high-density open clusters, and $\sim 1\,$Myr for embedded clusters. Stars in star clusters younger than $t_{\rm rlx}$ are in the midst of experiencing close encounters, in other words, a fraction of the planets that are candidates to be free-floaters are still gravitationally bound to their host stars in young clusters. As a result, younger open clusters show higher sensitivity for the detection of planets.

Open clusters may have evolved from denser clusters such as young massive clusters, as shown in Figure\,\ref{fig:core_evolution} in Appendix. In the central region of such dense clusters, stars are exposed to violent circumstances in which the detectability of planets should decreases because of enhanced stellar encounters, a strong gas expulsion, intense XUV irradiations from massive stars, frequent and subsequent collisions of stars, and effects of binary systems. In addition, a densely-packed region in a star cluster is often hardly resolved to individual stars within the field of view. Contamination of light from nearby stars obscures signals of planets imprinted in RVs and/or lightcurves of their host stars. High-density, massive clusters would have low priority for planet searches.

The observed occurrence rate of wide-orbit planets beyond 10--100\,AU in the field is extremely low, as mentioned before.
Such distant planets are supposed to be readily liberated from their host stars (see Fig. \,\ref{fig:survival_rate}). 
Planets are difficult to form (in situ) beyond 10\,AU in the core-accretion model because of the slow growth of cores in the outer region. 
On the other hand, disk instability may happen in a protoplanetary disk around metal-poor \citep[e.g.,][]{2006ApJ...636L.149C} and massive stars, and then planets are born beyond $\sim$10\,AU from their host stars.
In fact, many distant planets are found around metal-poor and/or massive stars\footnote{Although some planetary or substellar companions on wide orbits are orbiting around brown dwarfs,
these companions may form via fragmentation of molecular clouds like binary formation.}, except for the metal-rich GJ~504 \citep{2013ApJ...774...11K}.
Since massive stars are rarely seen in open clusters, the frequency of planets moving on wide orbits in open clusters is expected to be lower than that in the field.

\section{Conclusions}

We have performed a series of $N$-body simulations of star clusters,
modeling three open clusters (the Pleiades, Praesepe, and Hyades) and embedded clusters.
Stellar encounter rates increase on the core relaxation timescale ($t_{\rm rlx, c}$) in the initial state of a star cluster.
Eventually, the close encounters cease at $\sim 10\,t_{\rm rlx, c}$: 
$\sim1$\,Gyr for initially low-density open clusters and $<100$\,Myr for open clusters which initially had a density higher than those of present-day open clusters.
We also considered a merger model in which open clusters form via hierarchical merger of subclusters. In the merger case, stellar encounters increase on a relaxation timescale of the subclusters, and the cumulative encounter rate was comparable to those of the single high-density cluster model. 

Using the results of $N$-body simulations, we semi-analytically estimated the ejection rates of planets due to stellar close encounters in star clusters as a function of the semi-major axis of planets. The results are summarized as follows:

1) Close-in planets inside 1\,AU can be rarely liberated from any types of
host stars in open clusters;
the ejection fraction of such short-period planets is less than 1.5\,\%. We expect
no significant difference between the frequency of short-period planets in
open clusters and that in the field. 
This implies that non-detection of close-in planets in open cluster such as the Pleiades may be due to the incompleteness of planet surveys or suppression of planet formation in clustered environments.

2) The ejection rate of planets with semi-major axes ($a_{\rm p}$) of 1--10\,AU is at most 7\,\% in our high-density cluster models which initially has a core density higher than typical values of observed open clusters.
The ejection rates of planets orbiting more massive stars are higher; up to 29\,\% of planets within 10\,AU around B-stars experience orbital disruption. On the other hand, the ejection rate of planets around FGKM-type
stars is only a few \%.

3) The ejection rate of planets at 10--100\,AU around FGKM-type stars reaches a few tens \%.
If we extrapolate the probability distribution function (PDF) of observed giants planets within
10\,AU ($p(a_{\rm p})\propto a_{\rm p}^{-0.61}$) \citep{2008PASP..120..531C} to the outer region beyond 10\,AU, we found that the PDF of giant planets beyond 10\,AU changes due to close encounters in star clusters; 
$p(a_{\rm p})\propto a_{\rm p}^{-0.76}$ for our high-density cluster model.

4) If we assume that each star in a star cluster initially has one planet between 0.01--100\,AU based on the observed planet distribution ($\propto a_{\rm p}^{-0.61}$), the production rate of free-floating planet per star is 0.184 and 0.00960 for high- and low-density open cluster models and 0.0677 for embedded cluster models. These values are compatible with the observed fraction of free-floating planets, $\lesssim 0.25$ per main-sequence star \citep{2017Natur.548..183M}.

5) Distant planets with $a_{\rm p}>$ 100--1000\,AU such as directly-imaged ROXs 42B and DH Tau b can be ejected efficiently via stellar encounters in cluster environments.
However, the birthplace of planets are expected to be inside several 10AU
both in the core-accretion model \citep[e.g.,][]{1996Icar..124...62P} and in the disk instability scenario \citep[e.g.,][]{1997Sci...276.1836B}. This implies that most of planets around FGKM-stars are likely to survive against stellar encounters in open clusters.

\begin{acknowledgements}
We thank the referee, Richard Parker, for his useful comments.
Y.H. was supported by Grant-in-Aid for JSPS Fellows (No.25000465) from MEXT of Japan. M.F. was supprted by The University of Tokyo Excellent Young Researcher Program. We thank Timothy D. Brandt for providing us with useful information on distributions of wide-orbit substellar companions derived from imaging surveys.
Numerical computations were partially carried out on Cray XC30 (ATERUI) at the Center for Computational Astrophysics
(CfCA) of the National Astronomical Observatory of Japan.
\end{acknowledgements}

%
%
\bibliography{reference}

\begin{appendix} 

\section{Plummer model: Comparison with Spurzem et al. (2009)\label{Plummer}}
We show our semi-analytical treatment for ejection of planets is consistent with results of fully $N$-body simulations.
We consider a Plummer model with a single-mass component of $1\,M_{\odot}$ \citep{1911MNRAS..71..460P} as a star cluster. This is equivalent to the star cluster model adopted in previous work \citep{2009ApJ...697..458S} in which the orbital evolution of planets was numerically integrated as well as the stellar motions in the star cluster . 
We adopt the $N$-body unit ($G=M=4|E|=1$) for the gravitational
constant and the total mass and energy of a star cluster, respectively
\citep{1971Ap&SS..14..151H,1979ApJ...234.1036C,1986LNP...267..233H}. 
Assuming that the unit length ($R=GM/4|E|$) is 1\,pc \citep{2009ApJ...697..458S}, the core and half-mass radii are 0.417\,pc and 0.769\,pc.
We assume that the cluster has the total mass of $1.9 \times 10^4\,M_{\odot}$. With the number of particles of $1.9 \times 10^4$, the central density is $2.2 \times 10^4\,M_{\odot}\,{\rm pc}^{-3}$.

\citet{2009ApJ...697..458S} performed a series of direct $N$-body simulations of a star cluster containing 19,000 stars with mass of $1\,M_\odot$ up to 0.3--0.9 times half-mass relaxation time ($t_{\rm rh}$), 1,000 of which initially have a planet:
model 1 to 6 ($a_{\rm p} = 3 - 50\,{\rm AU}$, $e_{\rm p} = 0.01, 0.1, 0.3, 0.6, 0.9, {\rm and}\, 0.99$),
model E  ($a_{\rm p} = 3 - 50\,{\rm AU}$, $f(e) = 2e$), and
model EH ($a_{\rm p} = 0.03 - 5\,{\rm AU}$, $f(e) = 2e$),
where $a_{\rm p}$ and $e_{\rm p}$ are the initial semi-major axis and eccentricity of a planet and $f(e) = 2e$ means a thermal eccentricity distribution.
They also performed hybrid Monte Carlo (HMC) simulations 
of a star cluster for models E (soft) and EH (hard), but with an order of magnitude larger number of particles ($3.0 \times 10^5$ single-mass stars, $3.0 \times 10^4$ of which are planet-hosting stars).

We integrated the dynamical evolution of the star cluster up to 230 in $N$-body unit
(24.8\,Myr in physical unit), which is comparable to the two-body relaxation time ($t_{\rm rlx} = 234$ in $N$-body unit;
see Eq. (\ref{eq:t_rlx})). We note that the core-collapse occurs at a later stage for single component models, specifically, $\sim 20\,t_{\rm rlx}$,  \citep{1971ApJ...164..399S}.

We calculated the probability for a planet within a given distance to liberate from its host star ($P_{\rm ff, rlx}$) after one relaxation time (230 in $N$-body units) and found that 
$P_{\rm ff,rlx}$ are $0.0800 \pm 0.0034$ and $2.43 \pm 0.24 \times 10^{-3}$ for a planet at 12.2\,AU (soft) and 0.387\,AU (hard), which are the median of 3--50\,AU and 0.03--5\,AU in log-scale, respectively.
According to \citet{2009ApJ...697..458S}, these results are almost independent of the initial eccentricity distribution of planets (see their Table 3).
In \citet{2009ApJ...697..458S},
$P_{\rm ff, rlx}$ for models E (soft) and EH (hard) models are $0.112$ and $2.22 \times 10^{-3}$, respectively. 
Thus, we confirmed that our ejection rates of planets in a star cluster are in agreement with those of direct $N$-body simulations \citep{2009ApJ...697..458S}.

\begin{table*}
\caption{Plummer model with single-mass components \label{tb:models_pl}}
\centering
\begin{tabular}{lcccccccccc}
\hline \hline
Model     & $M$    &
$r_{\rm c}$  &
$r_{\rm h}$  & $r_{\rm t}$ &
$\rho_{\rm c}$   & $M_{\rm c}/M$    &
$N$ & 
$t_{\rm rlx}$ & $t_{\rm rlx,c}$ &
$N_{\rm run}$\\
           & $ (M_{\odot})$    &
$({\rm pc})$  &
$({\rm pc})$  & $({\rm pc})$ &
$(M_{\odot}\,{\rm pc}^{-3})$  &    &
   & 
$({\rm Myr})$ & $({\rm Myr})$ & \\
\hline
Plummer & $1.9\times 10^4$ & 0.42  & 0.77  & $\infty$  &  $2.2 \times 10^4$ &  0.192 & 19000  & 24 & 3.1 & 5 \\
\hline
\end{tabular}
\end{table*}

\section{Stellar evolution}\label{apdx_star}

\begin{figure}[ht]
\centering
\includegraphics[width=\hsize]{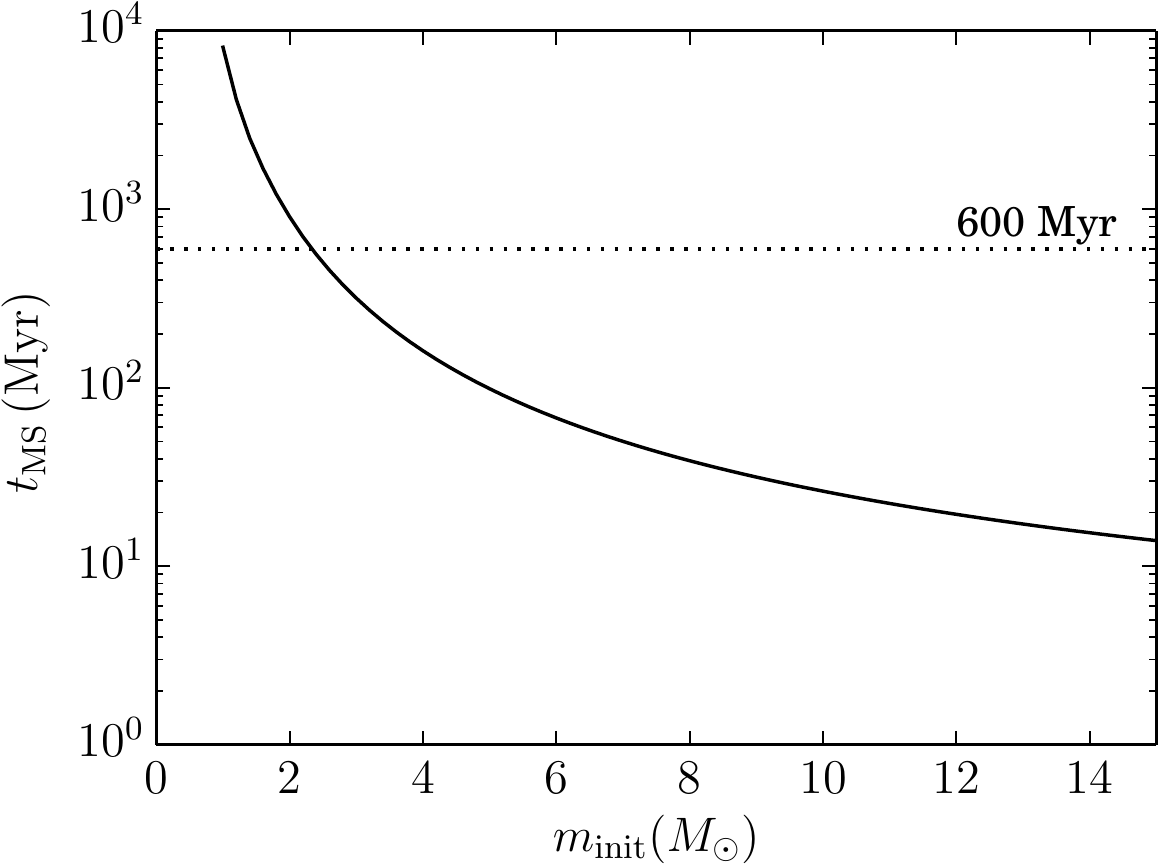}
\includegraphics[width=\hsize]{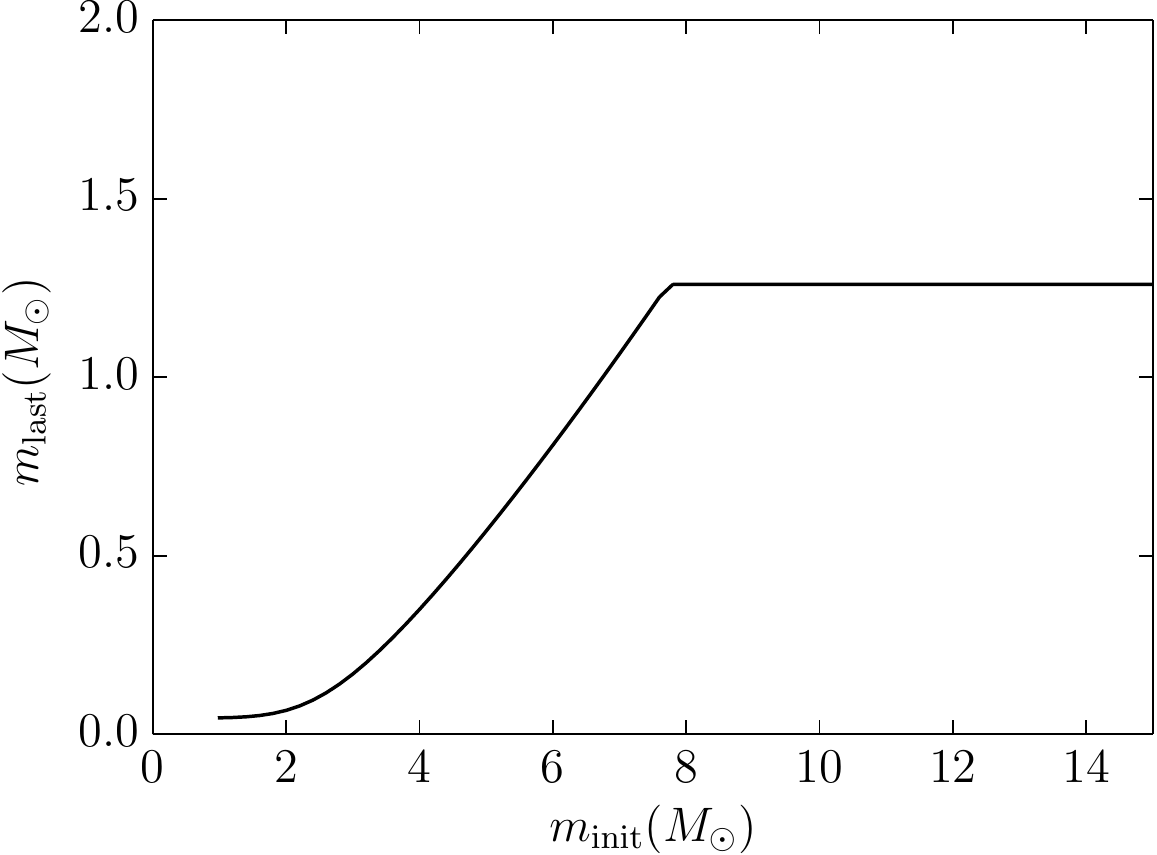}
\caption{Duration of a main-sequence phase (top) and the final mass of stars after stellar 
evolution ends (bottom) in our stellar evolution model.
\label{fig:stellar_evolution}}
\end{figure}

We adopt the stellar evolution model of \citet{2000MNRAS.315..543H}, 
simplifying their description: 1) mass loss is included when stars evolve to 
compact objects such as 
neutron stars and white dwarfs and 2) stellar evolution starts 
when the main-sequence phase ends. We assume that all the stars have the
solar metallicity. The duration of a main-sequence phase
as a function of initial mass of a star
is shown in the top panel of Figure \ref{fig:stellar_evolution}.
We present the final mass of a star after stellar evolution ends
as a function of initial mass of a star in the bottom panel of Figure \ref{fig:stellar_evolution}.
We use the same mass of a white dwarf as \citet{2000MNRAS.315..543H}, while
we adopt a neutron star's mass of $1.36M_{\odot}$ derived from \citet{2008ApJS..174..223B} whose model gives 
the same mass of a white dwarf as \citet{2000MNRAS.315..543H}.
In our simulations, when stars with masses of $\gtrsim 2.4M_{\odot}$ evolve
by $t=600$ Myr, $\sim 26$\% of the mass of a star cluster that has initially $\sim 1000M_{\odot}$ is lost.

\section{Evolution of merger models\label{merger}}
\begin{figure*}[ht]
\centering
\includegraphics[width=0.45\hsize]{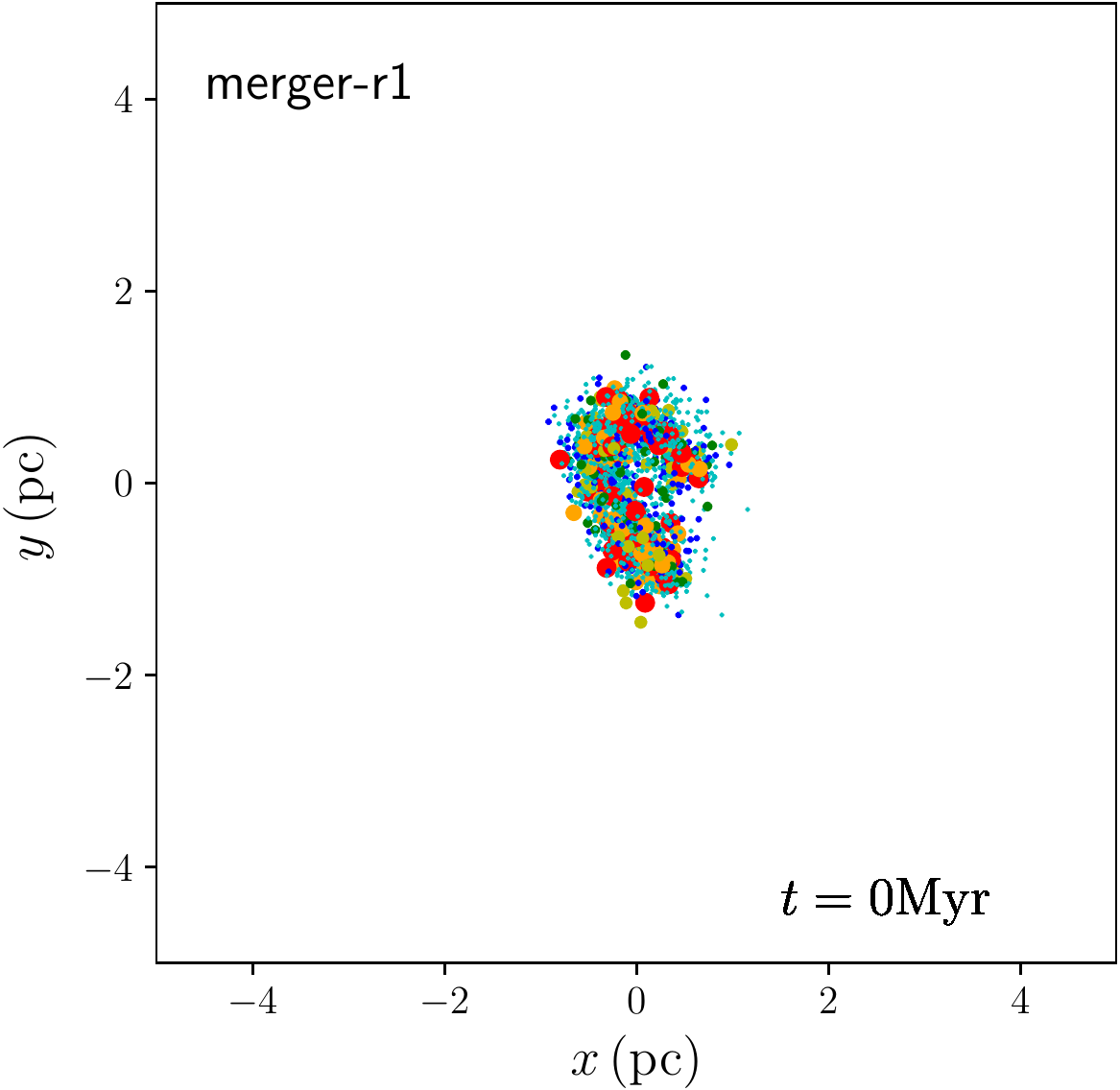}
\includegraphics[width=0.45\hsize]{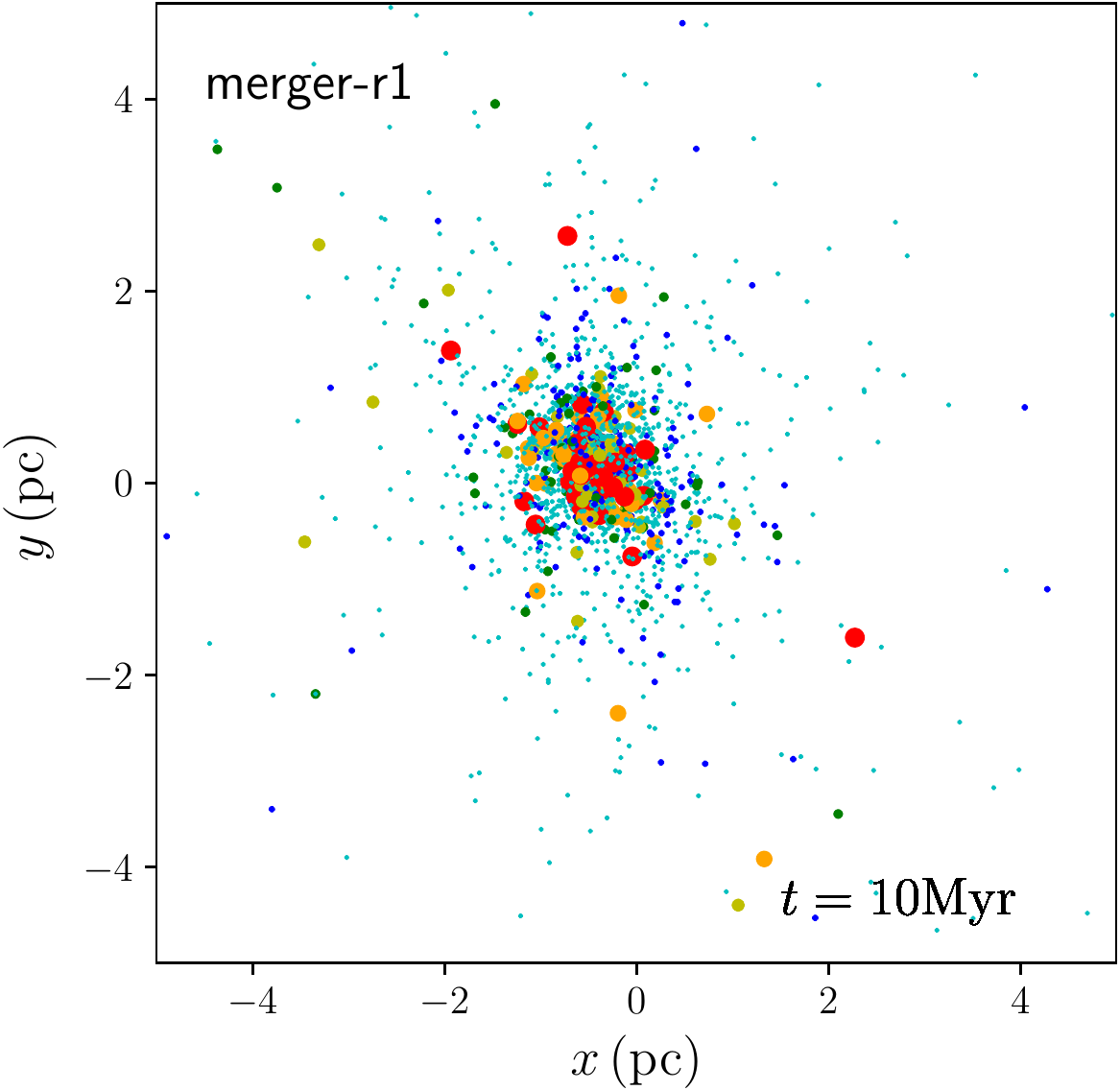}\\
\includegraphics[width=0.45\hsize]{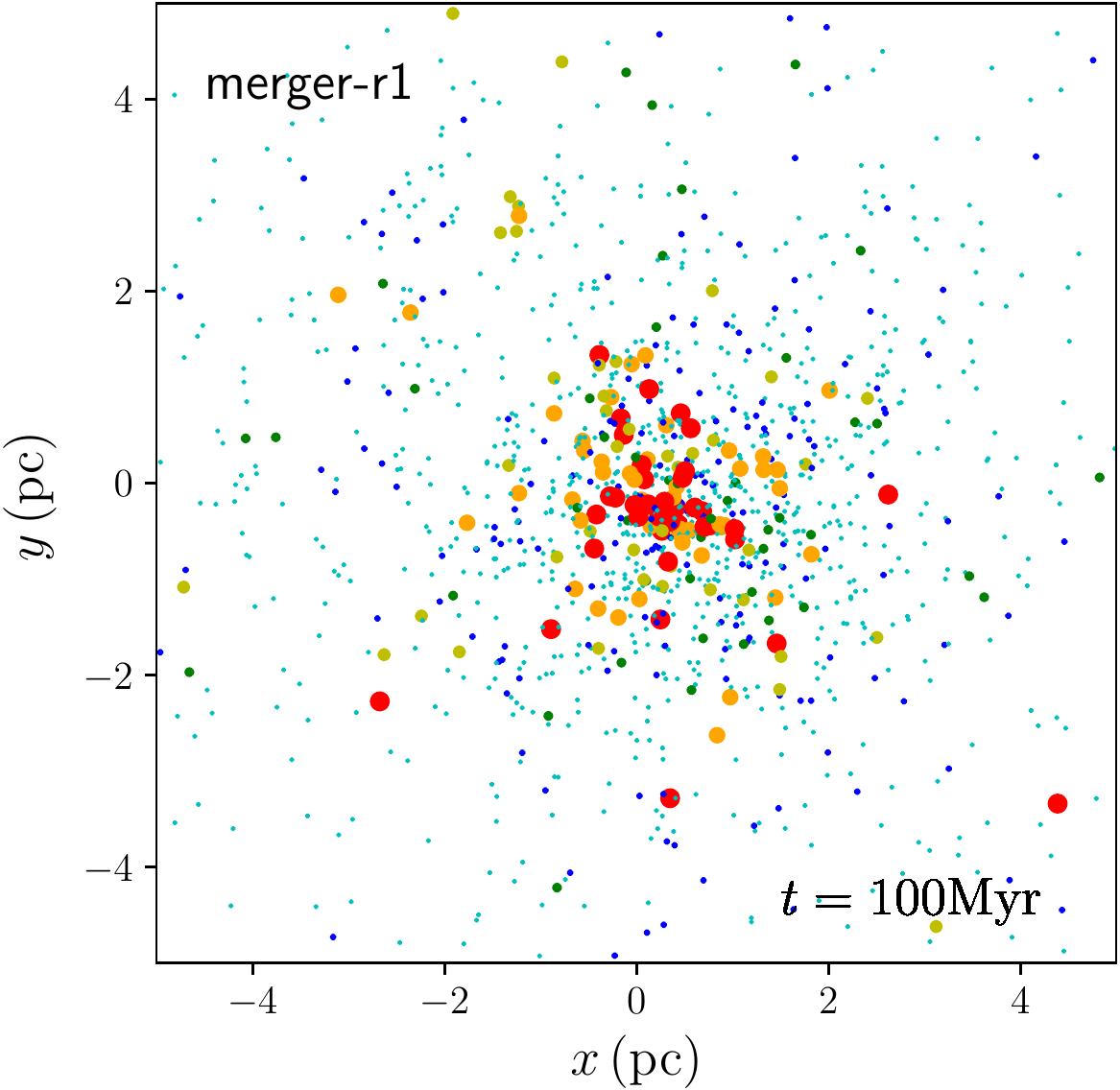}
\includegraphics[width=0.45\hsize]{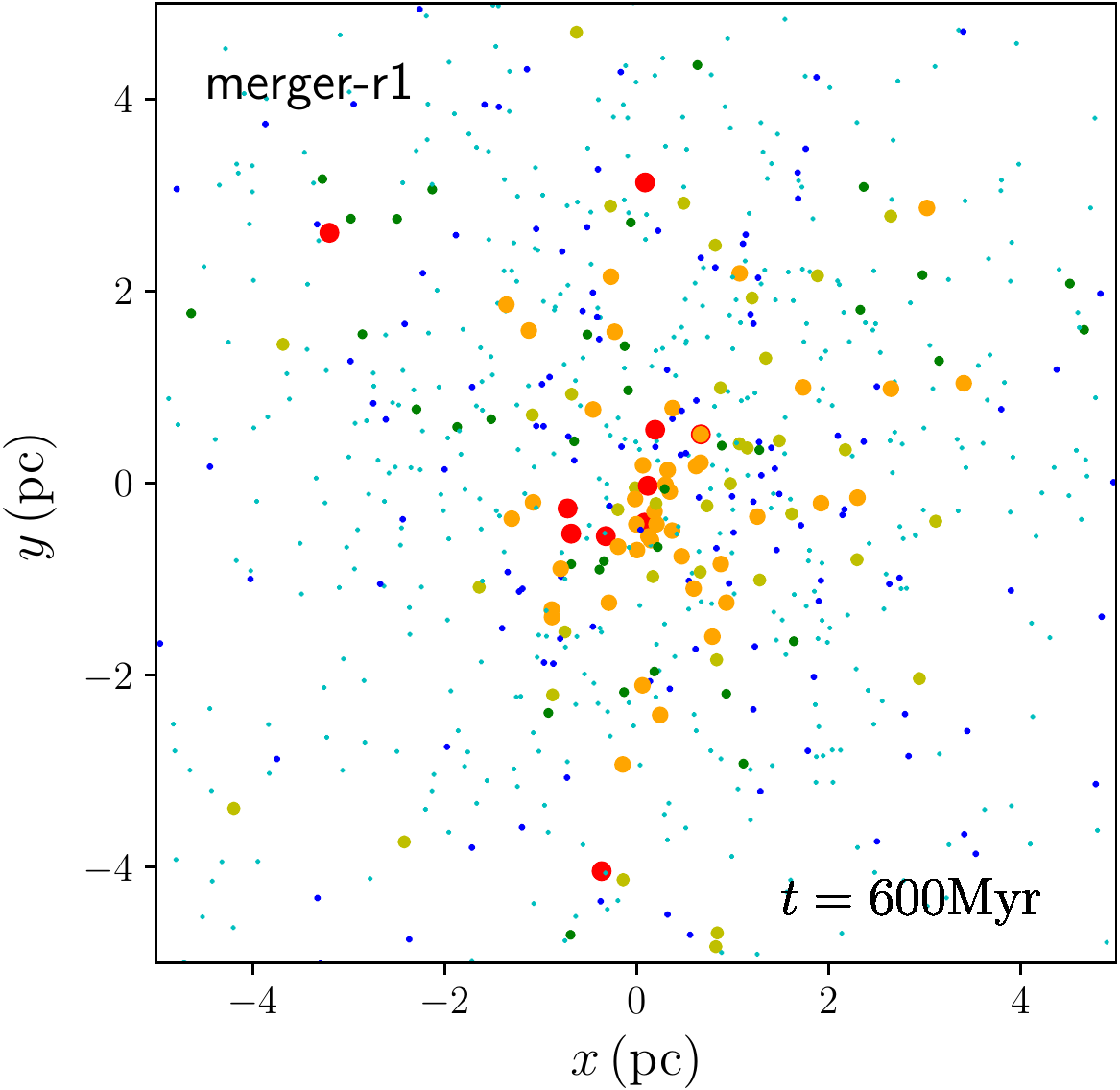}
\caption{Snapshots of model merger-r1. Red, orange, yellow, green, blue, and cyan indicate B-, A-, F-, G-, K-, and M-type stars, respectively.
\label{fig:snap_merger1}}
\end{figure*}

\begin{figure*}[ht]
\centering
\includegraphics[width=0.45\hsize]{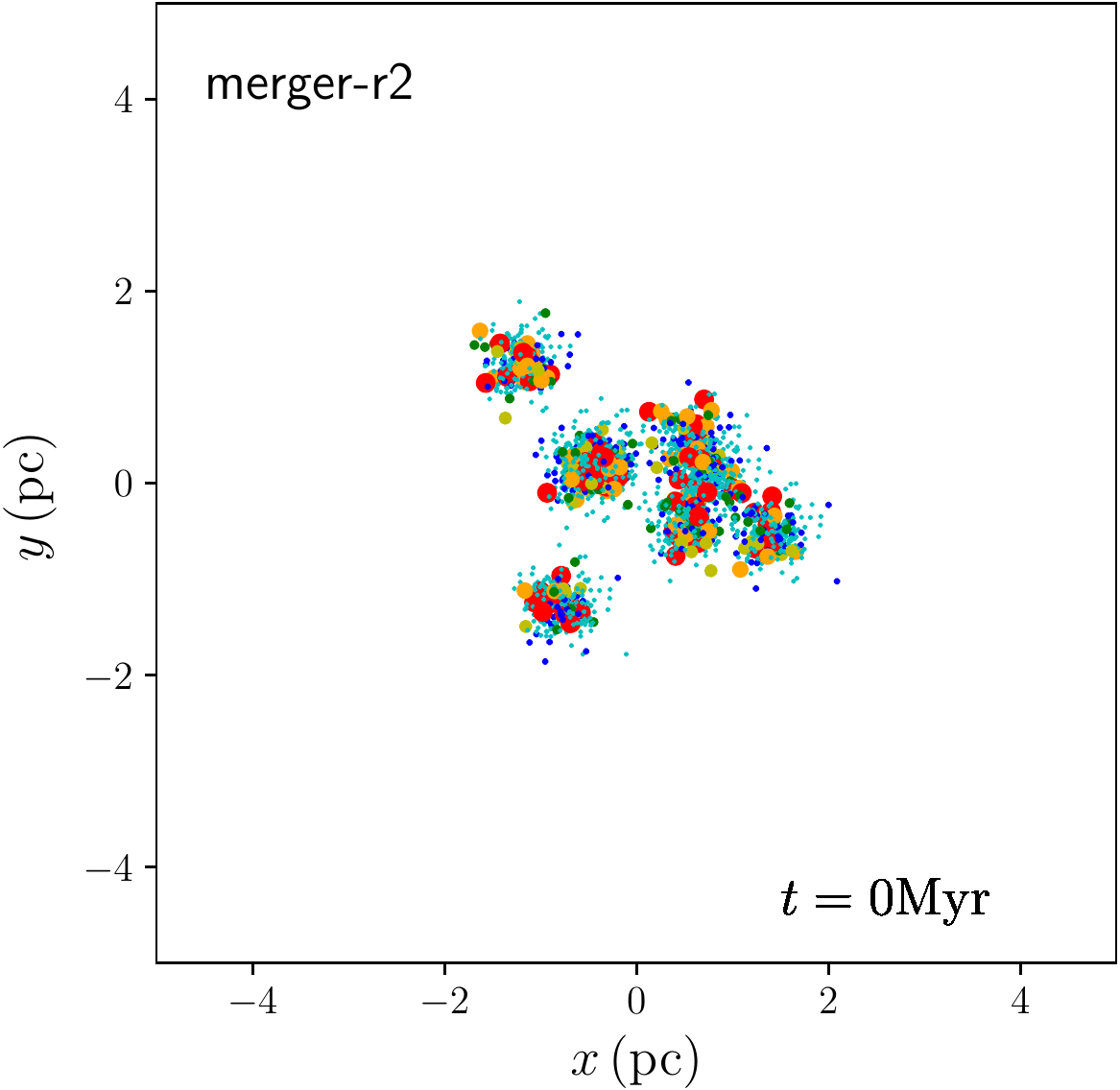}
\includegraphics[width=0.45\hsize]{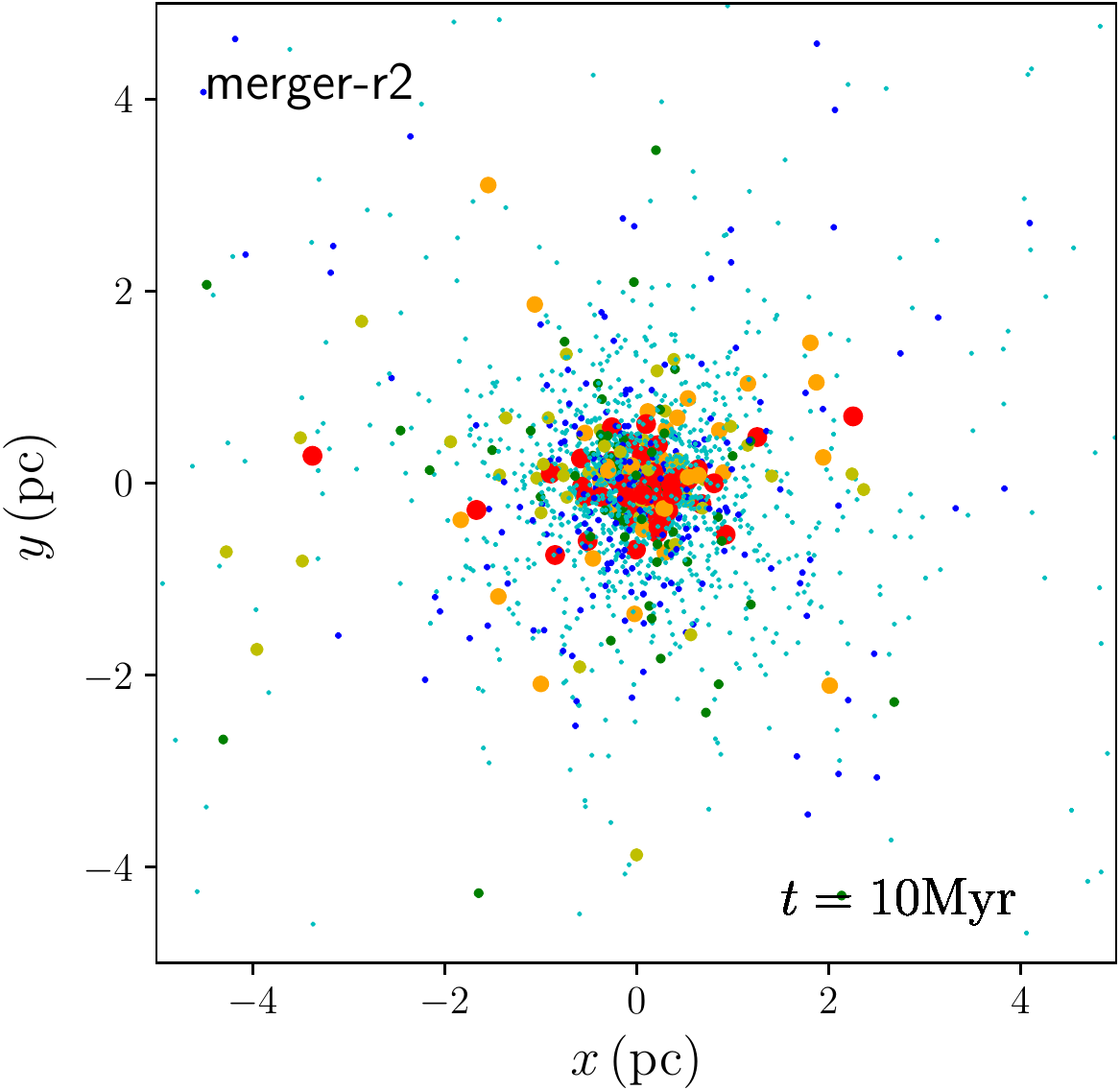}\\
\includegraphics[width=0.45\hsize]{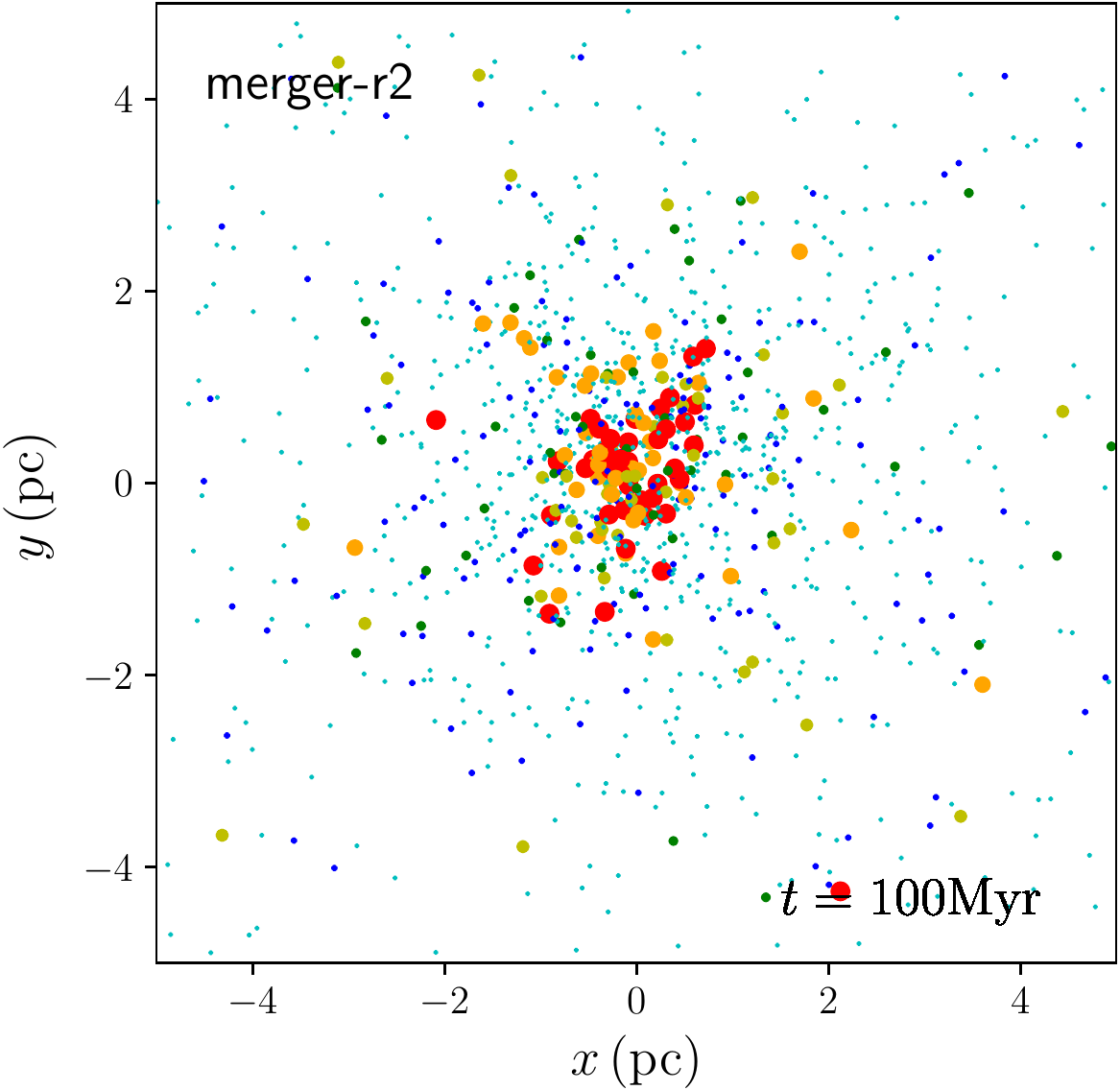}
\includegraphics[width=0.45\hsize]{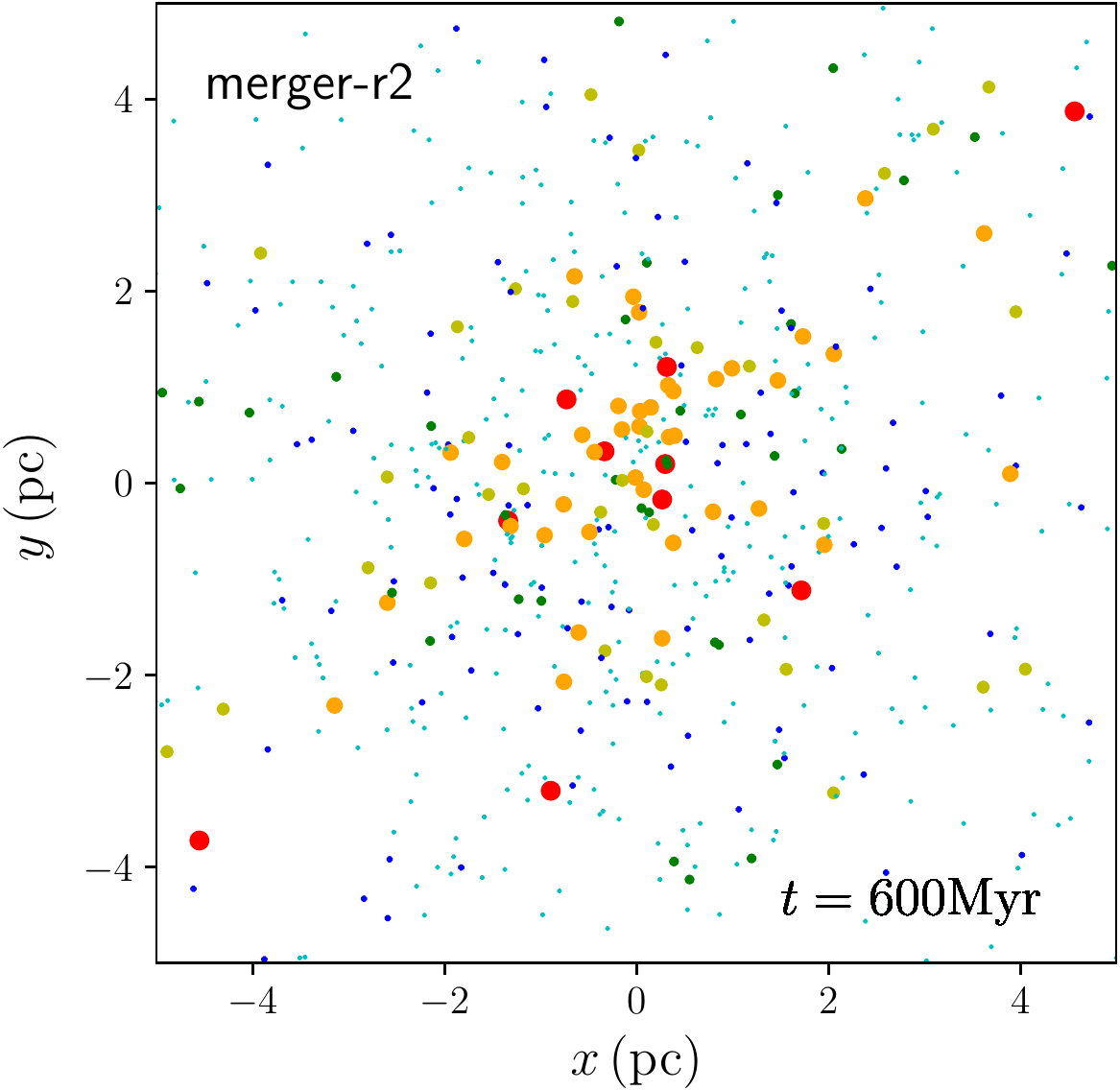}
\caption{Same as Fig. \ref{fig:snap_merger1} but for model merger-r2.
\label{fig:snap_merger2}}
\end{figure*}

\begin{figure*}[ht]
\centering
\includegraphics[width=0.45\hsize]{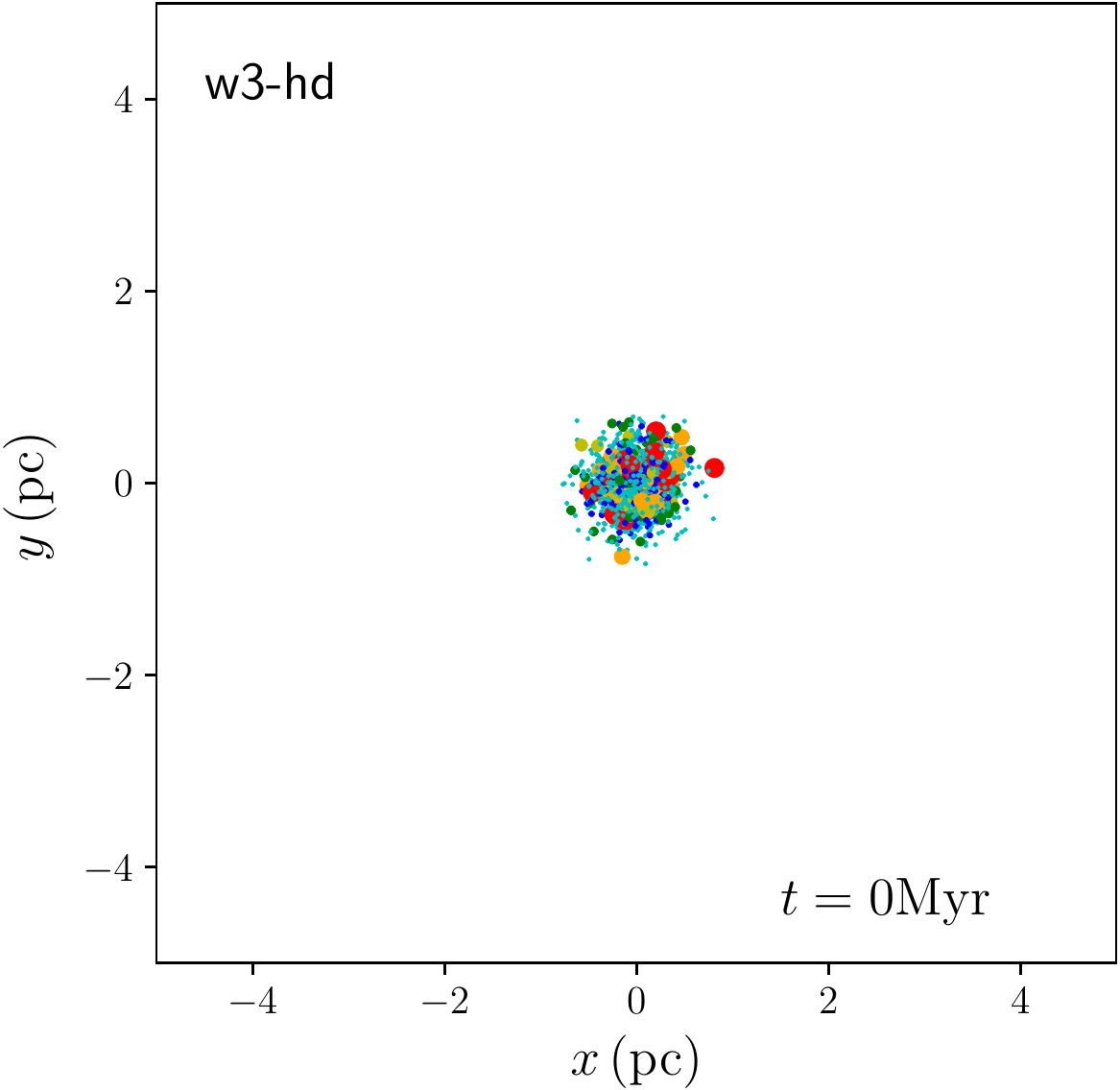}
\includegraphics[width=0.45\hsize]{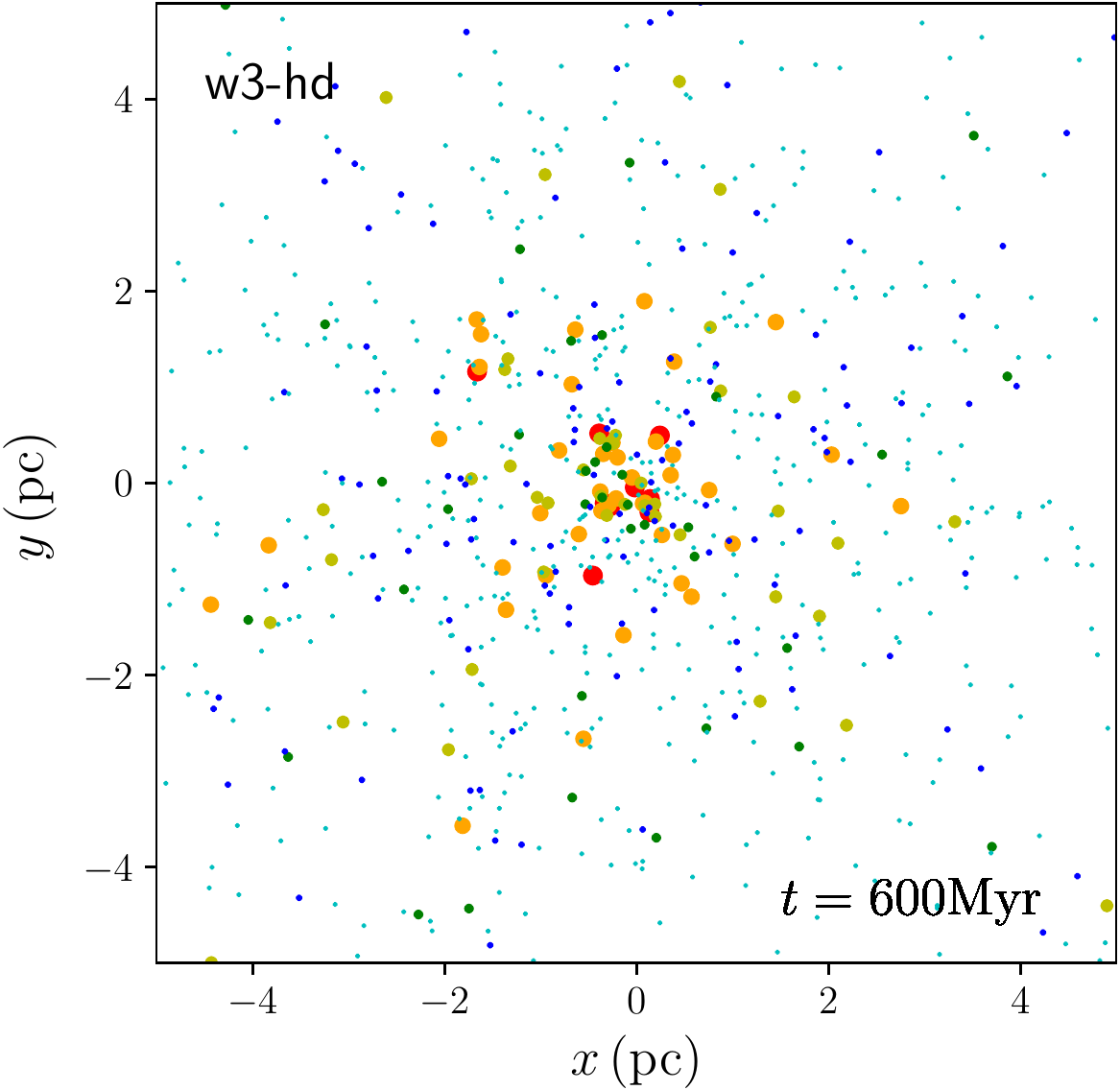}
\caption{Same as Fig. \ref{fig:snap_merger1} but for model w3-hd at 0 and 600\,Myr.
\label{fig:snap_hd}}
\end{figure*}

\begin{figure*}[ht]
\centering
\includegraphics[width=0.45\hsize]{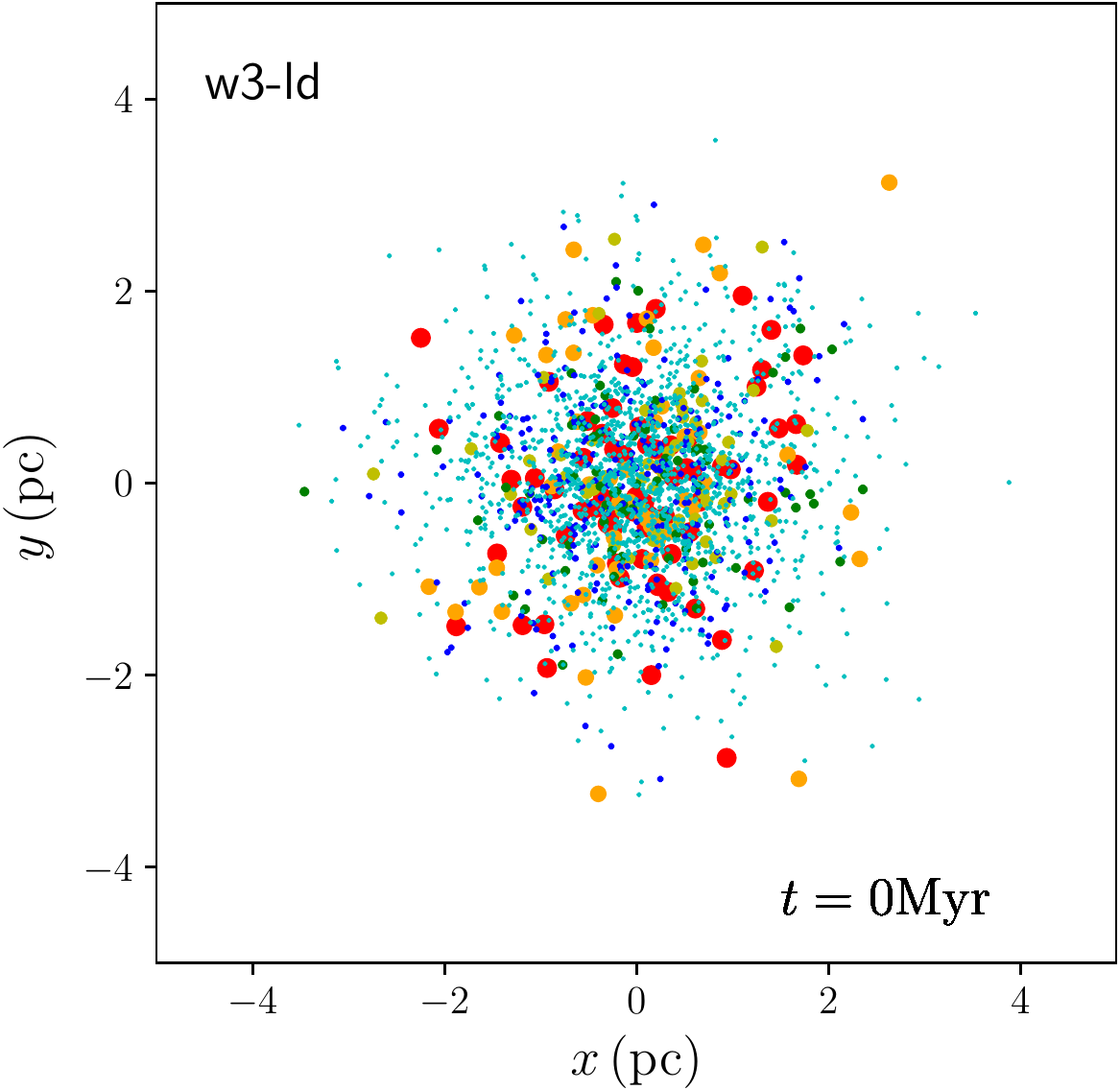}
\includegraphics[width=0.45\hsize]{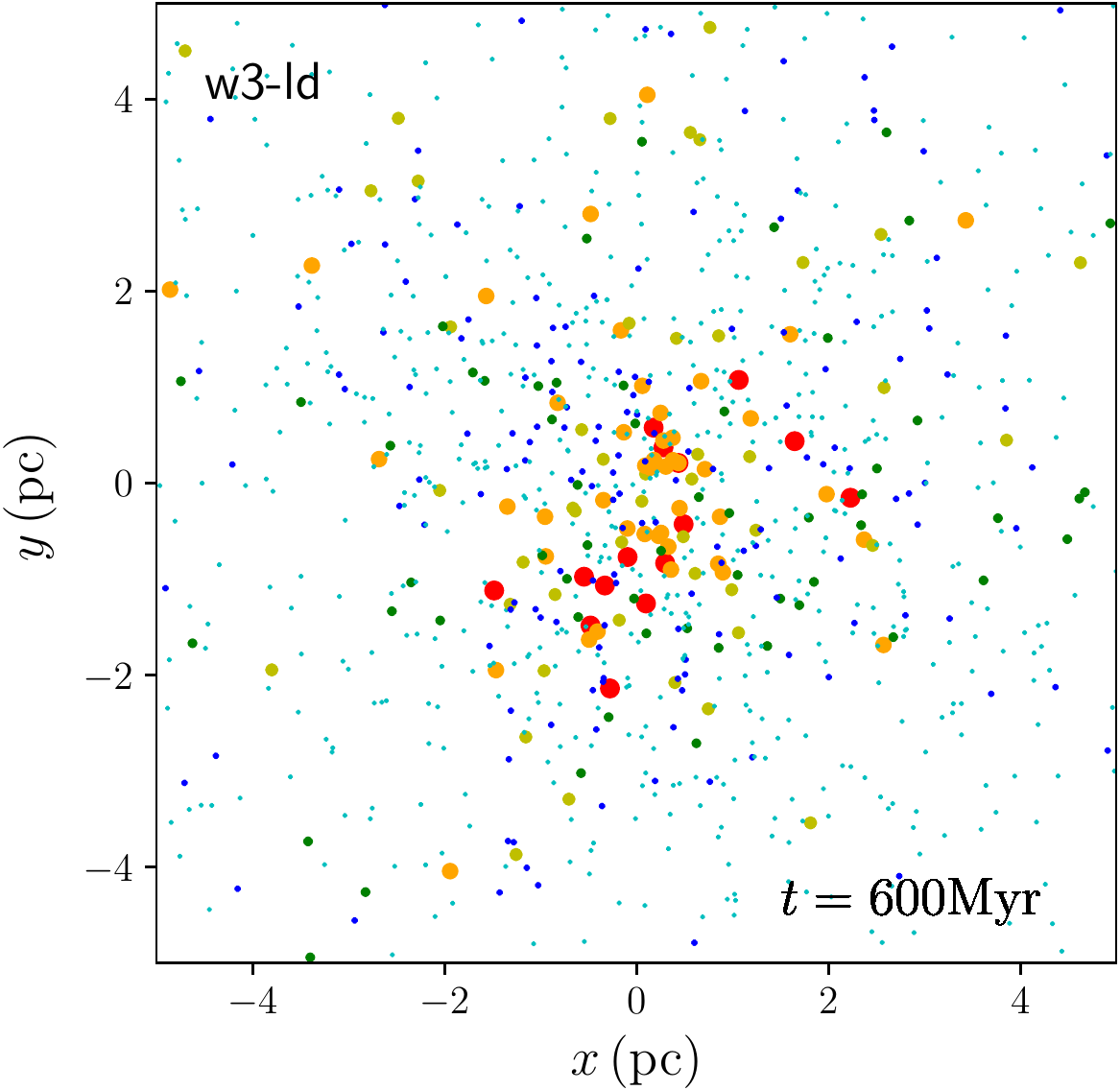}
\caption{Same as Fig. \ref{fig:snap_hd} but for model w3-ld. 
\label{fig:snap_ld}}
\end{figure*}

In this section, we describe the evolution of our merger models.
In Figure \ref{fig:snap_merger1} and \ref{fig:snap_merger2}, we present snapshots for one of models merger-r1 and merger-r2, which initially consist of eight subclusters, at $t=0, 10, 100,$ and 600\,Myr. The subclusters merge within a few Myr and form a cluster. This merger timescale is consistent with that obtained from $N$-body simulations of star clusters with a similar mass but starting from a fractal initial distribution of stars \citep{2014MNRAS.438..620P}. Once the subclusters merged, the cluster density gradually decreases (see Appendix. \ref{dynamical_ev} for the time evolution of the core density). For comparison, we also present the snapshots of one of models w3-hd and w3-ld at 0 and 600\,Myr in Figs. \ref{fig:snap_hd} and \ref{fig:snap_ld}. 

\section{Time evolution of core radius and density\label{dynamical_ev}}

\begin{figure*}
  \centering
  \includegraphics[width=0.45\hsize]{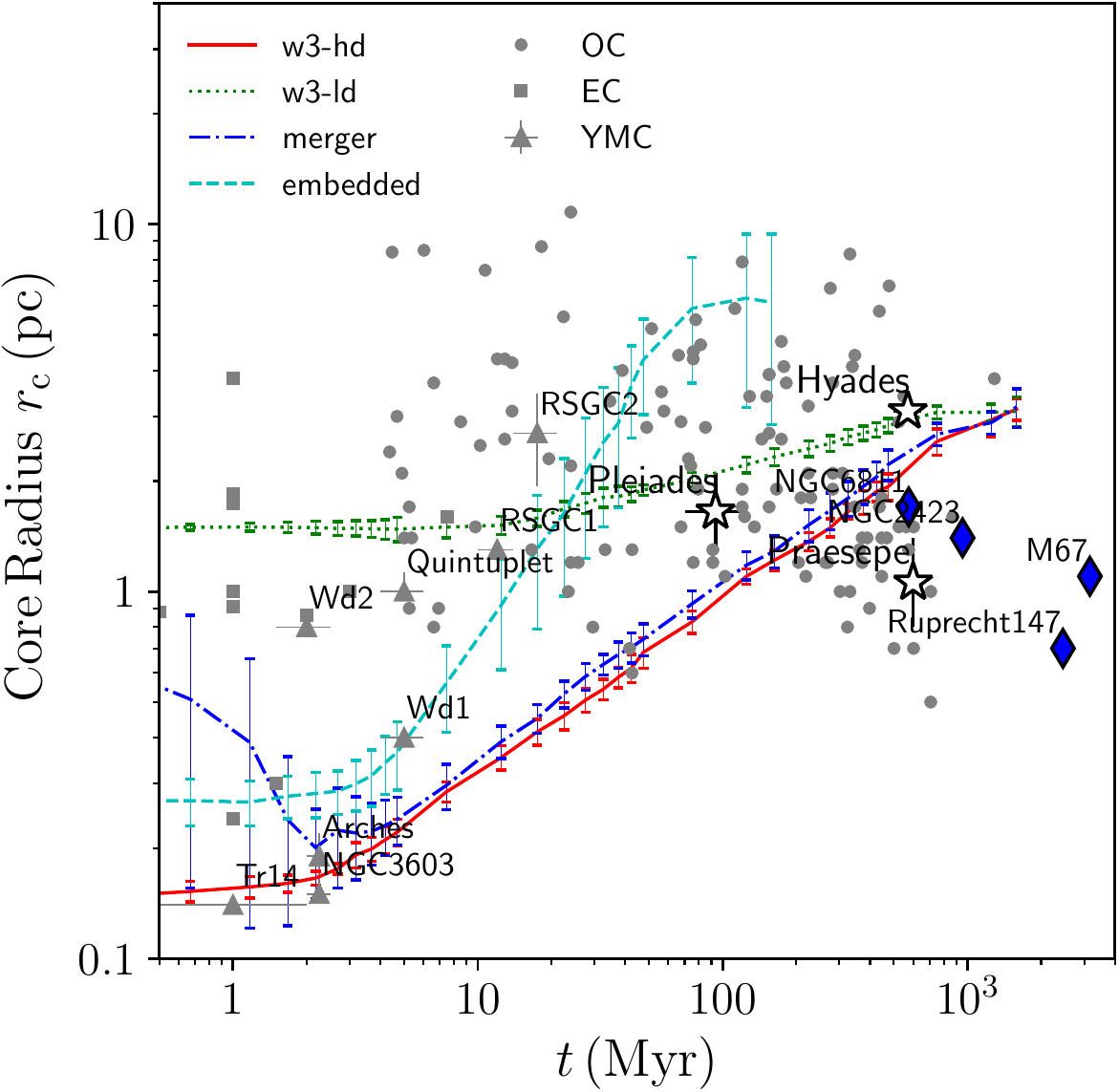}
  \includegraphics[width=0.45\hsize]{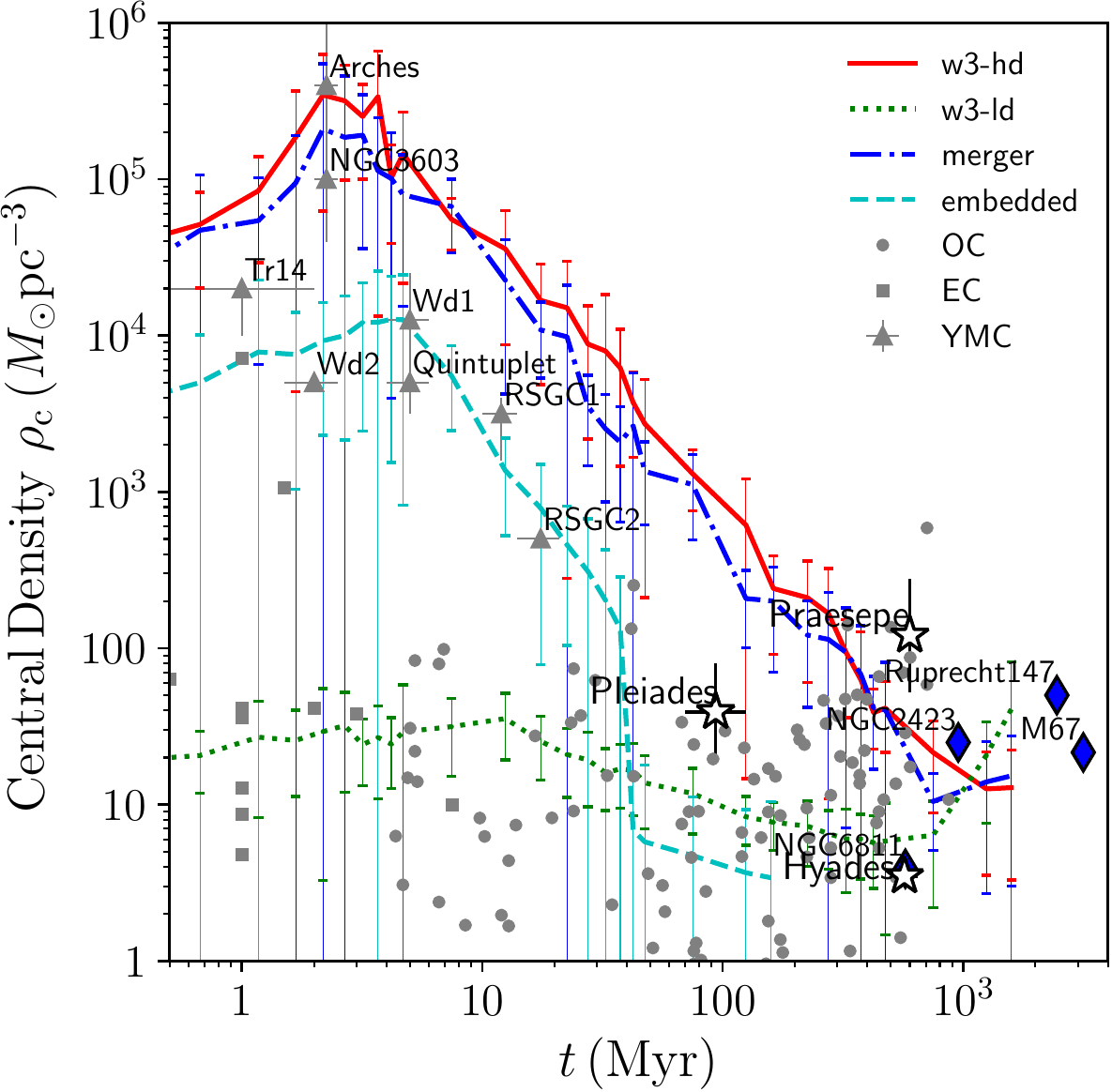}
  \caption{Time evolution of the core radius ($r_{\rm c}$) and central density ($\rho_{\rm c}$) for our cluster models.
    The central density and core radius are averaged over 0.5, 5, 50, and 500\,Myr for $t < 5\,{\rm Myr}, 
    5\,{\rm Myr}\,< t <50\,{\rm Myr}, 50\,{\rm Myr} < t < 500\,{\rm Myr}$, and $t>500$ Myr, respectively. 
    The error bars show the run-to-run variations.
    Each symbol indicates known star clusters near the Solar System: the present-day Pleiades, Praesepe, and Hyades, the other planet-hosting clusters (diamonds: \citet{2007yCat..34680151P}),
    open clusters (OC) with mass of $10^{2} < M <10^{3}\,M_{\odot}$ (circles: \citet{2007yCat..34680151P}),
    embedded clusters (EC) (squares: \citet{2003ARA&A..41...57L,2008IAUS..250..247F,2009A&A...498L..37P}), and
    young massive clusters (YMC) (triangles: \citet{2010ARA&A..48..431P}).
    Following \citet{2009A&A...498L..37P}, we use $\rho_{\rm c}=3M/(4\pi r_{\rm c}^3)$ for observed 
    open clusters.
    We use the average separation from the cluster center given in \citet{2008IAUS..250..247F} 
    as $r_{\rm c}$ because the two values are comparable \citep{2010ARA&A..48..431P}.
    For embedded clusters, we consider the observed radii as their cluster sizes. We therefore adopt the 
    half-mass radius as the radius for our embedded cluster model.
    \label{fig:core_evolution}}
\end{figure*}

Figure \ref{fig:core_evolution} shows the time evolution of the core radii 
and central densities of our cluster models as a function of time.
We also plot observed open clusters (OC) with mass of $10^{2} < M < 10^{3}\,M_{\odot}$ \citep{2007yCat..34680151P}, embedded clusters (EC) 
\citep{2003ARA&A..41...57L,2008IAUS..250..247F,2009A&A...498L..37P},
and young massive clusters (YMC) with $\sim 10^4\,M_{\odot}$
\citep{2010ARA&A..48..431P}.
We find that the Pleiades, Praesepe, Hyades clusters, and other planet-hosting star clusters are typical examples of open clusters, as seen in Figure \ref{fig:core_evolution}.
The behaviors over a few hundred Myr appear to be similar among all the models, 
whereas in the earlier phase of dynamical evolution, the central density of the high density model (w3-hd) increases rapidly and monotonously decreases after a few Myr.
The central density of model w3-hd is as high as those of YMCs during $t  \lesssim 100$\,Myr. This implies that realtively low-mass YMCs might dynamically evolve to open clusters in the future.

Our embedded cluster model (embedded) starts from a compact and dense state, evolving quickly
to a less dense open cluster at $\sim 100$\,Myr because of the short relaxation time of our embedded cluster
This suggests that observed embedded clusters are the ancestors of less massive open
clusters.

The merged cluster evolves dynamically in a manner similar to the high-density cluster and achieves a central density higher than those of individual subclusters, which is the embedded cluster model in our study.
The merged cluster is therefore dynamically more active than single clusters which have a similar initial central density \citep{2013MNRAS.430.1018F}. Indeed, the ejection rate of planets in our merger model is as high as that in our high-density single cluster model, although the initial central density of the subclusters is one order of magnitude lower than that of the high-density single model (see Fig. \ref{fig:f_a_min_type}).

\section{Tidal disruption\label{tidal}}

In this paper, the tidal disruption of star clusters due to the galactic potential is not included.
Stars in the outskirt of a cluster halo are stripped from the cluster due to the tidal effect. 
In this section, we demonstrate that the tidal effect on dynamical evolution of stars in our star clusters is limited. 

For a halo model, we adopt the NFW profile \citep{1997ApJ...490..493N}
with the concentrate parameter of the halo, $c=10$. 
The mass contained within the virial radius of the halo is $6.0 \times 10^{11}M_{\odot}$.
The circular speed at 8\,kpc away from the Galactic center is 200 \kms.

\begin{figure*}
\centering
\includegraphics[width=0.3\hsize]{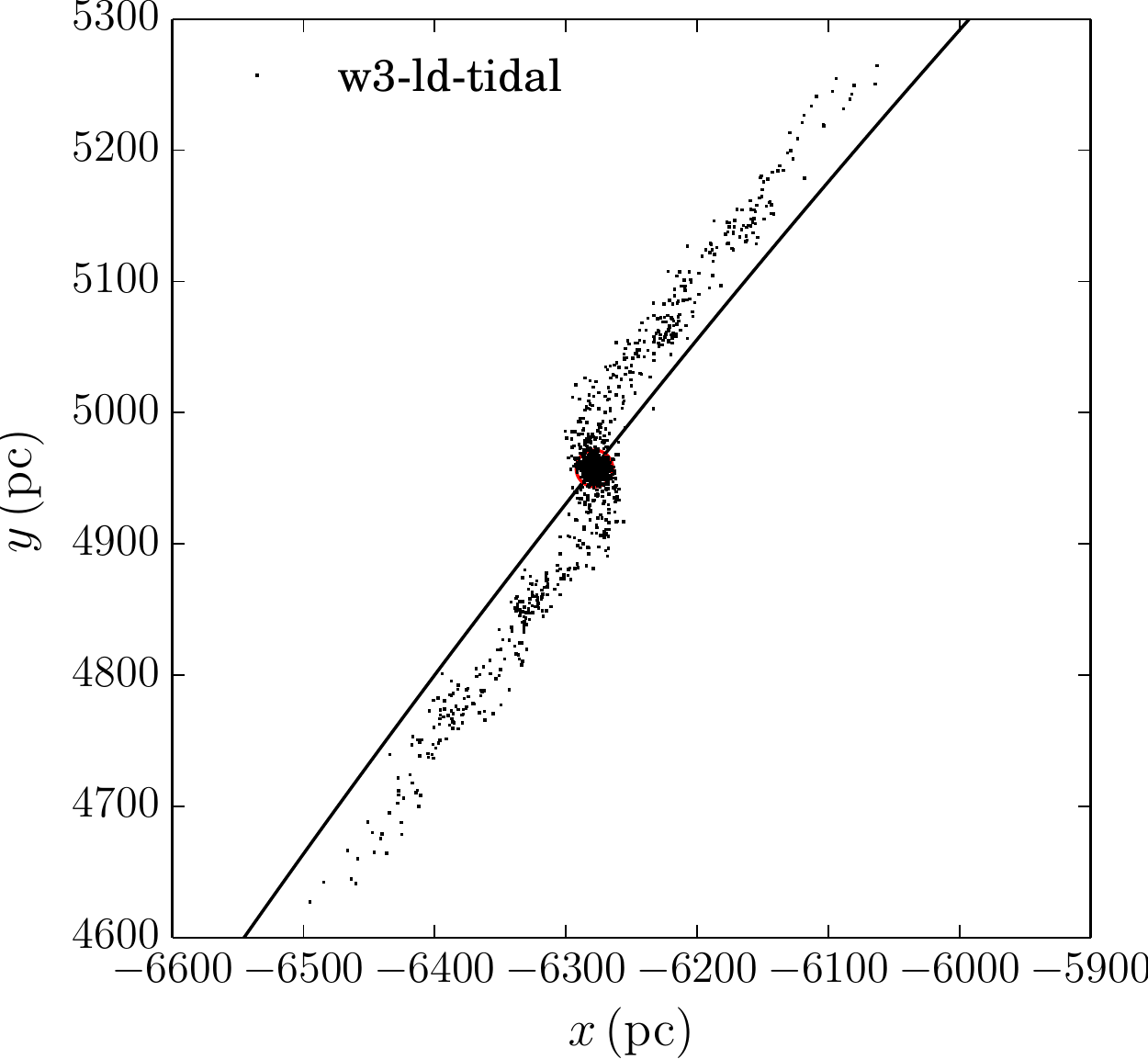}
\includegraphics[width=0.3\hsize]{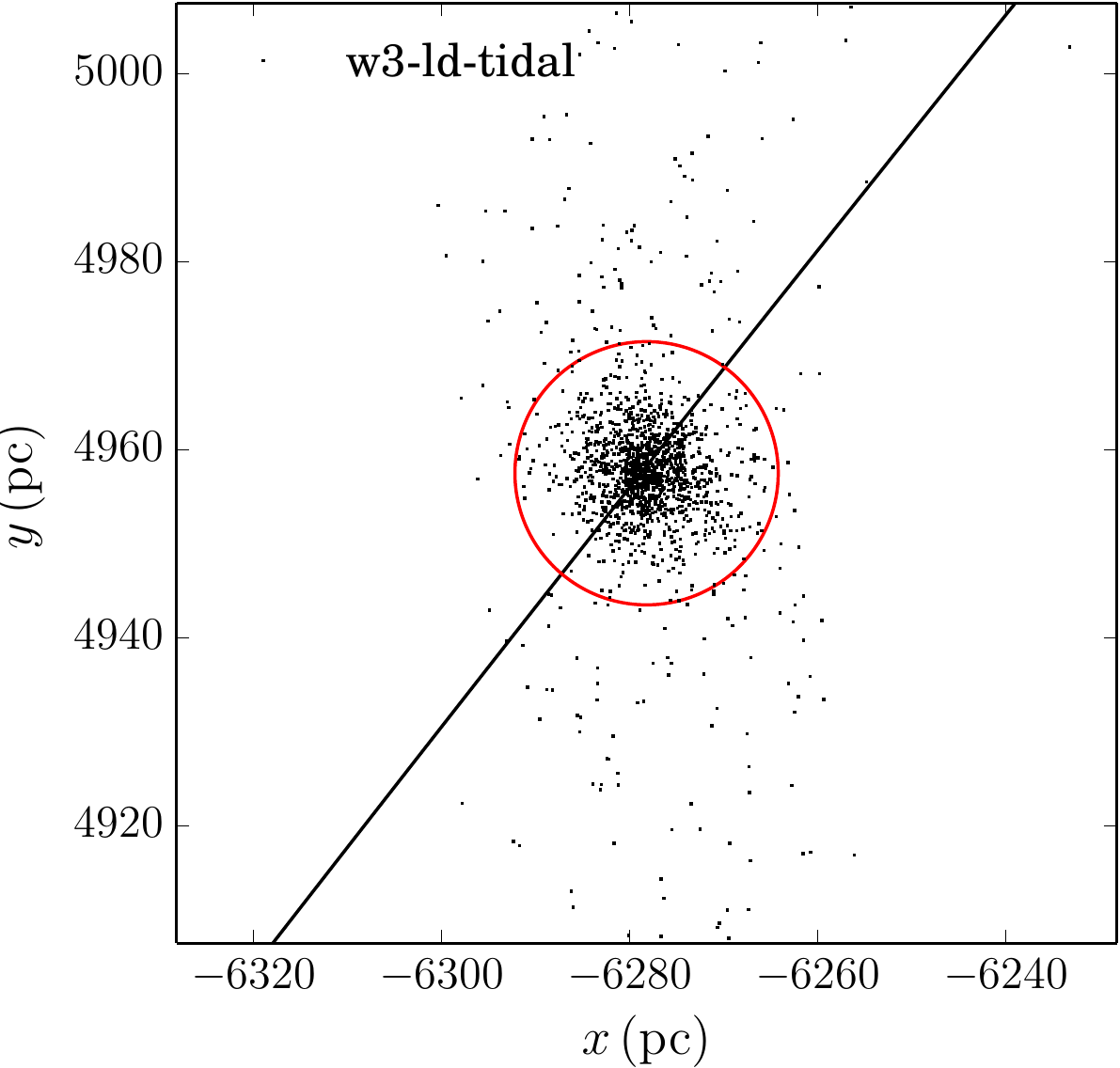}
\includegraphics[width=0.3\hsize]{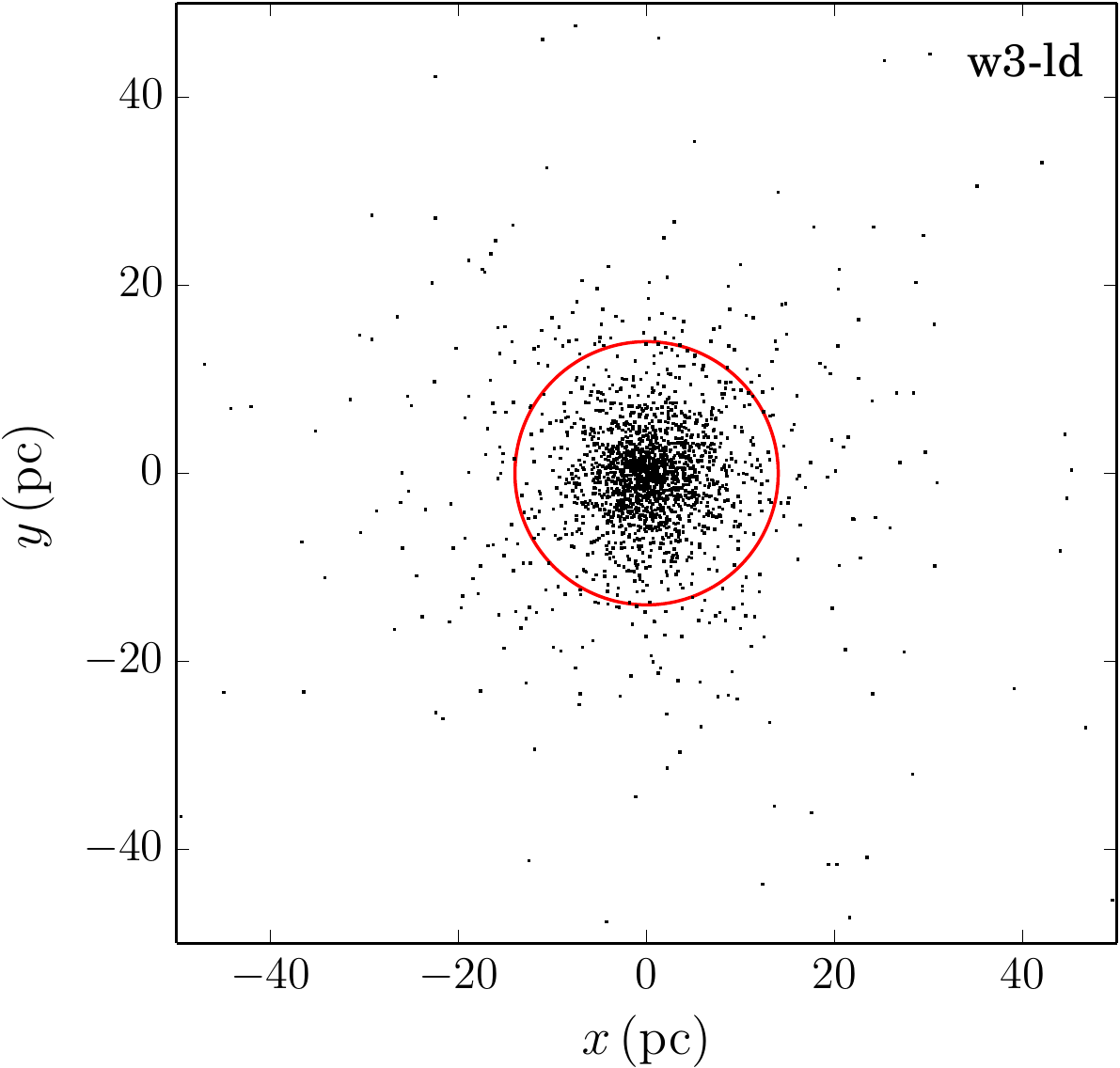}
\caption{
Projected positions of stars (black dots) at 600 Myr for models with the tidal force (w3-ld-tidal: 
left and middle) and without the tidal force (w3-ld: right). The $xy$ ranges for the middle and right panels are the same, while the range for the left panel is larger. Black lines indicate the cluster's circular orbit at 8\,kpc away from the Galactic center, and red circles indicate the tidal radius, 14\,pc.\label{fig:pos}}
\end{figure*}

We simulated dynamical evolution of stars in a star cluster, including the tidal field of the Galactic halo.
Figure \ref{fig:pos} presents projected positions of stars at 600 Myr in a low-density cluster model with the tidal force (w3-ld-tidal: left and middle panels.
We see tidal tails of stars along the cluster's orbit in the left panel of Figure \ref{fig:pos}. Nevertheless, most of stars still stay within the tidal radius (14\,pc) denoted by a red circle (see the middle panel
in Figure \ref{fig:pos}). For comparison, we also present the snapshot of 
the same model but without a tidal force (w3-ld: the right pane) in Figure 
\ref{fig:pos}). Although the outskirts of the halo overflow the tidal radius,
most of stars remain within the tidal radius. 

\begin{figure*}[ht]
\centering
\includegraphics[width=0.45\hsize]{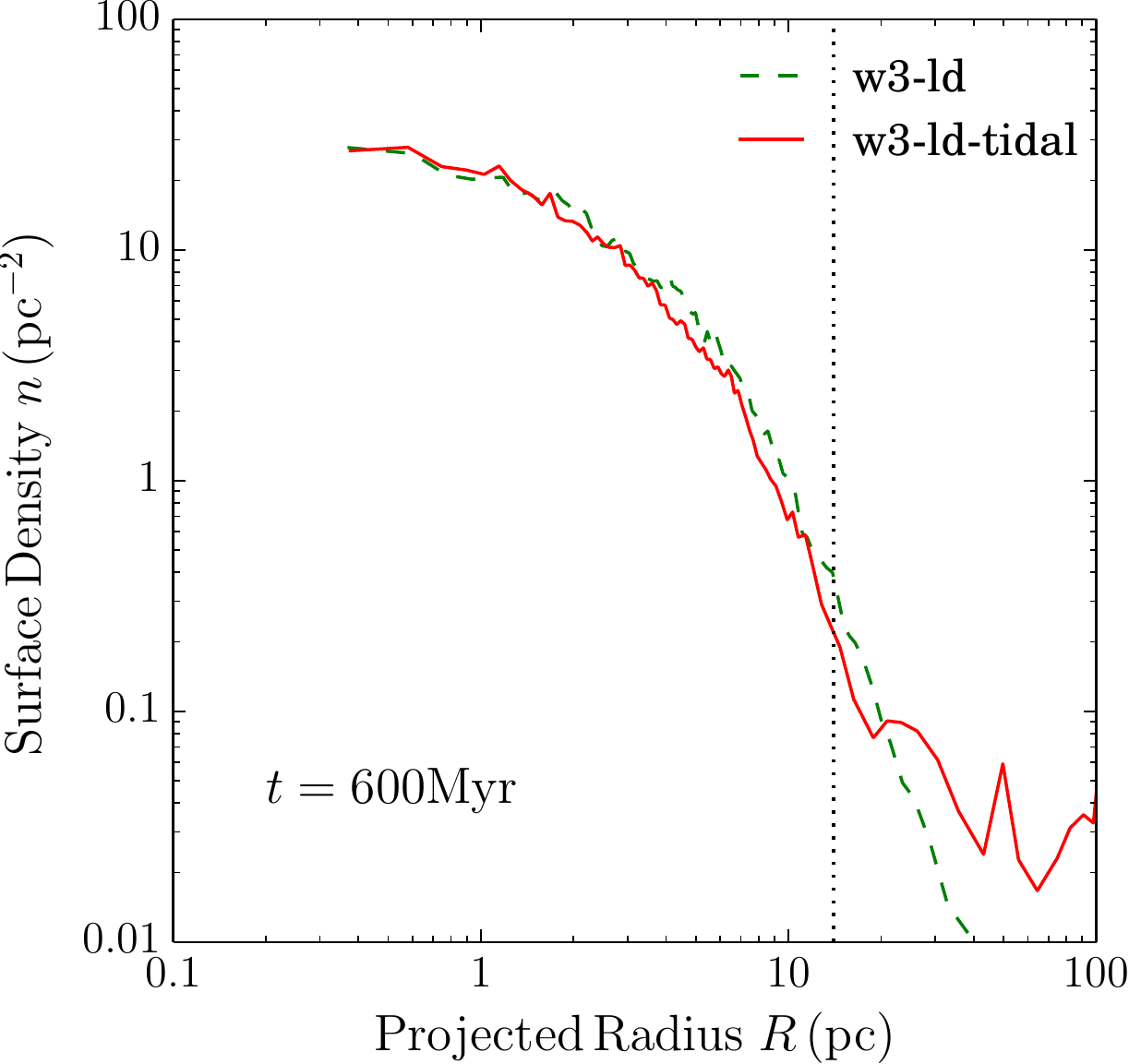}
\centering
\includegraphics[width=0.45\hsize]{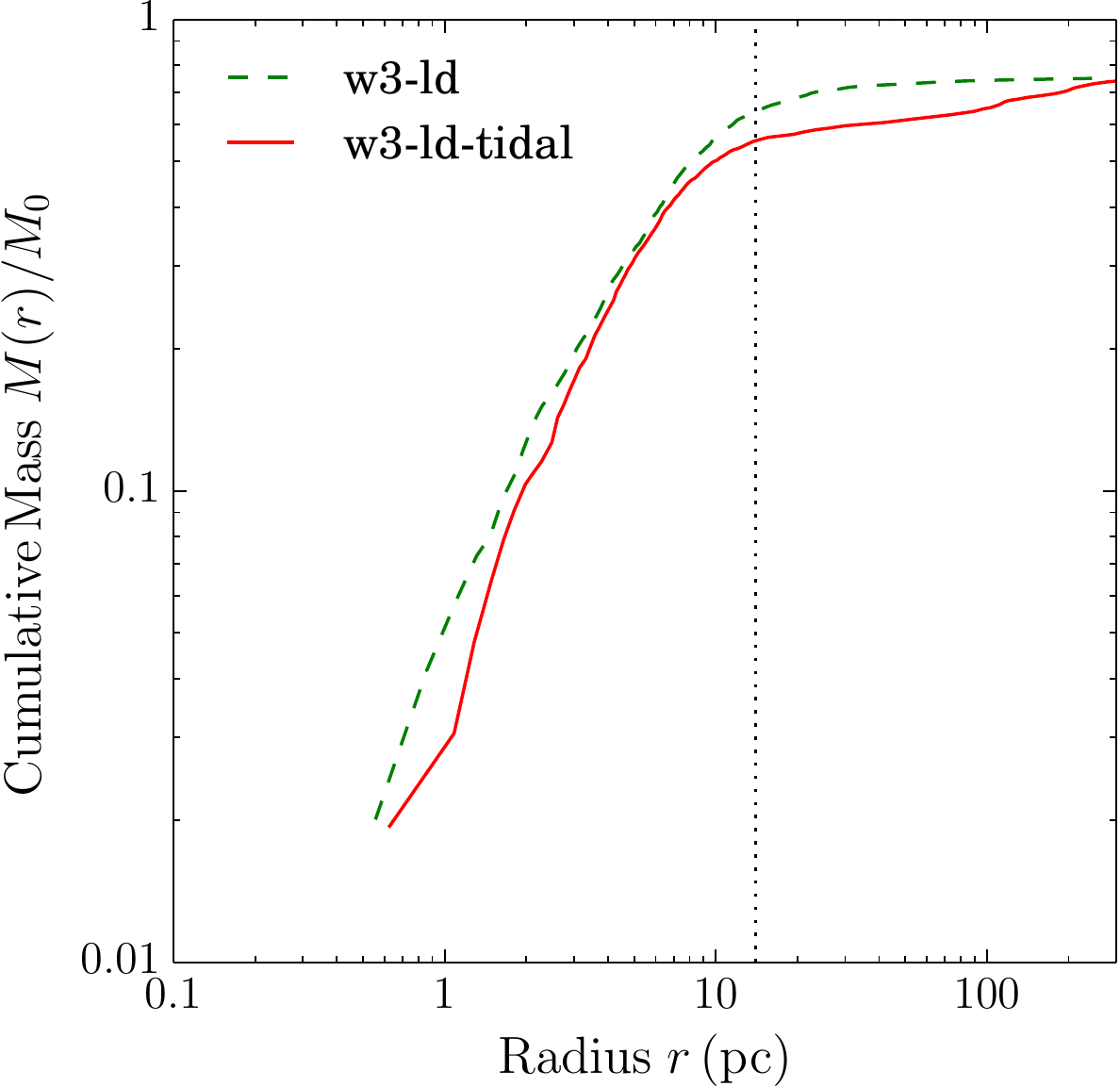}
\caption{Surface density profile (left) and
cumulative mass distribution form the cluster center (right) for models w3-ld (green dashed curve) and w3-ld-tidal (red solid curve). The cumulative mass distributions are normalized by the initial mass of the cluster. Vertical dotted line indicates the tidal radius (14 pc) at 8 kpc from the
Galactic center. The surface density profile are averaged over five snapshots around 600 Myr.
\label{fig:prof}}
\end{figure*}

We compare a radial distribution of stars within the tidal radius between the two models. 
Figure \ref{fig:prof} shows the the surface density profile (left) and cumulative mass (right) as a function of radius for models w3-ld and w3-ld-tidal at 600\,Myr.
The surface density and the cumulative mass distributions of the cluster halo for the model with the tidal field are slightly lower those for the isolated model, whereas both profiles are similar between the two clusters.
We note that about 20\% of the initial cluster mass is lost due to the stellar evolution.
Although a tidal effect enhances mass loss near the tidal radius, 
most of the cluster's mass are still loaded within the tidal radius.

\end{appendix}

\end{document}